\documentclass[a4paper,11pt]{article}

\usepackage{jheppub}

\usepackage[T1]{fontenc} 
\usepackage{dsfont}
\usepackage{framed}
\usepackage{mathrsfs}
\usepackage{makecell}

\usepackage{caption}
\usepackage{subcaption}
\usepackage{dsfont}
\usepackage{float}

\usepackage{graphicx}

\usepackage[compat=1.1.0]{tikz-feynman}
\usepackage{silence}
\usetikzlibrary{arrows.meta,decorations.markings}

\tikzfeynmanset{
  fermion/.style={
    line width=0.75pt,
    decoration={
      markings,
      mark=at position 0.55 with {
        \arrow{Stealth[length=2mm,width=2mm]}
      }
    },
    postaction=decorate,
  },
  every photon={line width=0.75pt},
  every scalar={line width=0.75pt},
}

\tikzfeynmanset{
  fermion shifted/.style={
    line width=0.75pt,
    decoration={
      markings,
      mark=at position #1 with {
        \arrow{Stealth[length=2mm,width=2mm]}
      }
    },
    postaction=decorate,
  },
  fermion shifted/.default=0.7,
}

\allowdisplaybreaks[1]

\def\({\left(}
\def\){\right)}
\def\[{\left[}
\def\]{\right]}
\def\<{\langle}
\def\>{\rangle}

\def\nn{\nonumber\\}

\def\bf{\mathbf}
\def\rm{\mathrm}

\def\cal{\mathcal}

\def\pa{\partial}
\def\ra{\rightarrow}

\def\a{\alpha}

\def\g{\gamma}
\def\G{\Gamma}
\def\d{\delta}
\def\D{\Delta}
\def\e{\epsilon}

\def\Th{\Theta}

\def\l{\lambda}
\def\L{\Lambda}
\def\m{\mu}
\def\n{\nu}
\def\x{\xi}

\def\p{\pi}

\def\f{\phi}

\def\r{\rho}

\def\s{\sigma}

\def\ps{\psi}
\def\Ps{\Psi}

\def\lra{\leftrightarrow}

\def\til{\tilde}

\begin{document}

\title{Central charges $C_J$ and $C_T$ in QED$_d$-GNY model and scalar QED$_d$}

\author{Yongwei Guo}
\author{and Zhijin Li}
\emailAdd{guoyw33@seu.edu.cn,  zhijin\_li@seu.edu.cn}

\affiliation{School of Physics, Southeast University, Nanjing, 210096, China}
\affiliation{Shing-Tung Yau Center, Southeast University, Nanjing, 210096, China}

\abstract{We compute the leading-order $1/N$ corrections to the central charges $C_J$ and $C_T$ in the conformal QED$_d$-Gross-Neveu-Yukawa (GNY) model and the scalar QED$_d$ in $d$ dimensions.
The scaling dimensions of the lowest adjoint bilinear scalars are obtained to order $O(1/N)$ for general $d$. In $d=3$, the $U(1)$ Abelian gauge theory possesses a topological $U(1)$ global symmetry, and we evaluate the central charge $C_J^{\text{top}}$ of the topological symmetry current  to  subleading order in the $1/N$ expansion. Our interest in these theories is primarily motivated by their potential connection to the $SO(5)$ symmetric deconfined quantum critical point (DQCP). We compare the large $N$ results for the central charges $C_J$ and $C_T$ with the conformal data of the $SO(5)$ DQCP obtained from fuzzy sphere and conformal bootstrap. The large $N$ predictions of the QED$_3$-GNY model are found to be in reasonable agreement with the nonperturbative estimates for the $SO(5)$ DQCP.} 
	
\maketitle 

\section{Introduction}

In a local conformal field theory (CFT) with a continuous global symmetry, the two-point functions of the conserved current $\<J_\mu J_\nu\>$ and stress-energy tensor $\<T_{\mu\nu}T_{\rho\sigma}\>$ have fixed functional forms \cite{Osborn:1993cr}.
The coefficients $C_J$ and $C_T$, namely the central charges in the two-point functions are physically meaningful and cannot be scaled away completely, since the insertions of $J_\mu$ or $T_{\mu\nu}$ into other correlation functions are constrained by Ward identities.
The central charges appear in a wide variety of contexts.
One of the most famous examples is Zamolodchikov's $c$-theorem \cite{Zamolodchikov:1986gt}, where the monotonic function of the renormalization scale coincides with $C_T$ at conformal fixed points.
Some fundamental applications of the central charges can be found in condensed matter systems, where $C_J$ and $C_T$ correspond to the universal electrical conductivity and viscosity in the high frequency regime near quantum critical points \cite{Cha:1991amy,Huh:2013vga,Witczak-Krempa:2013nua,Huh:2014eea,Katz:2014rla,Witczak-Krempa:2015pia,Lucas:2016fju,Lucas:2017dqa}.

We are interested in $C_J$ and $C_T$ in the IR fixed points of the QED$_d$-Gross-Neveu-Yukawa (GNY) model and scalar QED$_d$. 
In $d=3$ dimensions, these theories arise in the studies of deconfined quantum critical points (DQCPs) and dualities (see \cite{Senthil:2018cru} for a review).
In particular, the two-flavor QED$_3$-GNY$_+$ model\footnote{The $+$ sign in the subscript is to distinguish the model from a similar theory with a smaller flavor symmetry \cite{Boyack:2018zfx,Benvenuti:2018cwd,Benvenuti:2019ujm}.
See also the beginning of Section \ref{QED-GNY model}.
We will only make this distinction in $d=3$.} and two-flavor scalar QED$_3$ are conjectured to flow to the same IR fixed point with an $SO(5)$ symmetry enhancement \cite{Wang:2017txt}, which describes the N\'eel to valence-bond-solid phase transition of spin-$1/2$ quantum magnets on a square lattice \cite{Senthil:2003eed,Senthil:2004fuw}.
More broadly, 3D Abelian gauge theories have attracted considerable attention  \cite{Chester:2016wrc,Nakayama:2016jhq,Chester:2017vdh,Li:2018lyb,Reehorst:2019pzi,Li:2020bnb,Li:2021emd,Albayrak:2021xtd,He:2021xvg,He:2021sto,He:2023ewx,Chester:2023njo,Chester:2025uxb,li2026bootstrap} in the modern conformal bootstrap program \cite{Rattazzi:2008pe,Poland:2018epd}.
 
 Our understanding of the nature of DQCP has been improved dramatically. Originally the $SO(5)$ symmetric DQCP was proposed to have no relevant $SO(5)$ singlet scalar and the critical point can be reached by tuning the $SO(5)$ crossover parameter only \cite{Senthil:2003eed,Senthil:2004fuw}.  The numerical simulations did not find discontinuity at the DQCP, indicating the phase transition is continuous, though the critical indices are drifting with system sizes \cite{Nahum2015x}. The conformal bootstrap bounds provide strong restrictions on the putative CFT data of DQCP. In particular, the critical indices estimated from the numerical simulations are not consistent with the requirement that the lowest $SO(5)$ singlet scalar being irrelevant \cite{Nakayama:2016jhq,Poland:2018epd,Li:2018lyb}. Consequently, the $SO(5)$ DQCP was widely expected to be of the pseudo-criticality \cite{Nahum2015x,Wang:2017txt,Gorbenko:2018ncu}. This was further supported by a recent fuzzy sphere study, in which rich CFT data of the $SO(5)$ DQCP has been generated using the fuzzy sphere regularization \cite{Zhou:2023qfi}. However, it has been observed that in an alternative multicriticality scenario of the $SO(5)$ DQCP \cite{Zhao_2020} in which the lowest $SO(5)$ singlet scalar is relevant, part of the perturbative and numerical CFT data is consistent with the bootstrap constraints and nearly saturates the bootstrap bounds, indicating the critical indices of the DQCP may correspond to a full fledged unitary CFT \cite{Li:2018lyb,Chester:2023njo}.
The large scalar quantum Monte Carlo (QMC) study has reached unprecedent precision for the DQCP \cite{Takahashi2024}. The phase transition was observed along a weakly first-order line, which is conjectured to be ended at a multicritical fixed point. The critical indices were estimated near the presumed multicritical fixed point with tiny error bars. Surprisingly the QMC results and the fuzzy sphere data are unified in a special conic structure in the bootstrap bounds \cite{li2026bootstrap}. The results strongly support that the $SO(5)$ DQCP is related to a unitary CFT with a relevant $SO(5)$ singlet scalar, i.e., a multicritical fixed point.

The motivation of this work is to compute fundamental CFT data of the $SO(5)$ DQCP using large $N$ expansion to provide helpful inputs and cross-checks for the nonperturbative results obtained from bootstrap, QMC and fuzzy sphere approaches.
The QED$_3$-GNY model has been studied using the $4-\epsilon$ expansion in \cite{Janssen:2017eeu, Ihrig:2018ojl,Zerf:2018csr} and large $N$ expansion \cite{Gracey:2018fwq, Boyack:2018zfx}, in which the theory is shown to admit an infrared fixed point and the scaling dimensions of low-lying scalar operators have been estimated.
However it should be noted that the 3D theory obtained from the $4-\epsilon$ dimensional continuation is not exactly the QED$_3$-GNY$_+$ model appearing in the duality web \cite{Boyack:2018zfx}.
In this work, we will compute the central charges $C_J$ and $C_T$ of QED$_3$-GNY$_+$ model. For the $N_f=2$ scalar QED$_3$, which is conjectured to be dual to the QED$_3$-GNY$_+$ model, the central charge $C_J$ has been computed in \cite{Huh:2013vga}, and we will compute its central charge $C_T$ in this work. In $d=3$, the $U(1)$ gauge theory admits a topological conserved current 
$J_{\rm{top}}^\m\propto \epsilon^{\mu\nu\rho}F_{\nu\rho}$, where $F_{\nu\rho}$ is the field strength tensor. 
The coefficient in its two-point function $\langle J_{\rm{top}}^\m J_{\rm{top}}^\n \rangle$ is denoted by $C_J^{\text{top}}$ and we will compute its leading $1/N$ correction.

In the large $N$ expansion, the leading-order results are given by the generalized free field theories, of which the conformal data can be solved exactly.
To access the theories at finite $N$, we use Feynman diagrammatic methods\footnote{The relevant techniques for computing Feynman diagrams are described in the appendices of \cite{Diab:2016spb}.} to compute the $1/N$ corrections.
Although the most physically interesting case is in $d=3$, it is useful to keep the dimension $d$ general and study the central charges as functions of $d$ \cite{Petkou:1995vu,Diab:2016spb,Giombi:2016fct}.\footnote{See also \cite{Gracey:2017fzu,Gracey:2018fwq,Gracey:2018qba,Gracey:2021yyl} for recent work on critical exponents for various theories in general $d$.}
We will assume
\begin{align}
    2<d<4
\end{align}
throughout the paper.
The new results for $C_J$, $C_T$, and $C_J^{\text{top}}$ are summarized below.

\subsection{Summary of results}
\label{Summary of results}

The central charges $C_J$ and $C_T$ are defined by \cite{Osborn:1993cr}
\begin{align}
    \<J_\m^a(x)J_\n^b(0)\>=&\frac{C_J\,\d^{ab}}{x^{2(d-1)}}I_{\m\n}(x)\,,\label{CJ def} \\
    \<T_{\m\n}(x)T_{\r\l}(0)\>&=\frac{C_T}{x^{2d}}\cal I_{\m\n,\r\l}(x)
    \,,\label{CT def}
\end{align}
where the tensor structures are
\begin{align}
    I_{\m\n}&=\d_{\m\n}-2\frac{x_\m x_\n}{x^2}\,,\nn
    \cal I_{\m\n,\r\l}(x)&=\frac 1 2 \big(I_{\m\r}(x)I_{\n\l}(x)+I_{\m\l}(x)I_{\n\r}(x)\big)-\frac 1 d \d_{\m\n}\d_{\r\l}
    \,,
\end{align}
and the indices $a,b$ label the generators of the global symmetry group.

\subsection*{QED$_d$-GNY model:}
The QED$_d$-GNY model admits a $1/N$ expansion, where $N$ is the total number of fermionic components
\begin{align}
    N=N_f\, \rm{tr}\bf 1=4N_f
    \,.
\end{align}
Here $N_f$ is the number of four-component Dirac fermions and the symbol $\bf 1$ represents the $4\times 4$ identity matrix.
One can translate our results directly into the case of $N'_f$ two-component Dirac fermions by treating $\bf 1$ as the $2\times 2$ identity matrix and substituting
\begin{align}
    N=N'_f\, \rm{tr}\bf 1=2N'_f
    \,.\label{translate into two component}
\end{align}
The results for the central charges are
\begin{align}
    C_J&=\frac{\mathrm{tr}\mathbf{1}}{S_d^2}
    \Bigg[
        1+\frac{\gamma^\text{QED-GNY}}{N}\left(\frac{3(d-2)}{4}\Theta(d)+\frac{2(d-2)}{d}\right)+O(1/N^2)
    \Bigg]\,,\label{CJ results QED-GNY}\\
    C_T&=N\frac{d}{2S_d^2}
    \Bigg[
        1+\frac{\gamma^\text{QED-GNY}}{N}\left(\frac{3(d-2)}{4}\Theta(d)+\frac{(d-2)(2 d^2-d-2)}{(d-1)d(d+2)}\Psi(d)\right.\nonumber\\
        &\hspace{3.6cm}\left.-\frac{(d^2-4d+2)(d^2+2d-4)}{(d-1)^2 d^2 (d+2)}\right)+O(1/N^2)
    \Bigg]
    \,,\label{CJ CT results QED-GNY}
\end{align}
where $S_d=\frac{2\p^{d/2}}{\G(d/2)}$ is the volume of the $(d-1)$-sphere, and
\begin{align}
    \g^{\text{QED-GNY}}=\frac{4(d-1)\G(d-2)}{\G(1-\frac{d}{2})\G(\frac{d}{2})^3}
    \label{gamma GNY}
\end{align}
corresponds to the leading correction to the scaling dimension of the adjoint fermion bilinear operator in QED-GNY model
\begin{align}
    \D_{\bar{\ps}t^a\ps}=d-1+\frac{\g^{\text{QED-GNY}}}{N}+O(1/N^2)
    \,.\label{Delta adj}
\end{align}
The result of $\g^{\text{QED-GNY}}$ agrees with \cite{Gracey:2018fwq} for general $d$.
In (\ref{CJ results QED-GNY}) and (\ref{CJ CT results QED-GNY}) we have defined the functions
\begin{equation}
\begin{aligned}
    \Psi(d)&=\psi(1-d/2)+\psi(d-2)-\psi(d/2)-\psi(1)\,, \\
    \Theta(d)&=\psi'(d/2)-\psi'(1)
    \,,
\end{aligned}
\end{equation}
where $\ps(x)=\G'(x)/\G(x)$ is the digamma function and $\ps'(x)$ is its derivative.
Although the dimensional continuation of the results corresponds to the QED$_3$-GNY$_-$ model in $d=3$, the leading corrections to $C_J$, $C_T$, and $\D_{\bar{\ps}t^a\ps}$ coincide with those in the QED$_3$-GNY$_+$ model. The results \eqref{CJ results QED-GNY}, \eqref{CJ CT results QED-GNY} and \eqref{gamma GNY} apply to both models in $d=3$.
We have also obtained $C_J^{\text{top}}$ for the topological $U(1)$ current
\begin{align}
    C_{J,-}^{\text{top}}=\frac{1}{N'_f}\frac{8}{\p^4}\[1+\frac{1}{N'_f}\(4-\frac{112}{3\p^2}\)+O(1/N_f^{\prime 2})\]
    \,,\label{CJtop}
\end{align}
in which we have used the normalization \eqref{CJ def}.
The minus sign in the subscript indicates that the result corresponds to the QED$_3$-GNY$_-$ model.
For the QED$_3$-GNY$_+$ model we find 
\begin{align}
    C^{\text{top}}_{J,+}=\frac{1}{N'_f}\frac{8}{\p^4}\[1+\frac{1}{N'_f}\(4-\frac{368}{3\p^2}\)+O(1/N_f^{\prime 2})\]
    \,.\label{CJtop GNY+}
\end{align}
We have written $C^{\text{top}}_{J,\pm}$ in terms of $N'_f$ since we work in $d=3$ here.

\subsection*{Scalar QED$_d$:}
The large $N$ expansion in scalar QED$_d$ runs in the number $N$ of complex scalars.
We find the central charges\footnote{The $1/N$ correction to $C_J$ in $d=3$ has been studied in \cite{Huh:2013vga}.
Their result does not match the $d=3$ case of our result due to a discrepancy in the Feynman integral $D'_1$ in Fig. \ref{JJ diagrams sQED}, which is caused by either a typo or different renormalization scheme used in \cite{Huh:2013vga}.
A similar difference also appears in $C_J^{\text{top}}$ of scalar QED$_3$.}
\begin{align}
    C_J&=\frac{2}{(d-2)S_d^2}\Bigg[1+\frac{\gamma^\text{Scalar QED}}{N}\left(\frac{3(d-1)}{8}\,\Theta(d)-\frac{(d-1)(d^2+d-8)}{2(d-2)^2 d}\right)+O(1/N^2)\Bigg]\,,\label{CJ results scalar}\\
    C_T&=N\frac{2d}{(d-1)S_d^2}\Bigg[
        1+\frac{\gamma^\text{Scalar QED}}{N}\left(\frac{3(d-1)}{8}\,\Theta(d)+\frac{d^3-2 d^2+4}{2 (d-2)d(d+2)}\Psi(d)\right.\nonumber\\
        &\hspace{4cm}\left.+\frac{d^5-16 d^4+38 d^3+20 d^2-88 d+32}{4 (d-2)^2 d^2 (d+2)}\right)+O(1/N^2)
    \Bigg]
    \,,\label{CJ CT results scalar}
\end{align}
where
\begin{align}
    \g^{\text{Scalar QED}}=\frac{8\G(d-1)}{\G(1-\frac{d}{2})\G(\frac{d}{2}-1)\G(\frac{d}{2})^2}
    \label{gamma scalar}
\end{align}
gives the leading correction to the scaling dimension of the adjoint scalar bilinear operator\footnote{For the $N=2N'_f$ QED$_3$-GNY$_{\pm}$ models and $N=N'_f$ scalar QED$_3$, the leading anomalous dimensions of the adjoint bilinear operators agree, $\g^{\text{QED-GNY}}/(2N'_f)=\g^{\text{Scalar QED}}/N'_f=-16/(\p^2 N'_f)$.
This agreement extends to all (broken) higher spin currents in the adjoint sector \cite{Zhou:2022pah}.}
\begin{align}
    \D_{\f^*t^a\f}=d-2+\frac{\g^{\text{Scalar QED}}}{N}+O(1/N^2)
    \,.\label{Delta adj sQED}
\end{align} 
The $d=3$ case of $\g^{\text{Scalar QED}}$ agrees with \cite{Benvenuti:2019ujm}.
To the best of our knowledge, the result for general $d$ \eqref{gamma scalar} is new.
Moreover, $C_J^{\text{top}}$ for the topological $U(1)$ current in $d=3$ is
\begin{align}
    C_J^{\text{top}}=\frac{1}{N}\frac{8}{\p^4}\[1+\frac{1}{N}\(4-\frac{272}{3\p^2}\)+O(1/N^2)\]
    \,,\label{CJtop sQED}
\end{align}
in which we have used the normalization \eqref{CJ def}.

The rest of this paper is organized as follows.
We first consider the QED$_d$-GNY model in Section \ref{QED-GNY model}.
The setup of the large $N$ expansion is described in Section \ref{GNY large N expansion}, followed by the calculation of the central charges $C_J$ and $C_T$ in Section \ref{CJ CT GNY}.
We obtain $C_J^{\text{top}}$ for the topological $U(1)$ current in Section \ref{CJ top GNY}, where the Feynman diagrams are closely related to those for $C_J$.
An analogous procedure is carried out for scalar QED$_d$ in Section \ref{Scalar QED}.
We compare our results with the fuzzy sphere and bootstrap data of $SO(5)$ DQCP in Section \ref{The two-flavor theories in d=3}. Conclusions are made in Section \ref{sec5}.
In Appendices \ref{gamma QED-GNY appendix} and \ref{gamma Scalar appendix}, we obtain the anomalous dimensions of the lowest adjoint bilinear scalars.
Other appendices contain some intermediate steps and results for Feynman integrals.

\section{QED$_d$-GNY model}
\label{QED-GNY model}

In this section, we consider a $U(1)$ gauge theory with $N_f$ massless charged fermions coupled to a real massless scalar field $\f$ in Euclidean space.
The action takes the form\footnote{Here, the summations over flavor indices are written explicitly for clarity, but we will use Einstein summation convention below.}
\begin{align}\label{S QED-GNY}
    S=\int\mathrm{d}^d x\Big(
        \frac{1}{4e^2}F^{\mu\nu}F_{\mu\nu}-\sum_{j=1}^{N_f}\bar{\psi}_j\gamma^\mu D_\m\ps^j+\frac{1}{2g^2}\pa_\m\f\pa^\m\f+\sum_{j=1}^{N_f}\f\bar{\ps}_j\ps^j+\l\f^4
    \Big)
    \,,
\end{align}
where $e,g,\l$ are coupling constants, $\ps^j$ are four-component complex spinors, $A_\m$ is the gauge field, $D_\m=\pa_\m+iA_\m$ is the covariant derivative, $F_{\m\n}=\pa_\m A_\n-\pa_\n A_\m$ is the field strength, and the Yukawa interaction strength is set to one.
Following \cite{Giombi:2015haa,Giombi:2016fct}, the dimensional continuation is defined by holding fixed the number of fermionic components in each $\ps^j$, and $\g^\m$ are $4\times 4$ matrices satisfying $\{\g^\m,\g^\n\}=2\d^{\m\n}\bf 1$.
We will refer to this theory as the QED$_d$-GNY model.

Note that the dimensional continuation does not yield an $SU(2N_f)$ symmetric model in $d=3$ \cite{Boyack:2018zfx}.\footnote{See \cite{DiPietro:2015taa,DiPietro:2017kcd} for related discussions in the context of the $\e=4-d$ expansion.}
Instead, the flavor symmetry is $SU(N_f)\times SU(N_f)$.
This model is often dubbed the QED$_3$-GNY$_-$ model, as opposed to the $SU(2N_f)$ symmetric QED$_3$-GNY$_+$ model.
From the large $N$ expansion perspective, the two models are closely related.
For the leading corrections, the Feynman diagrams for $C_J$, $C_T$, and $\g^{\text{QED-GNY}}$ in the QED$_3$-GNY$_-$ model are identical to the QED$_3$-GNY$_+$ case, so our results for $C_J$, $C_T$, and $\g^{\text{QED-GNY}}$ apply to both cases in $d=3$.

\subsection{Large $N$ expansion}
\label{GNY large N expansion}

For $d<4$, the self interaction term $\l\f^4$ and the kinetic terms for $A_\m$ and $\f$ in \eqref{S QED-GNY} can be dropped since they are irrelevant.\footnote{See \cite{Janssen:2017eeu,Ihrig:2018ojl,Zerf:2018csr} for analyses of the renormalization group flow.}
Taking $N_f=\infty$ and integrating out the fermions, we find the effective action
\begin{align}
    S_{\text{eff}}=\int\mathrm{d}^d x\,\mathrm{d}^d y \;\frac 1 2\Big(
        & A^\m(x)\<\til J_\m(x)\til J_\n(y)\>_0 A^\n(y)+ \f(x)\<(\bar{\psi}_j\ps^j)(x)(\bar{\psi}_k\ps^k)(y)\>_0\f(y)
    \Big)
    \,,
\end{align}
where $\til J_\m=\bar{\ps}_j\g_\m\ps^j$ and $\<\ldots\>_0$ represents correlation functions in the free theory.
The effective propagators for the gauge field and scalar field can be found by inverting the effective kinetic terms above.
Considering a generalized Feynman gauge and using the free fermion propagator in momentum space
\begin{align}
    \<\ps^j(p)\bar{\ps}_k(-p)\>_0=\frac{i\g^\m p_\m}{p^2}\d^j_k
    \,,
\end{align}
we obtain the effective propagators
\begin{align}
    \<A_\m(p)A_\n(-p)\>_{\text{eff}}=\frac{1}{N}\frac{C_A}{p^{2(\frac{d}{2}-1+\D)}}\[\d_{\m\n}-(1-\x)\frac{p_\m p_\n}{p^2}\]\,, \quad \<\f(p)\f(-p)\>_{\text{eff}}=\frac{1}{N}\frac{C_\f}{p^{2(\frac{d}{2}-1+\D)}}
\,,\label{eff prop}
\end{align}
where $\x$ is a gauge parameter, with $\x=0$ corresponding to the Landau gauge $\pa_\m A^\m=0$, and the normalizations are
\begin{align}
    C_A=\frac{(4\p)^{\frac d 2} \Gamma (d)}{2\Gamma(2-\frac{d}{2}) \Gamma(\frac{d}{2})^2}\,,\qquad
    C_\f=\frac{2(4\p)^{\frac d 2} \Gamma (d-2)}{\Gamma(2-\frac{d}{2}) \Gamma(\frac{d}{2}-1)^2}
    \,.
\end{align}
As in \cite{Vasiliev:1975mq,Vasiliev:1981dg,Vasiliev:1981yc,Derkachov:1997ch,Diab:2016spb,Giombi:2016fct}, the dimensions of $A_\m$ and $\f$ are shifted by a small parameter $\D$ in order to regularize the theory in the $1/N$ expansion discussed below.

For large $N$, the effective propagators \eqref{eff prop} are suppressed by $1/N$, while fermion loops with flavor indices traced over contribute factors of $N$.
Therefore, the diagrammatic expansion at large $N$ is organized by the difference between the numbers of effective propagators and fermion loops that carry $N$ factors.
We summarize the Feynman rules as follows.
The propagators are represented by
\begin{align}
\hspace{-0.5cm}
\begin{tikzpicture}[baseline=(c.base)]
\begin{feynman}
  \vertex[label=above:$j$] (c) at (0,0) {};
  \vertex[label=above:$k$] (d) at (1.5,0) {};
  \diagram*{
    (c) -- [fermion, edge label=$p$] (d),
  };
\end{feynman}
\end{tikzpicture}
=\<\ps^j(p)\bar{\ps}_k(-p)\>_0
\,,\hspace{0.2cm}
\begin{tikzpicture}[baseline=(c.base)]
\begin{feynman}
  \vertex[label=above:$\m$] (a) at (0,0) {};
  \vertex[label=above:$\n$] (b) at (1.5,0) {};
  \diagram*{
    (a) -- [photon] (b),
  };
\end{feynman}
\end{tikzpicture}
=\<A_\m(p)A_\n(-p)\>_{\text{eff}}
\,,\hspace{0.2cm}
\begin{tikzpicture}[baseline=(c.base)]
\begin{feynman}
  \vertex (c) at (0,0) {};
  \vertex (d) at (1.6,0) {};
  \diagram*{
    (c) -- [scalar] (d),
  };
\end{feynman}
\end{tikzpicture}
=\<\f(p)\f(-p)\>_{\text{eff}}
\,.
\end{align}
The vertices correspond to
\begin{align}
\begin{tikzpicture}[baseline={(current bounding box.center)}]
\begin{feynman}
  \vertex[dot] (v) at (0,0) {};
  \vertex (l) at (-1.4,0) {};
  \vertex (u) at (1.2,1.2) {};
  \vertex (d) at (1.2,-1.2) {};
  \diagram*{
    (l) -- [scalar] (v),
    (u) -- [fermion] (v) -- [fermion] (d),
  };
\end{feynman}
\end{tikzpicture}
=-1
\,,
\qquad
\begin{tikzpicture}[baseline={(current bounding box.center)}]
\begin{feynman}
  \vertex[dot,label=below:$\m$] (v) at (0,0) {};
  \vertex (l) at (-1.4,0) {};
  \vertex (u) at (1.2,1.2) {};
  \vertex (d) at (1.2,-1.2) {};
  \diagram*{
    (l) -- [photon] (v),
    (u) -- [fermion] (v) -- [fermion] (d),
  };
\end{feynman}
\end{tikzpicture}
=i\g^\m
\,.
\end{align}
These rules are similar to those of QED$_d$ in \cite{Giombi:2016fct}, except that we have an extra propagator and an extra vertex due to the scalar field $\f$ and the Yukawa interaction.

\subsection{Calculation of $C_J$ and $C_T$}
\label{CJ CT GNY}

\begin{figure}[H]
    \centering
    \includegraphics[width=0.7\linewidth]{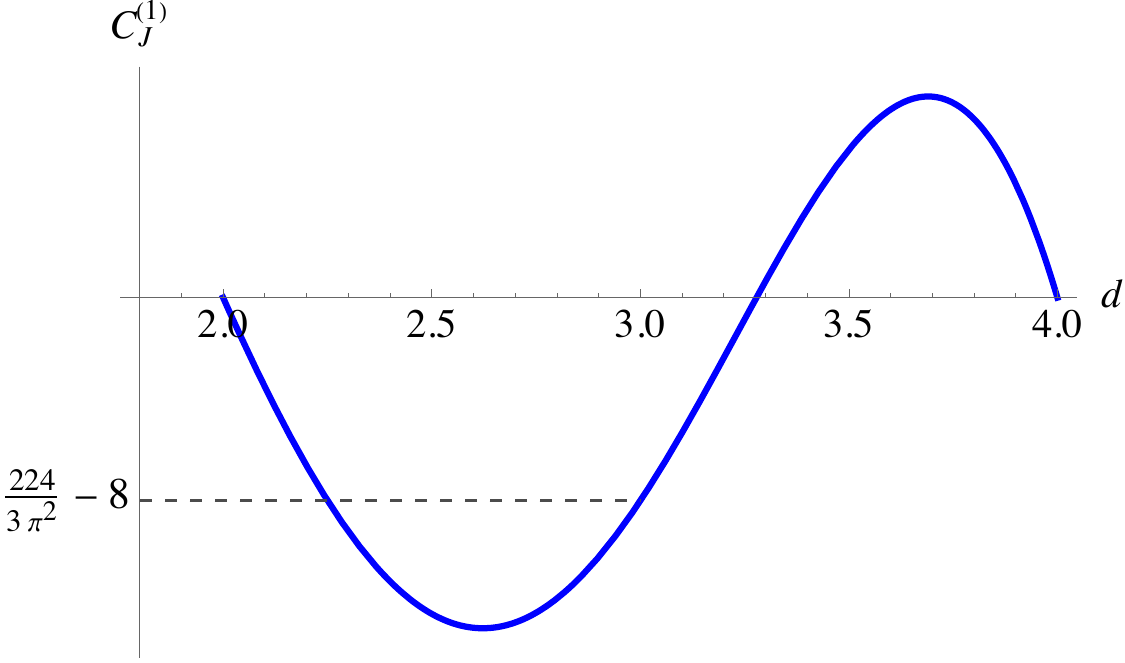}
    \caption{Plot of the coefficient $C_J^{(1)}$ associated with the leading correction to $C_J$.}
    \label{CJGNY plot}
\end{figure}

The global symmetry current takes the form
\begin{align}
    J^a_\m=-\bar{\ps}_j (t^a)^j_k \g_\m\ps^k
    \,,
\end{align}
where $t^a$ is a generator of $SU(N_f)$ satisfying $\rm{tr}(t^a t^b)=-\d^{ab}$.
In momentum space, the current can be represented diagrammatically as
\begin{align}
\begin{tikzpicture}[baseline={(current bounding box.center)}]
\begin{feynman}
  \vertex[dot,label=left:$J_\m^a(p)$] (v) at (0,0) {};
  \vertex (u) at (1.2,1) {};
  \vertex (d) at (1.2,-1) {};
  \diagram*{
    (u) -- [fermion] (v) -- [fermion] (d),
  };
\end{feynman}
\end{tikzpicture}
=-\g_\m\, t^a
\,.
\end{align}
The Feynman diagrams associated with $C_J$ are shown in Fig. \ref{JJ diagrams}, and the integrals are given in Appendix \ref{JJ TT QED-GNY appendix}.
We obtain the two-point function
\begin{align}
    \<J^a_\m(p)J^b_\n(-p)\>&=D_0+D_1+D'_1+D_2+D'_2 \nn
    &=\frac{\G(1-\frac d 2)\G(\frac d 2)^2\,\rm{tr}\bf 1}{(4\p)^{\frac d 2}(d-1)\G(d-2)}\rm{tr}(t^a t^b)\(\frac{p_\mu p_\nu}{p^2}-\delta_{\mu\nu}\)p^{2(\frac d 2-1)}\nn
    &\hspace{0.5cm}\times\Bigg[1+\frac{\gamma^\text{QED-GNY}}{N}\left(\frac{3(d-2)}{4}\Theta(d)+\frac{2(d-2)}{d}\right)+O(1/N^2)\Bigg]
    \,.\label{JJ GNY result}
\end{align}
The gauge parameter $\x$, the $\log p^2$ term, and the pole in $\D$ are all canceled.
Moreover, each order $1/N$ diagram contributes an extra $\d_{\m\n}$ term, but this extra term is canceled after summing over all diagrams.
Using the Fourier transform
\begin{equation}
    \begin{aligned}
\int \frac{d^d p}{(2\pi)^d} \frac{e^{ipx}}{p^{2\a}} &= \frac{\G(\frac{d}{2} - \alpha)}{4^{\alpha}\pi^{\frac{d}{2}}\Gamma(\alpha)} \frac{1}{x^{2(\frac d 2-\a)}}\,, \\
\int d^d x \frac{e^{-ipx}}{x^{2\a}} &= \frac{\pi^{\frac{d}{2}}\Gamma(\frac{d}{2} - \alpha)}{4^{\alpha-\frac d 2}\Gamma(\alpha)} \frac{1}{p^{2(\frac d 2-\a)}}
    \,,
\end{aligned}
\end{equation}
the tensor structure in (\ref{JJ GNY result}) leads to the position space tensor structure $I_{\mu\nu}$ in (\ref{CJ def}), and we  obtain the central charge $C_J$ presented in \eqref{CJ CT results QED-GNY}.
With the convention
\begin{align}
    C_J=C_J^{(0)}\(1+\frac{1}{N}C_J^{(1)}+O(1/N^2)\)
    \,,\label{CJ expansion}
\end{align}
the coefficient $C_J^{(1)}$ of the leading $1/N$ correction is plotted in Fig. \ref{CJGNY plot}.
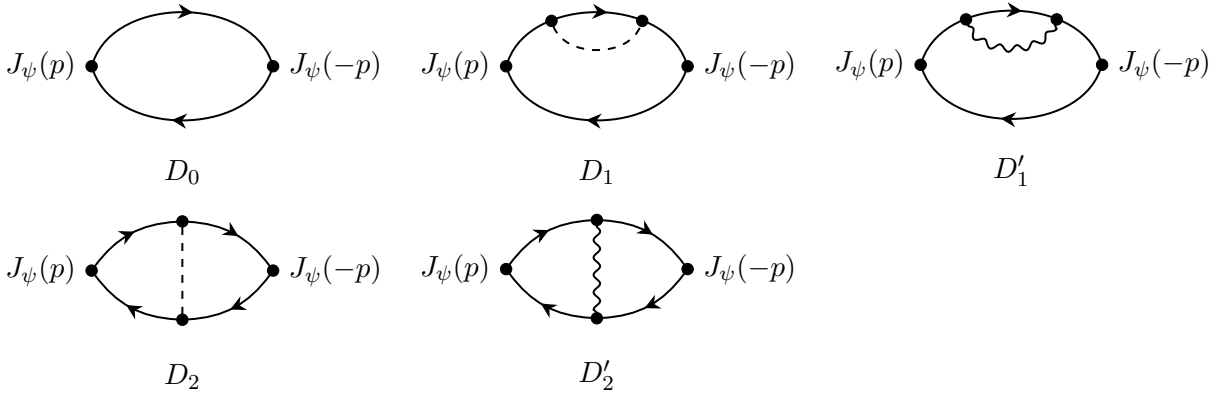
\begin{figure}[H]
\centering
\renewcommand{\arraystretch}{4} 
\begin{tabular}{ccc}

\hspace{-9mm}
\begin{tikzpicture}[baseline={(current bounding box.center)}]
  \begin{feynman}
    \vertex[dot, label=left:$J_\ps(p)$] (a) at (0,0) {};
    \vertex[dot, label=right:$J_\ps(-p)$] (b) at (2.4,0) {};
    \diagram* {
      (a) -- [fermion, bend left=70] (b) -- [fermion, bend left=70] (a)
    };
    \node at (1.2, -1.4) {$D_0$};
  \end{feynman}
\end{tikzpicture}

&

\hspace{-3mm}
\begin{tikzpicture}[baseline={(current bounding box.center)}]
  \begin{feynman}
    \vertex[dot, label=left:$J_\ps(p)$] (a) at (0,0) {};
    \vertex[dot, label=right:$J_\ps(-p)$] (b) at (2.4,0) {};
    \vertex[dot] (v1) at (0.6, 0.6) {};
    \vertex[dot] (v2) at (1.8, 0.6) {};
    \diagram* {
      (a)  -- [fermion, bend left=70] (b) -- [fermion, bend left=70] (a),
      (v1) -- [scalar, bend right=70] (v2)
    };
    \node at (1.2, -1.4) {$D_1$};
  \end{feynman}
\end{tikzpicture}

&

\hspace{-3mm}
\begin{tikzpicture}[baseline={(current bounding box.center)}]
  \begin{feynman}
    \vertex[dot, label=left:$J_\ps(p)$] (a) at (0,0) {};
    \vertex[dot, label=right:$J_\ps(-p)$] (b) at (2.4,0) {};
    \vertex[dot] (v1) at (0.6, 0.6) {};
    \vertex[dot] (v2) at (1.8, 0.6) {};
    \diagram* {
      (a)  -- [fermion, bend left=70] (b) -- [fermion, bend left=70] (a),
      (v1) -- [photon, bend right=70] (v2)
    };
    \node at (1.2, -1.4) {$D'_1$};
  \end{feynman}
\end{tikzpicture}

\\

\hspace{-9mm}
\begin{tikzpicture}[baseline={(current bounding box.center)}]
  \begin{feynman}
    \vertex[dot, label=left:$J_\ps(p)$] (a) at (0,0) {};
    \vertex[dot, label=right:$J_\ps(-p)$] (b) at (2.4,0) {};
    \vertex[dot] (v1) at (1.2, 0.65) {};
    \vertex[dot] (v2) at (1.2, -0.65) {};
    \diagram* {
      (a) -- [fermion, bend left=25] (v1) -- [fermion, bend left=25] (b) -- [fermion, bend left=25] (v2) -- [fermion, bend left=25] (a),
      (v1) -- [scalar] (v2)
    };
    \node at (1.2, -1.4) {$D_2$};
  \end{feynman}
\end{tikzpicture}

&

\hspace{-3mm}
\begin{tikzpicture}[baseline={(current bounding box.center)}]
  \begin{feynman}
    \vertex[dot, label=left:$J_\ps(p)$] (a) at (0,0) {};
    \vertex[dot, label=right:$J_\ps(-p)$] (b) at (2.4,0) {};
    \vertex[dot] (v1) at (1.2, 0.65) {};
    \vertex[dot] (v2) at (1.2, -0.65) {};
    \diagram* {
      (a) -- [fermion, bend left=25] (v1) -- [fermion, bend left=25] (b) -- [fermion, bend left=25] (v2) -- [fermion, bend left=25] (a),
      (v1) -- [photon] (v2)
    };
    \node at (1.2, -1.4) {$D'_2$};
  \end{feynman}
\end{tikzpicture}

&

\end{tabular}
\caption{Diagrams contributing to $C_J$ to order $1/N$.}
\label{JJ diagrams}
\end{figure}

\begin{figure}[H]
    \centering
    \includegraphics[width=0.7\linewidth]{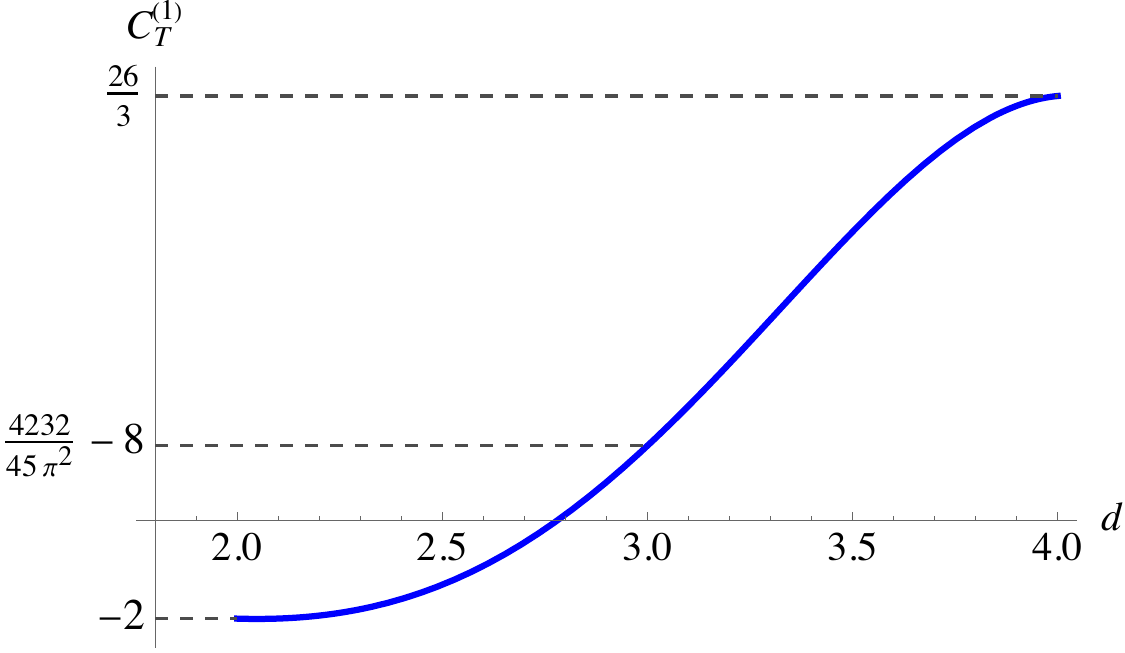}
    \caption{Plot of the coefficient $C_T^{(1)}$ in the leading correction to $C_T$.}
    \label{CTGNY plot}
\end{figure}

The stress-energy tensor is given by
\begin{align}
    T_{\m\n}=-\frac 1 4\Big[\big(\bar{\ps}_j\g_\m D_\n\ps^j-D_\m^*\bar{\ps}_j\g_\n\ps^j\big)+(\m\lra\n)\Big]-\text{trace}
    \,.
\end{align}
The diagrammatic representation is
\begin{align}
\begin{tikzpicture}[baseline={(current bounding box.center)}]
\begin{feynman}
  \vertex[dot,label=left:$(T_\ps)_{\m\n}(p)$] (v) at (0,0) {};
  \vertex (u) at (1.2,1) {};
  \vertex (d) at (1.2,-1) {};
  \diagram*{
    (u) -- [fermion, edge label'=$k$] (v) -- [fermion, edge label'=$k+p$] (d),
  };
\end{feynman}
\end{tikzpicture}
&=-\frac 1 4 i\Big[(2k_\m+p_\m)\g_\n+(\m\lra\n)\Big]-\text{trace}
\,,\nn
\begin{tikzpicture}[baseline={(current bounding box.center)}]
\begin{feynman}
  \vertex[dot,label=left:$(T_A)_{\m\n}(p)$,label=above:$\r$] (v) at (0,0) {};
  \vertex (i) at (1.2,0) {};
  \vertex (u) at (1.2,1) {};
  \vertex (d) at (1.2,-1) {};
  \diagram*{
     (i) -- [photon] (v),
    (u) -- [fermion] (v) -- [fermion] (d),
  };
\end{feynman}
\end{tikzpicture}
&=-\frac 1 2 i\Big[\g_\m \d_{\n\r}+(\m\lra\n)\Big]-\text{trace}
\,.
\end{align}
As pointed out in \cite{Diab:2016spb,Giombi:2016fct}, a ``renormalization'' factor $Z_T$ is needed in order for the Ward identity to hold.
This is computed in Appendix \ref{ZT QED-GNY appendix},\footnote{In the $1/N$ correction to $Z_T$, the term multiplying $\g^{\text{QED-GNY}}/N$ is a rational function of $d$.
It is not obvious a priori that this should be the case, since the correlation functions related by the Ward identity contain transcendental functions $\Ps(d)$ and $\Th(d)$.
But these transcendental functions match exactly without a $Z_T$ factor, so $Z_T$ only needs to compensate for the rational part multiplied by $\g^{\text{QED-GNY}}/N$.
This phenomenon also occurs in the O($N$) model, GNY model, QED$_d$ \cite{Diab:2016spb,Giombi:2016fct}, and scalar QED$_d$ discussed in Section \ref{Scalar QED}.}
\begin{align}
    Z_T=1-\frac{\g^{\text{QED-GNY}}}{N}\(\frac{(2 d^2-d-2) (d-2)}{2 (d-1) d (d+2) }\frac{1}{\D}+\frac{2 (d-2)}{(d-1)(d+2)}\)+O(1/N^2)
    \,.\label{ZT QED-GNY}
\end{align}
To find $C_T$, we calculate the Feynman diagrams in Fig. \ref{TT diagrams}.
The results are presented in Appendix \ref{JJ TT QED-GNY appendix}.
Collecting these results, we get
\begin{align}
    Z_T^2\<T_{\m\n}(p)T_{\r\l}(-p)\>=&N\frac{\Gamma(1-\frac{d}{2}) \Gamma
   (\frac{d}{2}+1)^2}{2(4\p)^{\frac d 2}\Gamma (d+2)}\times \nn 
   &\Bigg[1+\frac{\g^{\text{QED-GNY}}}{N}\(\frac{3(d-2)}{4}\,\Theta(d)+\frac{(d-2)(2 d^2-d-2)}{(d-1) d (d+2)}\Psi(d)\right. \nn  
    &\left.~ -\frac{(d^2-4d+2)(d^2+2d-4)}{(d-1)^2 d^2 (d+2)}\)+O(1/N^2)\Bigg]\bf{T}_{\m\n,\r\l}
    \,, \label{TT result}
\end{align}
where $\<T_{\m\n}(p)T_{\r\l}(-p)\>=D_0+D_1+D'_1+D_2+D'_2+D_3+D'_3+D_4+D_5+D_6+D_7+D_8$, and we have defined
\begin{align}
    \bf T_{\m\n,\r\l} &=\frac{2 \delta _{\lambda  \rho } \delta _{\mu  \nu }}{d}-\frac{(d-1) \delta _{\lambda  \nu } \delta _{\mu  \rho }}{d}-\frac{(d-1) \delta _{\lambda  \mu } \delta
   _{\nu  \rho }}{d}-\frac{2 p_{\lambda } p_{\rho } \delta _{\mu  \nu }}{d p^2}-\frac{2 p_{\mu } p_{\nu } \delta _{\lambda  \rho}}{d p^2}+\frac{(d-1) p_{\nu } p_{\rho } \delta _{\lambda  \mu }}{d p^2}
   \nn &\hspace{4mm}+\frac{(d-1)p_{\mu } p_{\rho } \delta _{\lambda  \nu }}{d p^2}+\frac{(d-1) p_{\lambda } p_{\nu } \delta _{\mu  \rho }}{d p^2}+\frac{(d-1) p_{\lambda } p_{\mu } \delta _{\nu  \rho }}{d p^2}-\frac{2 (d-2) p_{\lambda } p_{\mu } p_{\nu } p_{\rho }}{d p^4}
   \,.
\end{align}
Note that the poles in $\D$ are all canceled after  the factor $Z_T$ is taken into account, thus providing a nontrivial check of $Z_T$.
Like the loop computations for the correlator $\<JJ\>$, the extra tensor structures in (\ref{TT result}) besides $\bf{T}_{\m\n,\r\l}$ are canceled as well.
The Fourier transformation of the tensor structure $\bf{T}_{\m\n,\r\l}$ reproduces the position space tensor structure $\cal I_{\m\n,\r\l}(x)$ in (\ref{CT def}). 
From (\ref{TT result}) we extract the central charge $C_T$ given in \eqref{CJ CT results QED-GNY}. The results can be written in the form
\begin{align}
    C_T=N C_T^{(0)}\(1+\frac{1}{N}C_T^{(1)}+O(1/N^2)\)
    \,.\label{CT expansion}
\end{align}
and we plot $C_T^{(1)}$ as a function of $d$ in Fig. \ref{CTGNY plot}.
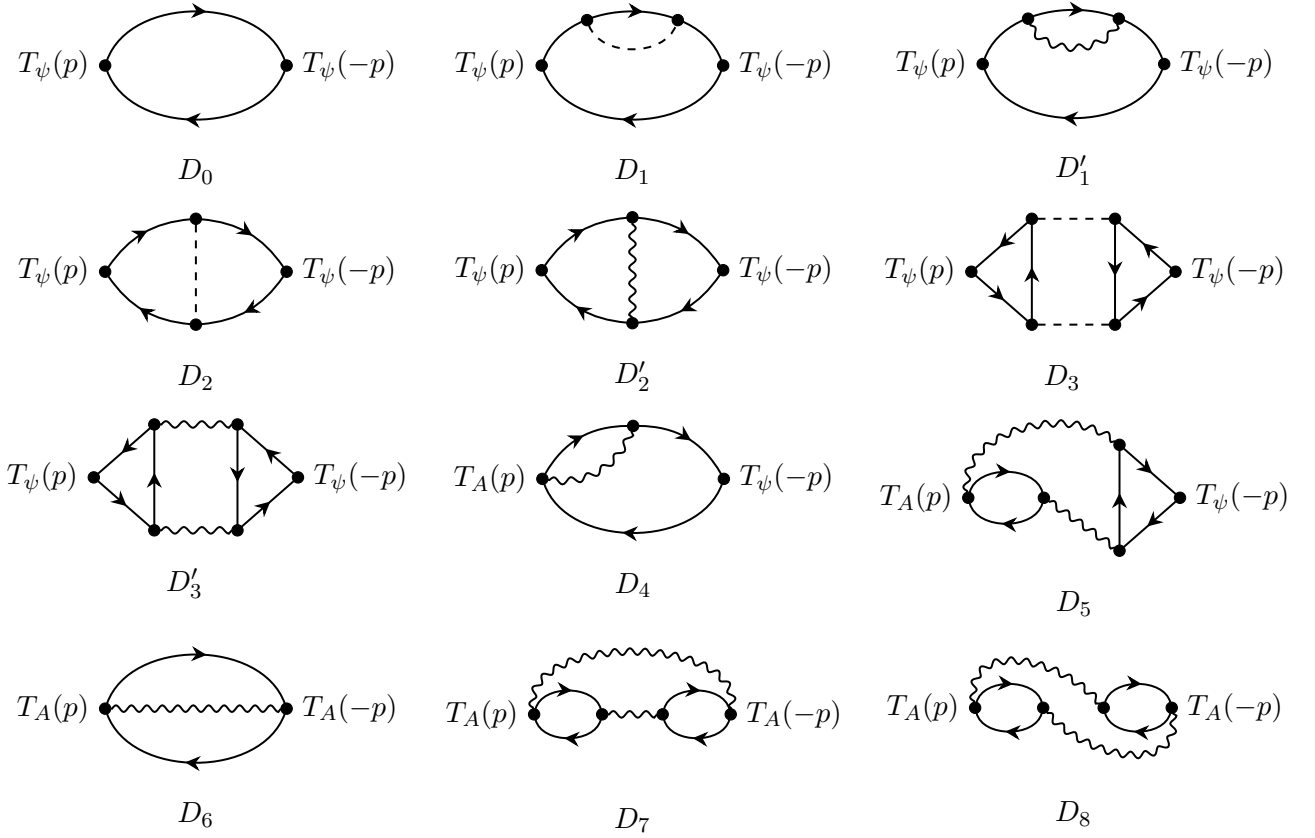
\begin{figure}[H]
\centering
\renewcommand{\arraystretch}{4} 
\begin{tabular}{ccc}

\hspace{-12mm}
\begin{tikzpicture}[baseline={(current bounding box.center)}]
  \begin{feynman}
    \vertex[dot, label=left:$T_\ps(p)$] (a) at (0,0) {};
    \vertex[dot, label=right:$T_\ps(-p)$] (b) at (2.4,0) {};
    \diagram* {
      (a) -- [fermion, bend left=70] (b) -- [fermion, bend left=70] (a)
    };
    \node at (1.2, -1.4) {$D_0$};
  \end{feynman}
\end{tikzpicture}

&

\hspace{-3mm}
\begin{tikzpicture}[baseline={(current bounding box.center)}]
  \begin{feynman}
    \vertex[dot, label=left:$T_\ps(p)$] (a) at (0,0) {};
    \vertex[dot, label=right:$T_\ps(-p)$] (b) at (2.4,0) {};
    \vertex[dot] (v1) at (0.6, 0.6) {};
    \vertex[dot] (v2) at (1.8, 0.6) {};
    \diagram* {
      (a)  -- [fermion, bend left=70] (b) -- [fermion, bend left=70] (a),
      (v1) -- [scalar, bend right=70] (v2)
    };
    \node at (1.2, -1.4) {$D_1$};
  \end{feynman}
\end{tikzpicture}

&

\hspace{-3mm}
\begin{tikzpicture}[baseline={(current bounding box.center)}]
  \begin{feynman}
    \vertex[dot, label=left:$T_\ps(p)$] (a) at (0,0) {};
    \vertex[dot, label=right:$T_\ps(-p)$] (b) at (2.4,0) {};
    \vertex[dot] (v1) at (0.6, 0.6) {};
    \vertex[dot] (v2) at (1.8, 0.6) {};
    \diagram* {
      (a)  -- [fermion, bend left=70] (b) -- [fermion, bend left=70] (a),
      (v1) -- [photon, bend right=70] (v2)
    };
    \node at (1.2, -1.4) {$D'_1$};
  \end{feynman}
\end{tikzpicture}

\\

\hspace{-12mm}
\begin{tikzpicture}[baseline={(current bounding box.center)}]
  \begin{feynman}
    \vertex[dot, label=left:$T_\ps(p)$] (a) at (0,0) {};
    \vertex[dot, label=right:$T_\ps(-p)$] (b) at (2.4,0) {};
    \vertex[dot] (v1) at (1.2, 0.7) {};
    \vertex[dot] (v2) at (1.2, -0.7) {};
    \diagram* {
      (a) -- [fermion, bend left=25] (v1) -- [fermion, bend left=25] (b) -- [fermion, bend left=25] (v2) -- [fermion, bend left=25] (a),
      (v1) -- [scalar] (v2)
    };
    \node at (1.2, -1.4) {$D_2$};
  \end{feynman}
\end{tikzpicture}

&

\hspace{-3mm}
\begin{tikzpicture}[baseline={(current bounding box.center)}]
  \begin{feynman}
    \vertex[dot, label=left:$T_\ps(p)$] (a) at (0,0) {};
    \vertex[dot, label=right:$T_\ps(-p)$] (b) at (2.4,0) {};
    \vertex[dot] (v1) at (1.2, 0.7) {};
    \vertex[dot] (v2) at (1.2, -0.7) {};
    \diagram* {
      (a) -- [fermion, bend left=25] (v1) -- [fermion, bend left=25] (b) -- [fermion, bend left=25] (v2) -- [fermion, bend left=25] (a),
      (v1) -- [photon] (v2)
    };
    \node at (1.2, -1.4) {$D'_2$};
  \end{feynman}
\end{tikzpicture}

&

\hspace{-3mm}
\begin{tikzpicture}[baseline={(current bounding box.center)}]
  \begin{feynman}
    \vertex[dot, label=left:$T_\ps(p)$] (a) at (0,0) {};
    \vertex[dot] (tl) at (0.8, 0.7) {};
    \vertex[dot] (bl) at (0.8, -0.7) {};
    \vertex[dot] (tr) at (1.9, 0.7) {};
    \vertex[dot] (br) at (1.9, -0.7) {};
    \vertex[dot, label=right:$T_\ps(-p)$] (b) at (2.7,0) {};
    \diagram* {
      (a) -- [fermion] (bl) -- [fermion] (tl) -- [fermion] (a),
      (b) -- [fermion] (tr) -- [fermion] (br) -- [fermion] (b),
      (tl) -- [scalar] (tr),
      (bl) -- [scalar] (br)
    };
    \node at (1.2, -1.4) {$D_3$};
  \end{feynman}
\end{tikzpicture}

\\

\hspace{-12mm}
\begin{tikzpicture}[baseline={(current bounding box.center)}]
  \begin{feynman}
    \vertex[dot, label=left:$T_\ps(p)$] (a) at (0,0) {};
    \vertex[dot] (tl) at (0.8, 0.7) {};
    \vertex[dot] (bl) at (0.8, -0.7) {};
    \vertex[dot] (tr) at (1.9, 0.7) {};
    \vertex[dot] (br) at (1.9, -0.7) {};
    \vertex[dot, label=right:$T_\ps(-p)$] (b) at (2.7,0) {};
    \diagram* {
      (a) -- [fermion] (bl) -- [fermion] (tl) -- [fermion] (a),
      (b) -- [fermion] (tr) -- [fermion] (br) -- [fermion] (b),
      (tl) -- [photon] (tr),
      (bl) -- [photon] (br)
    };
    \node at (1.2, -1.4) {$D'_3$};
  \end{feynman}
\end{tikzpicture}

&

\hspace{-3mm}
\begin{tikzpicture}[baseline={(current bounding box.center)}]
  \begin{feynman}
    \vertex[dot, label=left:$T_A(p)$] (a) at (0,0) {};
    \vertex[dot, label=right:$T_\ps(-p)$] (b) at (2.4,0) {};
    \vertex[dot] (t1) at (1.2, 0.7) {};
    \diagram* {
      (a) -- [fermion, bend left=30] (t1) -- [fermion, bend left=30] (b) -- [fermion, bend left=70] (a),
      (a) -- [photon, bend right=40] (t1)
    };
    \node at (1.2, -1.4) {$D_4$};
  \end{feynman}
\end{tikzpicture}

&

\hspace{-3mm}
\begin{tikzpicture}[baseline={(current bounding box.center)}]
  \begin{feynman}
    \vertex[dot, label=left:$T_A(p)$] (a) at (0,0) {};
    \vertex[dot] (m1) at (1.0, 0) {};
    \vertex[dot] (t2) at (2.0, 0.7) {};
    \vertex[dot] (b2) at (2.0, -0.7) {};
    \vertex[dot, label=right:$T_\ps(-p)$] (b) at (2.8, 0) {};
    \diagram* {
      (a) -- [fermion, bend left=75] (m1) -- [fermion, bend left=75] (a),
      (b) -- [fermion] (b2) -- [fermion] (t2) -- [fermion] (b),
      (a) -- [photon, out=100, in=140] (t2),
      (m1) -- [photon] (b2)
    };
    \node at (1.4, -1.4) {$D_5$};
  \end{feynman}
\end{tikzpicture}

\\

\hspace{-12mm}
\begin{tikzpicture}[baseline={(current bounding box.center)}]
  \begin{feynman}
    \vertex[dot, label=left:$T_A(p)$] (a) at (0,0) {};
    \vertex[dot, label=right:$T_A(-p)$] (b) at (2.4,0) {};
    \diagram* {
      (a) -- [fermion, bend left=70] (b) -- [fermion, bend left=70] (a),
      (a) -- [photon] (b)
    };
    \node at (1.2, -1.4) {$D_6$};
  \end{feynman}
\end{tikzpicture}

&

\hspace{-3mm}
\begin{tikzpicture}[baseline={(current bounding box.center)}]
  \begin{feynman}
    \vertex[dot, label=left:$T_A(p)$] (a) at (0,0) {};
    \vertex[dot] (m1) at (0.9, 0) {};
    \vertex[dot] (m2) at (1.7, 0) {};
    \vertex[dot, label=right:$T_A(-p)$] (b) at (2.6, 0) {};
    \diagram* {
      (a) -- [fermion, bend left=75] (m1) -- [fermion, bend left=75] (a),
      (m2) -- [fermion, bend left=75] (b) -- [fermion, bend left=75] (m2),
      (m1) -- [photon] (m2),
      (a) -- [photon, out=100, in=80] (b)
    };
    \node at (1.3, -1.4) {$D_7$};
  \end{feynman}
\end{tikzpicture}

&

\hspace{-3mm}
\begin{tikzpicture}[baseline={(current bounding box.center)}]
  \begin{feynman}
    \vertex[dot, label=left:$T_A(p)$] (a) at (0,0) {};
    \vertex[dot] (m1) at (0.9, 0) {};
    \vertex[dot] (m2) at (1.7, 0) {};
    \vertex[dot, label=right:$T_A(-p)$] (b) at (2.6, 0) {};
    \diagram* {
      (a) -- [fermion, bend left=75] (m1) -- [fermion, bend left=75] (a),
      (m2) -- [fermion, bend left=75] (b) -- [fermion, bend left=75] (m2),
      (a) -- [photon, out=100, in=130, looseness=1.3] (m2),
      (m1) -- [photon, out=-50, in=-80, looseness=1.3] (b)
    };
    \node at (1.3, -1.4) {$D_8$};
  \end{feynman}
\end{tikzpicture}

\end{tabular}
\caption{Diagrams corresponding to the two-point function of $T=T_\ps+T_A$ to order $N^0$.
For some topologies, we need to account for fermion loops with different directions, which are not explicitly drawn in the figure.}
\label{TT diagrams}
\end{figure}

\subsection{$C_J^{\text{top}}$ for the topological current in $d=3$}
\label{CJ top GNY}

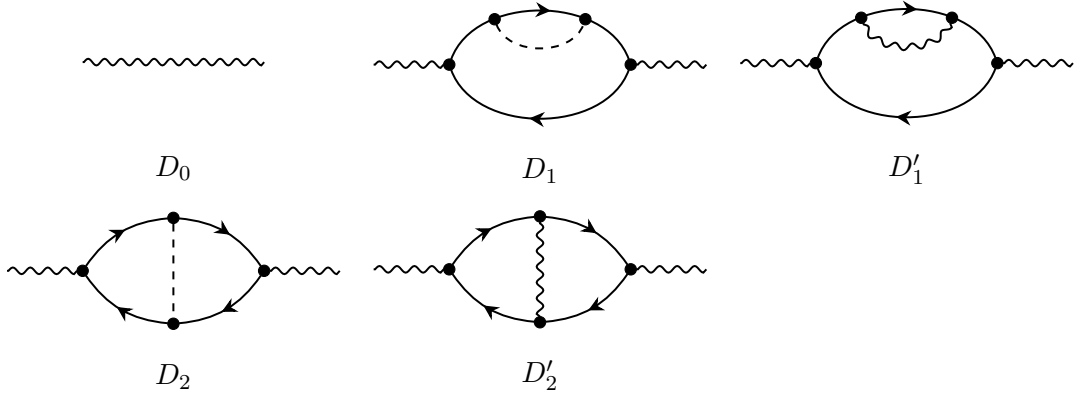
\begin{figure}[H]
\centering
\renewcommand{\arraystretch}{4}
\begin{tabular}{ccc}

\begin{tikzpicture}[baseline={(current bounding box.center)}]
  \begin{feynman}
    \vertex (in) at (-1.2,0);
    \vertex (out) at (1.2,0);
    \vertex (phantom) at (0, 0.75) {};
    \diagram* {
      (in) -- [photon] (out)
    };
    \node at (0, -1.4) {$D_0$};
  \end{feynman}
\end{tikzpicture}

&

\begin{tikzpicture}[baseline={(current bounding box.center)}]
  \begin{feynman}
    \vertex (in) at (-1,0);
    \vertex[dot] (a) at (0,0) {};
    \vertex[dot] (b) at (2.4,0) {};
    \vertex[dot] (v1) at (0.6, 0.6) {};
    \vertex[dot] (v2) at (1.8, 0.6) {};
    \vertex (out) at (3.4,0);
    \diagram* {
      (in) -- [photon] (a),
      (a) -- [fermion, bend left=70] (b) -- [fermion, bend left=70] (a),
      (v1) -- [scalar, bend right=70] (v2),
      (b) -- [photon] (out)
    };
    \node at (1.2, -1.4) {$D_1$};
  \end{feynman}
\end{tikzpicture}

&

\begin{tikzpicture}[baseline={(current bounding box.center)}]
  \begin{feynman}
    \vertex (in) at (-1,0);
    \vertex[dot] (a) at (0,0) {};
    \vertex[dot] (b) at (2.4,0) {};
    \vertex[dot] (v1) at (0.6, 0.6) {};
    \vertex[dot] (v2) at (1.8, 0.6) {};
    \vertex (out) at (3.4,0);
    \diagram* {
      (in) -- [photon] (a),
      (a) -- [fermion, bend left=70] (b) -- [fermion, bend left=70] (a),
      (v1) -- [photon, bend right=70] (v2),
      (b) -- [photon] (out)
    };
    \node at (1.2, -1.4) {$D'_1$};
  \end{feynman}
\end{tikzpicture}

\\

\begin{tikzpicture}[baseline={(current bounding box.center)}]
  \begin{feynman}
    \vertex (in) at (-1,0);
    \vertex[dot] (a) at (0,0) {};
    \vertex[dot] (b) at (2.4,0) {};
    \vertex[dot] (v1) at (1.2, 0.7) {};
    \vertex[dot] (v2) at (1.2, -0.7) {};
    \vertex (out) at (3.4,0);
    \diagram* {
      (in) -- [photon] (a),
      (a) -- [fermion, bend left=25] (v1) -- [fermion, bend left=25] (b) -- [fermion, bend left=25] (v2) -- [fermion, bend left=25] (a),
      (v1) -- [scalar] (v2),
      (b) -- [photon] (out)
    };
    \node at (1.2, -1.4) {$D_2$};
  \end{feynman}
\end{tikzpicture}

&

\begin{tikzpicture}[baseline={(current bounding box.center)}]
  \begin{feynman}
    \vertex (in) at (-1,0);
    \vertex[dot] (a) at (0,0) {};
    \vertex[dot] (b) at (2.4,0) {};
    \vertex[dot] (v1) at (1.2, 0.7) {};
    \vertex[dot] (v2) at (1.2, -0.7) {};
    \vertex (out) at (3.4,0);
    \diagram* {
      (in) -- [photon] (a),
      (a) -- [fermion, bend left=25] (v1) -- [fermion, bend left=25] (b) -- [fermion, bend left=25] (v2) -- [fermion, bend left=25] (a),
      (v1) -- [photon] (v2),
      (b) -- [photon] (out)
    };
    \node at (1.2, -1.4) {$D'_2$};
  \end{feynman}
\end{tikzpicture}

\end{tabular}
\caption{Diagrams associated with $\<A_\m(p)A_\n(-p)\>$ to order $1/N^2$.}
\label{top JJ diagram}
\end{figure}

In $d=3$, one can define the topological $U(1)$ current
\begin{align}
    J_{\rm{top}}^\m=\frac{i}{4\p}\e^{\m\n\r}F_{\n\r}
    \,,\label{Jtop def}
\end{align}
where $\e^{\m\n\r}$ is the totally antisymmetric tensor.
The Feynman diagrams for $C_J^{\text{top}}$ are shown in Fig. \ref{top JJ diagram}.
The diagrams have the same structures as those in Fig. \ref{JJ diagrams}, so they can be calculated directly based on the previous results. We get the two-point correlator
\begin{align}
    \<J_{\rm{top}}^\m(p)J_{\rm{top}}^\n(-p)\>
    =\frac{1}{N}\frac{8}{\p^2}\[1+\frac{1}{N}\(8-\frac{224}{3\p^2}\)+O(1/N^2)\]\(\frac{p^\m p^\n}{p^2}-\d^{\m\n}\)(p^2)^{\frac 1 2}
    \,.
\end{align}
The $\D$ pole and the terms depending on the gauge parameter $\x$ and $\log p^2$ are all canceled, as in \eqref{JJ GNY result}.
From the above result we extract the coefficient $C_{J,-}^{\text{top}}$, which is given in \eqref{CJtop}.

\begin{figure}[H]
    \centering
\begin{tikzpicture}[baseline={(current bounding box.center)}]
  \begin{feynman}
    \vertex (in) at (-1,0);
    \vertex[dot] (a) at (0,0) {};
    \vertex[dot] (tl) at (1, 0.9) {};
    \vertex[dot] (bl) at (1, -0.9) {};
    \vertex[dot] (tr) at (2.3, 0.9) {};
    \vertex[dot] (br) at (2.3, -0.9) {};
    \vertex[dot] (b) at (3.3,0) {};
    \vertex (out) at (4.3,0);
    \diagram* {
      (in) -- [photon] (a),
      (a) -- [fermion] (bl) -- [fermion] (tl) -- [fermion] (a),
      (b) -- [fermion] (tr) -- [fermion] (br) -- [fermion] (b),
      (tl) -- [scalar] (tr),
      (bl) -- [photon] (br),
      (b) -- [photon] (out)
    };
  \end{feynman}
\end{tikzpicture}
    \caption{The Aslamazov-Larkin diagram $D_+$ in the QED$_3$-GNY$_+$ model.}
    \label{extra AL diagram}
\end{figure}
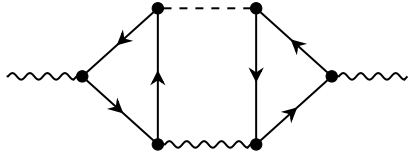

In the QED$_3$-GNY$_+$ model, there is an extra contribution to $C_{J,+}^{\text{top}}$ from the Aslamazov-Larkin three-loop diagram shown in Fig. \ref{extra AL diagram}.
This Aslamazov-Larkin diagram is absent in Fig. \ref{top JJ diagram} because it involves traces of an odd number of 4D gamma matrices.
(Recall that the dimensional continuation holds the gamma matrices fixed.)
However, traces of this type are nonzero if we use 3D gamma matrices $\s^\m$.
In particular, we have
\begin{align}
    \rm{tr}(\s^\m\s^\n\s^\r)&=2i\e^{\m\n\r}\,, \qquad
    \\
    \rm{tr}(\sigma^{\mu}\sigma^{\nu}\sigma^{\rho}\sigma^{\alpha}\sigma^{\beta}) &= 2i(\delta^{\nu\rho}\epsilon^{\mu\alpha\beta} - \delta^{\mu\rho}\epsilon^{\nu\alpha\beta} + \delta^{\alpha\beta}\epsilon^{\mu\nu\rho} + \delta^{\mu\nu}\epsilon^{\rho\alpha\beta})
    \,.
\end{align}
Working directly in $d=3$, the Aslamazov-Larkin diagram in Fig. \ref{extra AL diagram} is\footnote{To evaluate the Aslamazov-Larkin integral we use the $d=3$ integral formula
\begin{align}
    \int \frac{\rm{d}^3 k}{(2\p)^3}\frac{1}{k^2(p-k)^2(q-k)^2}=\frac{1}{8(p^2 q^2 (p-q)^2)^{1/2}}
    \,,
\end{align}
and then integrate over each internal momentum sequentially.}
\begin{align}
    D_{+}=\frac{1}{N_f^{\prime 2}}\frac{4096}{3\p^2}\(\frac{p_\m p_\n}{p^2}-\d_{\m\n}\)\frac{1}{(p^2)^{\frac 1 2}}
    \,.
\end{align}
Together with the diagrams in Fig. \ref{top JJ diagram}, we obtain $C_{J,+}^{\text{top}}$ in \eqref{CJtop GNY+}, where the $1/N'_f$ correction differs from \eqref{CJtop} by a large amount due to Fig. \ref{extra AL diagram}.
The large effects of additional Aslamazov-Larkin diagrams have also been noted in \cite{Boyack:2018zfx}.

\section{Scalar QED$_d$}
\label{Scalar QED}

In this section, we consider the Maxwell theory coupled to $N$ massless charged complex scalars with $SU(N)$ global symmetry.
The action for this theory reads
\begin{align}
    S=\int \rm{d}^d x \(\frac{1}{4e^2}F^{\m\n}F_{\m\n}+(D_\m\f_j)^* (D^\m\f^j)+\frac{\l}{2}(\f_j^*\f^j)^2\)
    \,,\label{S scalar QED}
\end{align}
where $e,\l$ are coupling constants, $\f$ is a complex scalar field, $A_\m$ is the $U(1)$ gauge field, $D_\m=\pa_\m+iA_\m$ denotes the covariant derivative, and $F_{\m\n}=\pa_\m A_\n-\pa_\n A_\m$ is the field strength.
We will refer to this theory as scalar QED$_d$.
In $d=3$, the central charges $C_J$ and $C_J^{\text{top}}$ of scalar QED$_3$ have been computed in \cite{Huh:2013vga}. Here we study the central charges $C_J$ and $C_J^{\text{top}}$ in general $d$, and compare our results in $d=3$ with the results in \cite{Huh:2013vga}. 
In addition, we present a new large $N$ result for the central charge $C_T$ of the scalar QED$_d$.

\subsection{Large $N$ expansion}

We rewrite the action in terms of a Hubbard-Stratonovich auxiliary field $\s$,
\begin{align}
    S=\int \rm{d}^d x \(\frac{1}{4e^2}F^{\m\n}F_{\m\n}+(D_\m\f_j)^*(D^\m\f^j)+\s\f_j^*\f^j-\frac{1}{2\l}\s^2\)
    \,.
\end{align}
The equation of motion $\s=\l \f_j^*\f^j$ leads to the original action \eqref{S scalar QED}.
For $d<4$, the Maxwell term and the $-\frac{1}{2\l}\s^2$ term can be dropped as they are irrelevant.
At $N=\infty$, integrating out the complex scalars leads to the effective action
\begin{align}
    S_{\text{eff}}=\int\mathrm{d}^d x\,\mathrm{d}^d y\Big(
        &\frac 1 2 A^\m(x)\<\til J_\m(x)\til J_\n(y)\>_0 A^\n(y)+\frac 1 2 \s(x)\<(\f_j^*\f^j)(x)(\f_k^*\f^k)(y)\>_0\s(y)
    \Big)
    \,,
\end{align}
where $\til{J}_\m=i(\f^*_j\pa_\m\f^j-\pa_\m\f^*_j\f^j)$.
In a generalized Feynman gauge, we invert the effective kinetic terms to obtain the effective propagators
\begin{align}
    \<A_\m(p)A_\n(-p)\>_{\text{eff}}&=\frac{1}{N}\frac{C_A}{p^{2(\frac{d}{2}-1+\D)}}\[\d_{\m\n}-(1-\x)\frac{p_\m p_\n}{p^2}\]\,, \\
    \<\s(p)\s(-p)\>_{\text{eff}}&=\frac{1}{N}\frac{C_\s}{p^{2(\frac{d}{2}-2+\D)}}
    \,,
\end{align}
where we have used the free propagator
\begin{align}
    \<\f^j(p)\f_k^*(-p)\>_0=\frac{1}{p^2}\d^j_k
    \,.
\end{align}
The normalizations for the effective propagators are
\begin{align}
    C_A=\frac{(4\p)^{\frac d 2}(d-2)\Gamma (d)}{4\Gamma(2-\frac{d}{2}) \Gamma(\frac{d}{2})^2}\,,\qquad
    C_\s=-\frac{(4\p)^{\frac{d}{2}}\G(d-2)}{\G(2-\frac{d}{2})\G(\frac{d}{2}-1)^2}
    \,.
\end{align}
Moreover, the dimensions of $A_\m$ and $\s$ are shifted by a regulator $\D$, similar to \eqref{eff prop}.

As in the QED$_d$-GNY model, the effective propagators carry $1/N$ factors, and complex scalar loops with $SU(N)$ indices traced over contribute factors of $N$.
The Feynman rules are summarized below.
The propagators are denoted as
\begin{align}
\hspace{-0.5cm}
\begin{tikzpicture}[baseline=(c.base)]
\begin{feynman}
  \vertex[label=above:$j$] (c) at (0,0) {};
  \vertex[label=above:$k$] (d) at (1.5,0) {};
  \diagram*{
    (c) -- [fermion, edge label=$p$] (d),
  };
\end{feynman}
\end{tikzpicture}
=\<\f^j(p)\f_k^*(-p)\>_0
\,,\hspace{0.2cm}
\begin{tikzpicture}[baseline=(c.base)]
\begin{feynman}
  \vertex[label=above:$\m$] (a) at (0,0) {};
  \vertex[label=above:$\n$] (b) at (1.5,0) {};
  \diagram*{
    (a) -- [photon] (b),
  };
\end{feynman}
\end{tikzpicture}
=\<A_\m(p)A_\n(-p)\>_{\text{eff}}
\,,\hspace{0.2cm}
\begin{tikzpicture}[baseline=(c.base)]
\begin{feynman}
  \vertex (c) at (0,0) {};
  \vertex (d) at (1.6,0) {};
  \diagram*{
    (c) -- [scalar] (d),
  };
\end{feynman}
\end{tikzpicture}
=\<\s(p)\s(-p)\>_{\text{eff}}
\,.
\end{align}
The vertices are
\begin{align}
\begin{tikzpicture}[baseline={(current bounding box.center)}]
\begin{feynman}
  \vertex[dot] (v) at (0,0) {};
  \vertex (l) at (-1.4,0) {};
  \vertex (u) at (1.2,1.2) {};
  \vertex (d) at (1.2,-1.2) {};
  \diagram*{
    (l) -- [scalar] (v),
    (u) -- [fermion] (v) -- [fermion] (d),
  };
\end{feynman}
\end{tikzpicture}
=-1
\,,
\qquad
\begin{tikzpicture}[baseline={(current bounding box.center)}]
\begin{feynman}
  \vertex[dot,label=below:$\m$] (v) at (0,0) {};
  \vertex (l) at (-1.4,0) {};
  \vertex (u) at (1.2,1.2) {};
  \vertex (d) at (1.2,-1.2) {};
  \diagram*{
    (l) -- [photon] (v),
    (u) -- [fermion, edge label'=$k$] (v) -- [fermion, edge label'=$q$] (d),
  };
\end{feynman}
\end{tikzpicture}
=-(k_\m+q_\m)
\,,
\qquad
\begin{tikzpicture}[baseline={(current bounding box.center)}]
\begin{feynman}
  \vertex[dot,label=above:$\m$,label=below:$\n$] (v) at (0,0) {};
  \vertex (l) at (-1.2,1.2) {};
  \vertex (o) at (-1.2,-1.2) {};
  \vertex (u) at (1.2,1.2) {};
  \vertex (d) at (1.2,-1.2) {};
  \diagram*{
    (l) -- [photon] (v),
    (o) -- [photon] (v),
    (u) -- [fermion] (v) -- [fermion] (d),
  };
\end{feynman}
\end{tikzpicture}
=-\d_{\m\n}
\,.
\end{align}
In $d=3$, these Feynman rules are equivalent to those given in \cite{Huh:2013vga}, which considered the 3D $\bf{CP}^{N-1}$ model in the large $N$ expansion.

\subsection{Calculation of $C_J$ and $C_T$}

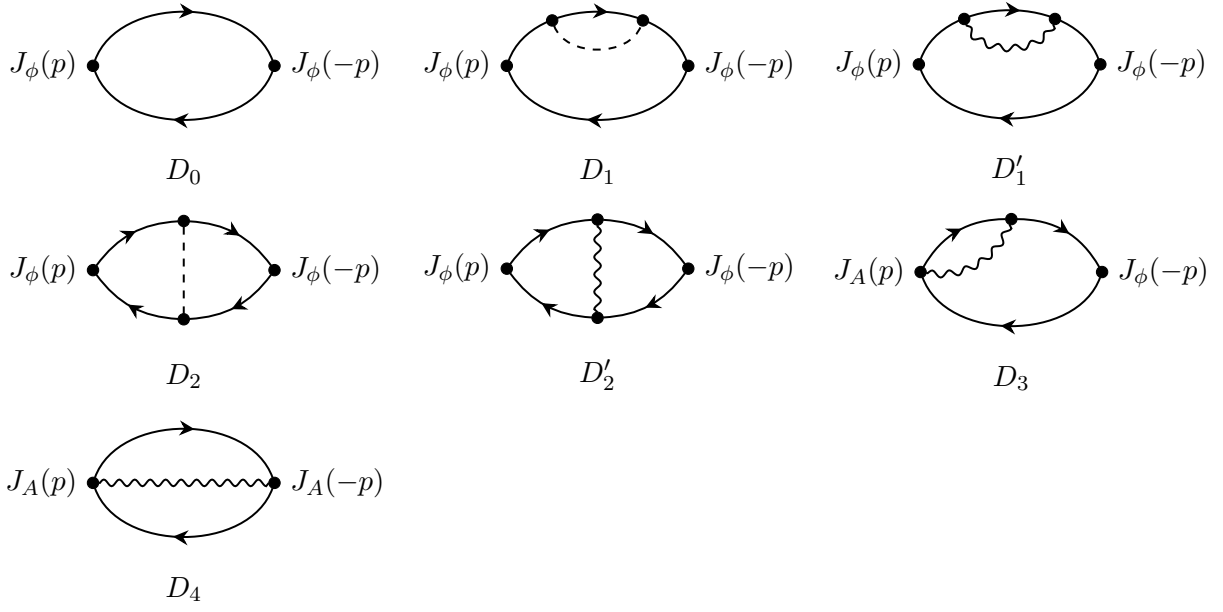
\begin{figure}[H]
    \centering
    \renewcommand{\arraystretch}{4} 
\begin{tabular}{ccc}

\hspace{-9mm}
\begin{tikzpicture}[baseline={(current bounding box.center)}]
  \begin{feynman}
    \vertex[dot, label=left:$J_\f(p)$] (a) at (0,0) {};
    \vertex[dot, label=right:$J_\f(-p)$] (b) at (2.4,0) {};
    \diagram* {
      (a) -- [fermion, bend left=70] (b) -- [fermion, bend left=70] (a)
    };
    \node at (1.2, -1.4) {$D_0$};
  \end{feynman}
\end{tikzpicture}

&

\hspace{-3mm}
\begin{tikzpicture}[baseline={(current bounding box.center)}]
  \begin{feynman}
    \vertex[dot, label=left:$J_\f(p)$] (a) at (0,0) {};
    \vertex[dot, label=right:$J_\f(-p)$] (b) at (2.4,0) {};
    \vertex[dot] (v1) at (0.6, 0.6) {};
    \vertex[dot] (v2) at (1.8, 0.6) {};
    \diagram* {
      (a)  -- [fermion, bend left=70] (b) -- [fermion, bend left=70] (a),
      (v1) -- [scalar, bend right=70] (v2)
    };
    \node at (1.2, -1.4) {$D_1$};
  \end{feynman}
\end{tikzpicture}

&

\hspace{-3mm}
\begin{tikzpicture}[baseline={(current bounding box.center)}]
  \begin{feynman}
    \vertex[dot, label=left:$J_\f(p)$] (a) at (0,0) {};
    \vertex[dot, label=right:$J_\f(-p)$] (b) at (2.4,0) {};
    \vertex[dot] (v1) at (0.6, 0.6) {};
    \vertex[dot] (v2) at (1.8, 0.6) {};
    \diagram* {
      (a)  -- [fermion, bend left=70] (b) -- [fermion, bend left=70] (a),
      (v1) -- [photon, bend right=70] (v2)
    };
    \node at (1.2, -1.4) {$D'_1$};
  \end{feynman}
\end{tikzpicture}

\\

\hspace{-9mm}
\begin{tikzpicture}[baseline={(current bounding box.center)}]
  \begin{feynman}
    \vertex[dot, label=left:$J_\f(p)$] (a) at (0,0) {};
    \vertex[dot, label=right:$J_\f(-p)$] (b) at (2.4,0) {};
    \vertex[dot] (v1) at (1.2, 0.65) {};
    \vertex[dot] (v2) at (1.2, -0.65) {};
    \diagram* {
      (a) -- [fermion, bend left=25] (v1) -- [fermion, bend left=25] (b) -- [fermion, bend left=25] (v2) -- [fermion, bend left=25] (a),
      (v1) -- [scalar] (v2)
    };
    \node at (1.2, -1.4) {$D_2$};
  \end{feynman}
\end{tikzpicture}

&

\hspace{-3mm}
\begin{tikzpicture}[baseline={(current bounding box.center)}]
  \begin{feynman}
    \vertex[dot, label=left:$J_\f(p)$] (a) at (0,0) {};
    \vertex[dot, label=right:$J_\f(-p)$] (b) at (2.4,0) {};
    \vertex[dot] (v1) at (1.2, 0.65) {};
    \vertex[dot] (v2) at (1.2, -0.65) {};
    \diagram* {
      (a) -- [fermion, bend left=25] (v1) -- [fermion, bend left=25] (b) -- [fermion, bend left=25] (v2) -- [fermion, bend left=25] (a),
      (v1) -- [photon] (v2)
    };
    \node at (1.2, -1.4) {$D'_2$};
  \end{feynman}
\end{tikzpicture}

&

\hspace{-3mm}
\begin{tikzpicture}[baseline={(current bounding box.center)}]
  \begin{feynman}
    \vertex[dot, label=left:$J_A(p)$] (a) at (0,0) {};
    \vertex[dot, label=right:$J_\f(-p)$] (b) at (2.4,0) {};
    \vertex[dot] (t1) at (1.2, 0.7) {};
    \diagram* {
      (a) -- [fermion, bend left=30] (t1) -- [fermion, bend left=30] (b) -- [fermion, bend left=70] (a),
      (a) -- [photon, bend right=40] (t1)
    };
    \node at (1.2, -1.4) {$D_3$};
  \end{feynman}
\end{tikzpicture}

\\

\hspace{-9mm}
\begin{tikzpicture}[baseline={(current bounding box.center)}]
  \begin{feynman}
    \vertex[dot, label=left:$J_A(p)$] (a) at (0,0) {};
    \vertex[dot, label=right:$J_A(-p)$] (b) at (2.4,0) {};
    \diagram* {
      (a) -- [fermion, bend left=70] (b) -- [fermion, bend left=70] (a),
      (a) -- [photon] (b)
    };
    \node at (1.2, -1.4) {$D_4$};
  \end{feynman}
\end{tikzpicture}

\end{tabular}
    \caption{Diagrams contributing to $C_J$ to order $1/N$.}
    \label{JJ diagrams sQED}
\end{figure}

\begin{figure}[H]
    \centering
    \includegraphics[width=0.7\linewidth]{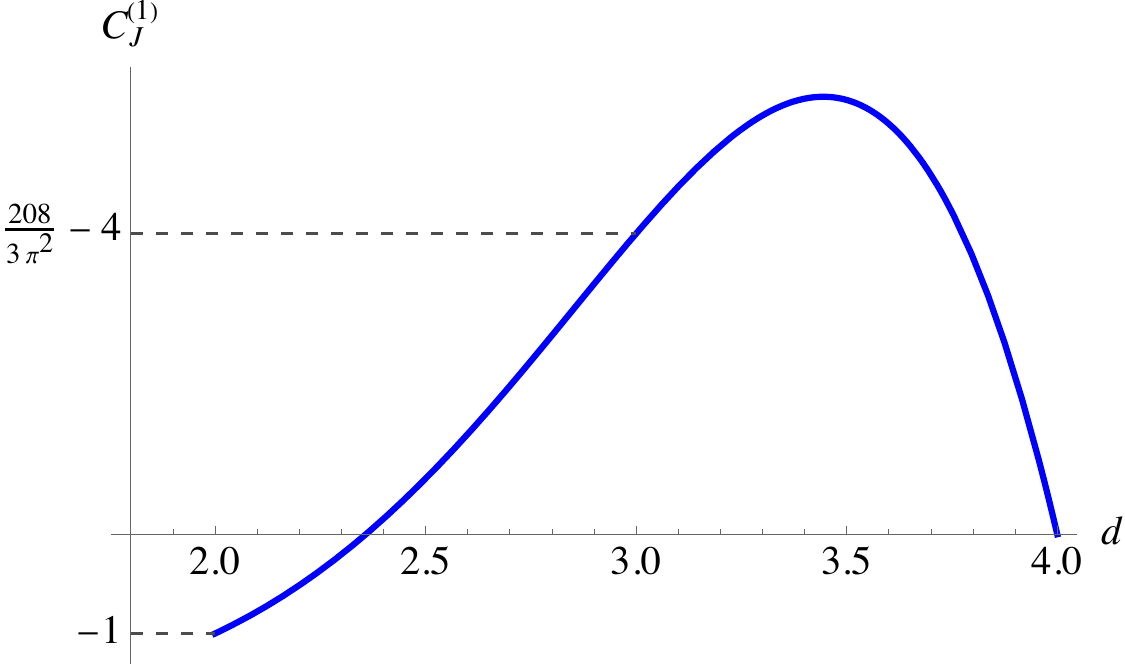}
    \caption{Plot of the coefficient $C_J^{(1)}$ in the leading correction to $C_J$.}
    \label{CJscalar Plot}
\end{figure}

The $SU(N)$ current is of the form
\begin{align}
    J^a_\m=-i\f_j^* (t^a)^j_k D_\m\f^k+i(D_\m\f_j)^* (t^a)^j_k \f^k
    \,,
\end{align}
where $t^a$ is a generator of $SU(N)$ with $\rm{tr}(t^a t^b)=\d^{ab}$.
We represent this current diagrammatically as
\begin{align}
    \begin{tikzpicture}[baseline={(current bounding box.center)}]
\begin{feynman}
  \vertex[dot,label=left:$(J_\f)_\m(p)$] (v) at (0,0) {};
  \vertex (u) at (1.2,1) {};
  \vertex (d) at (1.2,-1) {};
  \diagram*{
    (u) -- [fermion, edge label'=$k$] (v) -- [fermion, edge label'=$k+p$] (d),
  };
\end{feynman}
\end{tikzpicture}
=2k_\m+p_\m
\,,
\qquad
\begin{tikzpicture}[baseline={(current bounding box.center)}]
\begin{feynman}
  \vertex[dot,label=left:$(J_A)_\m(p)$,label=above:$\n$] (v) at (0,0) {};
  \vertex (i) at (1.2,0) {};
  \vertex (u) at (1.2,1) {};
  \vertex (d) at (1.2,-1) {};
  \diagram*{
     (i) -- [photon] (v),
    (u) -- [fermion] (v) -- [fermion] (d),
  };
\end{feynman}
\end{tikzpicture}
=2\d_{\m\n}
\,.
\end{align}
Using the Feynman diagrams shown in Fig. \ref{JJ diagrams sQED}, we find the two-point function
\begin{align}
    \<J^a_\m(p)J^b_\n(-p)\>&=D_0+D_1+D'_1+D_2+D'_2+D_3+D_4 \nn
    &=\frac{\G(2-\frac{d}{2})\G(\frac{d}{2}-1)^2}{(4\p)^{\frac{d}{2}}(d-1)\G(d-2)}\rm{tr}(t^a t^b)\(\frac{p_\mu p_\nu}{p^2}-\delta_{\mu\nu}\)p^{2(\frac d 2-1)} \label{JJ scalar result}\\
    &\hspace{0.5cm}\times\Bigg[1+\frac{\gamma^\text{Scalar QED}}{N}\left(\frac{3(d-1)}{8}\,\Theta(d)-\frac{(d-1)(d^2+d-8)}{2(d-2)^2 d}\right)+O(1/N^2)\Bigg]
    \,, \nonumber
\end{align}
where the results for individual diagrams are given in Appendix \ref{JJ TT JJ top sQED appendix}.
Again, the $\D$ poles, the terms with $\log p^2$ and/or the gauge parameter $\x$, and the extra $\d_{\m\n}$ terms are all canceled.
The result for $C_J$ is given in \eqref{CJ CT results scalar}.
In Fig. \ref{CJscalar Plot} we show the leading-order correction $C_J^{(1)}$ to the central charge $C_J$, which is determined according to the normalization  \eqref{CJ expansion}.

In $d=3$, the leading-order correction $C_J^{(1)}$ has also been computed in \cite{Huh:2013vga}.
However, their result does not agree with the $d=3$ case of our $C_J^{(1)}$.
This is due to a discrepancy in the regular part of the Feynman integral $D'_1$ in $d=3$:
\begin{equation}
\begin{aligned}
    D'_{1,\text{ours}}\Big|_{\x=0}&=\frac{1}{N}\[\(\frac{4}{3\p^2}\(\frac 1 \D-\log p^2\)-\frac{14}{9\p^2}\)\(\frac{p_\mu p_\nu}{p^2}-\delta_{\mu\nu}\)-\frac{4}{3\p^2}\d_{\m\n}\](p^2)^{\frac 1 2}\,,
    \\
    D'_{1,\text{theirs}}\Big|_{\x=0}&=\frac{1}{N}\[\(\frac{4}{3\p^2}\log \frac{\L^2}{p^2}-\frac{2}{\p^2}\)\(\frac{p_\mu p_\nu}{p^2}-\delta_{\mu\nu}\)-\frac{4}{3\p^2}\d_{\m\n}\](p^2)^{\frac 1 2}
    \,,\label{D1p compare}
\end{aligned}
\end{equation}
where $\L$ is the UV  momentum cutoff and the Landau gauge $\xi=0$ is assumed.  Note in \cite{Huh:2013vga} the loop integrals are regularized using UV momentum cutoff $\Lambda$ instead of the $\Delta$-regularization employed in this work.
We have verified that in the momentum cutoff regularization, by taking literal hard momentum cutoff the  diagram $D'_1$ can produce the same regular part as our result, so the discrepancy is caused either by a typo or different renormalization scheme employed in \cite{Huh:2013vga}. We have checked that for all the other Feynman diagrams our results agree with those in \cite{Huh:2013vga}.
A similar disagreement  appears in the topological central charge $C_J^{\text{top}}$, which will be studied in Subsection \ref{CJt for scalar QED3}.


The stress-energy tensor in scalar QED$_d$ takes the form
\begin{align}
    T_{\m\n}=\[\((D_\m\f_j)^* (D_\n\f^j)-\frac{d-2}{4(d-1)}\pa_\m\pa_\n(\f_j^*\f^j)\)+(\m\lra\n)\]-\text{trace}
    \,.
\end{align}
In Feynman diagrams, this is represented as
\begin{align}
\begin{tikzpicture}[baseline={(current bounding box.center)}]
\begin{feynman}
  \vertex[dot,label=left:$(T_\f)_{\m\n}(p)$] (v) at (0,0) {};
  \vertex (u) at (1.2,1) {};
  \vertex (d) at (1.2,-1) {};
  \diagram*{
    (u) -- [fermion, edge label'=$k$] (v) -- [fermion, edge label'=$k+p$] (d),
  };
\end{feynman}
\end{tikzpicture}
&=\[\((k_\m+p_\m)k_\n+\frac{d-2}{4(d-1)}p_\m p_\n\)+(\m\lra\n)\]-\rm{trace}
\,,\nn
\begin{tikzpicture}[baseline={(current bounding box.center)}]
\begin{feynman}
  \vertex[dot,label=left:$(T_A)_{\m\n}(p)$,label=above:$\r$] (v) at (0,0) {};
  \vertex (i) at (1.2,0) {};
  \vertex (u) at (1.2,1) {};
  \vertex (d) at (1.2,-1) {};
  \diagram*{
     (i) -- [photon] (v),
    (u) -- [fermion, edge label'=$k$] (v) -- [fermion, edge label'=$k+q+p$] (d),
  };
\end{feynman}
\end{tikzpicture}
&=\Big[(2k_\m+q_\m+p_\m)\d_{\n\r}+(\m\lra\n)\Big]-\text{trace}
\,.
\end{align}
The representation associated with $A_\m A_\n \f_j^*\f^j$ is not shown, since it does not contribute to the leading correction in the large $N$ expansion.
As mentioned in Section \ref{CJ CT GNY}, we need to include a $Z_T$ factor for the stress-energy tensor:
\begin{align}
    Z_T=1+\frac{\g^{\text{Scalar QED}}}{N}\(-\frac{d^3-2d^2+4}{4(d-2)d(d+2)}\frac{1}{\D}+\frac{1}{(d-2)(d+2)}\)+O(1/N^2)
    \,,\label{ZT scalar QED}
\end{align}
which is computed in Appendix \ref{ZT sQED appendix}.
The Feynman diagrams for $C_T$ are given in Fig. \ref{TT diagrams sQED}, and the results are listed in Appendix \ref{JJ TT JJ top sQED appendix}.
We find the stress-energy tensor two-point function
\begin{align}
    Z_T^2\<T_{\m\n}(p)T_{\r\l}(-p)\>=& N\frac{\Gamma(1-\frac{d}{2}) \Gamma
   (\frac{d}{2}+1)^2}{2(4\p)^{\frac d 2}(d-1)\Gamma (d+2)}\times \nn
   &\Bigg[1+\frac{\g^{\text{Scalar QED}}}{N}\left(\frac{3(d-1)}{8}\,\Theta(d)+\frac{(d^3-2 d^2+4)}{2 (d-2) d (d+2)}\Psi(d)\right.\nn
   &~~\left.+\frac{d^5-16 d^4+38 d^3+20 d^2-88 d+32}{4 (d-2)^2 d^2 (d+2)}\right)+O(1/N^2)\Bigg]\bf{T}_{\m\n,\r\l}
   \,,
\end{align}
where $\<T_{\m\n}(p)T_{\r\l}(-p)\>=D_0+D_1+D'_1+D_2+D'_2+D_3+D'_3+D_4+D_5+D_6+D_7+D_8$.
Besides the cancellations of the $\log p^2$ term and extra tensor structures, we emphasize that the $\D$ pole is canceled due to the $Z_T$ factor.
From $Z_T^2\<T_{\m\n}(p)T_{\r\l}(-p)\>$ we extract the central charge $C_T$ shown in \eqref{CJ CT results scalar}.
With the definition \eqref{CT expansion}, we plot the leading correction as a function of $d$ in Fig. \ref{CTscalar Plot}.

\begin{figure}[H]
    \centering
    \renewcommand{\arraystretch}{4} 
\begin{tabular}{ccc}

\hspace{-12mm}
\begin{tikzpicture}[baseline={(current bounding box.center)}]
  \begin{feynman}
    \vertex[dot, label=left:$T_\f(p)$] (a) at (0,0) {};
    \vertex[dot, label=right:$T_\f(-p)$] (b) at (2.4,0) {};
    \diagram* {
      (a) -- [fermion, bend left=70] (b) -- [fermion, bend left=70] (a)
    };
    \node at (1.2, -1.4) {$D_0$};
  \end{feynman}
\end{tikzpicture}

&

\hspace{-3mm}
\begin{tikzpicture}[baseline={(current bounding box.center)}]
  \begin{feynman}
    \vertex[dot, label=left:$T_\f(p)$] (a) at (0,0) {};
    \vertex[dot, label=right:$T_\f(-p)$] (b) at (2.4,0) {};
    \vertex[dot] (v1) at (0.6, 0.6) {};
    \vertex[dot] (v2) at (1.8, 0.6) {};
    \diagram* {
      (a)  -- [fermion, bend left=70] (b) -- [fermion, bend left=70] (a),
      (v1) -- [scalar, bend right=70] (v2)
    };
    \node at (1.2, -1.4) {$D_1$};
  \end{feynman}
\end{tikzpicture}

&

\hspace{-3mm}
\begin{tikzpicture}[baseline={(current bounding box.center)}]
  \begin{feynman}
    \vertex[dot, label=left:$T_\f(p)$] (a) at (0,0) {};
    \vertex[dot, label=right:$T_\f(-p)$] (b) at (2.4,0) {};
    \vertex[dot] (v1) at (0.6, 0.6) {};
    \vertex[dot] (v2) at (1.8, 0.6) {};
    \diagram* {
      (a)  -- [fermion, bend left=70] (b) -- [fermion, bend left=70] (a),
      (v1) -- [photon, bend right=70] (v2)
    };
    \node at (1.2, -1.4) {$D'_1$};
  \end{feynman}
\end{tikzpicture}

\\

\hspace{-12mm}
\begin{tikzpicture}[baseline={(current bounding box.center)}]
  \begin{feynman}
    \vertex[dot, label=left:$T_\f(p)$] (a) at (0,0) {};
    \vertex[dot, label=right:$T_\f(-p)$] (b) at (2.4,0) {};
    \vertex[dot] (v1) at (1.2, 0.7) {};
    \vertex[dot] (v2) at (1.2, -0.7) {};
    \diagram* {
      (a) -- [fermion, bend left=25] (v1) -- [fermion, bend left=25] (b) -- [fermion, bend left=25] (v2) -- [fermion, bend left=25] (a),
      (v1) -- [scalar] (v2)
    };
    \node at (1.2, -1.4) {$D_2$};
  \end{feynman}
\end{tikzpicture}

&

\hspace{-3mm}
\begin{tikzpicture}[baseline={(current bounding box.center)}]
  \begin{feynman}
    \vertex[dot, label=left:$T_\f(p)$] (a) at (0,0) {};
    \vertex[dot, label=right:$T_\f(-p)$] (b) at (2.4,0) {};
    \vertex[dot] (v1) at (1.2, 0.7) {};
    \vertex[dot] (v2) at (1.2, -0.7) {};
    \diagram* {
      (a) -- [fermion, bend left=25] (v1) -- [fermion, bend left=25] (b) -- [fermion, bend left=25] (v2) -- [fermion, bend left=25] (a),
      (v1) -- [photon] (v2)
    };
    \node at (1.2, -1.4) {$D'_2$};
  \end{feynman}
\end{tikzpicture}

&

\hspace{-3mm}
\begin{tikzpicture}[baseline={(current bounding box.center)}]
  \begin{feynman}
    \vertex[dot, label=left:$T_\f(p)$] (a) at (0,0) {};
    \vertex[dot] (tl) at (0.8, 0.7) {};
    \vertex[dot] (bl) at (0.8, -0.7) {};
    \vertex[dot] (tr) at (1.9, 0.7) {};
    \vertex[dot] (br) at (1.9, -0.7) {};
    \vertex[dot, label=right:$T_\f(-p)$] (b) at (2.7,0) {};
    \diagram* {
      (a) -- [fermion] (bl) -- [fermion] (tl) -- [fermion] (a),
      (b) -- [fermion] (tr) -- [fermion] (br) -- [fermion] (b),
      (tl) -- [scalar] (tr),
      (bl) -- [scalar] (br)
    };
    \node at (1.2, -1.4) {$D_3$};
  \end{feynman}
\end{tikzpicture}

\\

\hspace{-12mm}
\begin{tikzpicture}[baseline={(current bounding box.center)}]
  \begin{feynman}
    \vertex[dot, label=left:$T_\f(p)$] (a) at (0,0) {};
    \vertex[dot] (tl) at (0.8, 0.7) {};
    \vertex[dot] (bl) at (0.8, -0.7) {};
    \vertex[dot] (tr) at (1.9, 0.7) {};
    \vertex[dot] (br) at (1.9, -0.7) {};
    \vertex[dot, label=right:$T_\f(-p)$] (b) at (2.7,0) {};
    \diagram* {
      (a) -- [fermion] (bl) -- [fermion] (tl) -- [fermion] (a),
      (b) -- [fermion] (tr) -- [fermion] (br) -- [fermion] (b),
      (tl) -- [photon] (tr),
      (bl) -- [photon] (br)
    };
    \node at (1.2, -1.4) {$D'_3$};
  \end{feynman}
\end{tikzpicture}

&

\hspace{-3mm}
\begin{tikzpicture}[baseline={(current bounding box.center)}]
  \begin{feynman}
    \vertex[dot, label=left:$T_A(p)$] (a) at (0,0) {};
    \vertex[dot, label=right:$T_\f(-p)$] (b) at (2.4,0) {};
    \vertex[dot] (t1) at (1.2, 0.7) {};
    \diagram* {
      (a) -- [fermion, bend left=30] (t1) -- [fermion, bend left=30] (b) -- [fermion, bend left=70] (a),
      (a) -- [photon, bend right=40] (t1)
    };
    \node at (1.2, -1.4) {$D_4$};
  \end{feynman}
\end{tikzpicture}

&

\hspace{-3mm}
\begin{tikzpicture}[baseline={(current bounding box.center)}]
  \begin{feynman}
    \vertex[dot, label=left:$T_A(p)$] (a) at (0,0) {};
    \vertex[dot] (m1) at (1.0, 0) {};
    \vertex[dot] (t2) at (2.0, 0.7) {};
    \vertex[dot] (b2) at (2.0, -0.7) {};
    \vertex[dot, label=right:$T_\f(-p)$] (b) at (2.8, 0) {};
    \diagram* {
      (a) -- [fermion, bend left=75] (m1) -- [fermion, bend left=75] (a),
      (b) -- [fermion] (b2) -- [fermion] (t2) -- [fermion] (b),
      (a) -- [photon, out=100, in=140] (t2),
      (m1) -- [photon] (b2)
    };
    \node at (1.4, -1.4) {$D_5$};
  \end{feynman}
\end{tikzpicture}

\\

\hspace{-12mm}
\begin{tikzpicture}[baseline={(current bounding box.center)}]
  \begin{feynman}
    \vertex[dot, label=left:$T_A(p)$] (a) at (0,0) {};
    \vertex[dot, label=right:$T_A(-p)$] (b) at (2.4,0) {};
    \diagram* {
      (a) -- [fermion, bend left=70] (b) -- [fermion, bend left=70] (a),
      (a) -- [photon] (b)
    };
    \node at (1.2, -1.4) {$D_6$};
  \end{feynman}
\end{tikzpicture}

&

\hspace{-3mm}
\begin{tikzpicture}[baseline={(current bounding box.center)}]
  \begin{feynman}
    \vertex[dot, label=left:$T_A(p)$] (a) at (0,0) {};
    \vertex[dot] (m1) at (0.9, 0) {};
    \vertex[dot] (m2) at (1.7, 0) {};
    \vertex[dot, label=right:$T_A(-p)$] (b) at (2.6, 0) {};
    \diagram* {
      (a) -- [fermion, bend left=75] (m1) -- [fermion, bend left=75] (a),
      (m2) -- [fermion, bend left=75] (b) -- [fermion, bend left=75] (m2),
      (m1) -- [photon] (m2),
      (a) -- [photon, out=100, in=80] (b)
    };
    \node at (1.3, -1.4) {$D_7$};
  \end{feynman}
\end{tikzpicture}

&

\hspace{-3mm}
\begin{tikzpicture}[baseline={(current bounding box.center)}]
  \begin{feynman}
    \vertex[dot, label=left:$T_A(p)$] (a) at (0,0) {};
    \vertex[dot] (m1) at (0.9, 0) {};
    \vertex[dot] (m2) at (1.7, 0) {};
    \vertex[dot, label=right:$T_A(-p)$] (b) at (2.6, 0) {};
    \diagram* {
      (a) -- [fermion, bend left=75] (m1) -- [fermion, bend left=75] (a),
      (m2) -- [fermion, bend left=75] (b) -- [fermion, bend left=75] (m2),
      (a) -- [photon, out=100, in=130, looseness=1.3] (m2),
      (m1) -- [photon, out=-50, in=-80, looseness=1.3] (b)
    };
    \node at (1.3, -1.4) {$D_8$};
  \end{feynman}
\end{tikzpicture}

\end{tabular}
    \caption{Diagrams contributing to the two-point function of $T=T_\f+T_A$ to order $N^0$.
    For some topologies, we need to take into account the complex scalar loops with different directions, which are not explicitly drawn in the figure.}
    \label{TT diagrams sQED}
\end{figure}
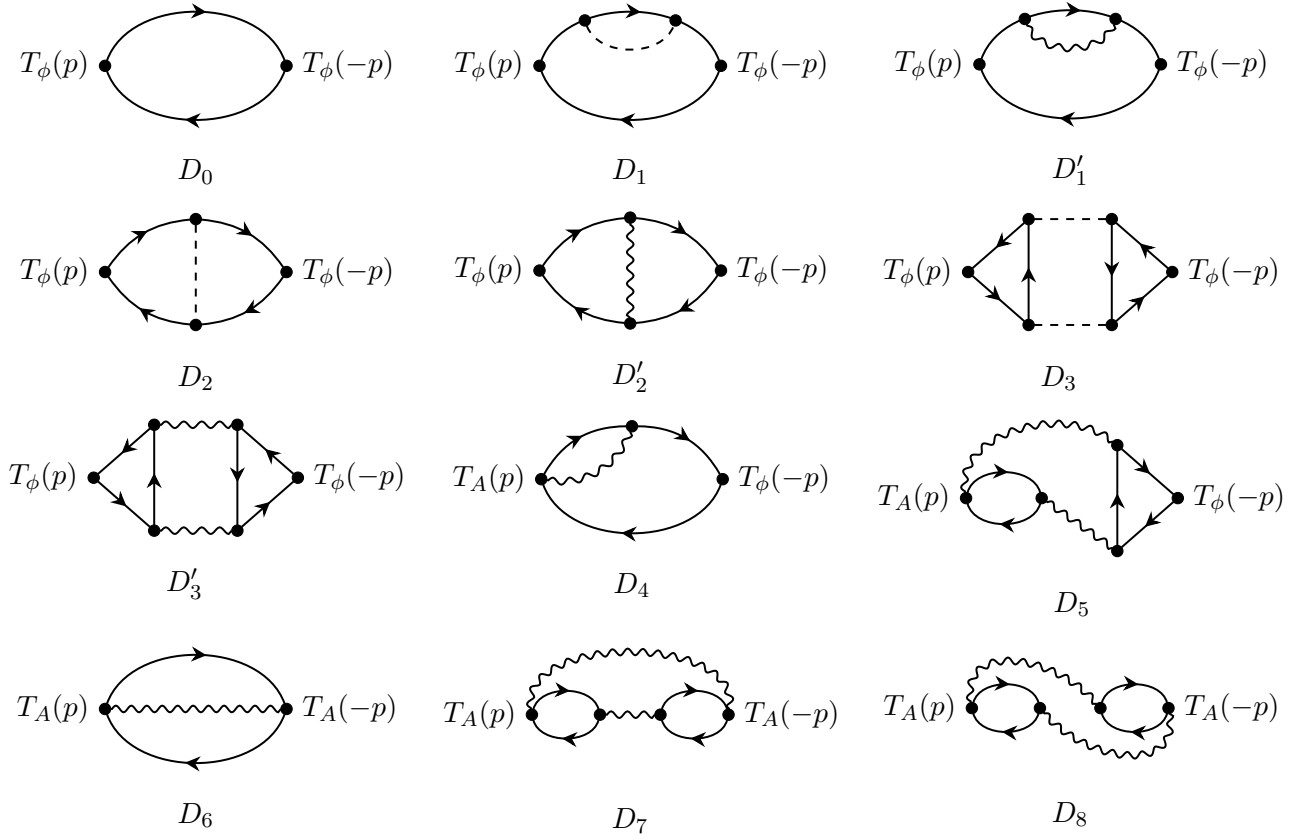
\begin{figure}[H]
    \centering
    \includegraphics[width=0.7\linewidth]{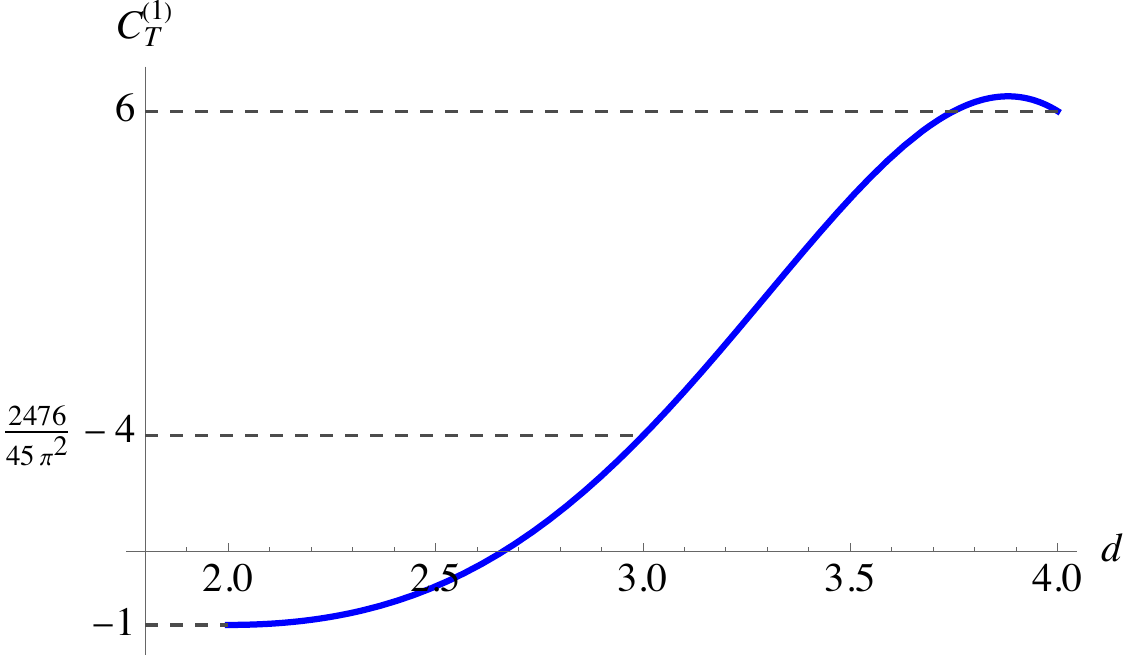}
    \caption{Plot of the coefficient $C_T^{(1)}$.}
    \label{CTscalar Plot}
\end{figure}

\subsection{$C_J^{\text{top}}$ for the topological current in $d=3$} \label{CJt for scalar QED3}

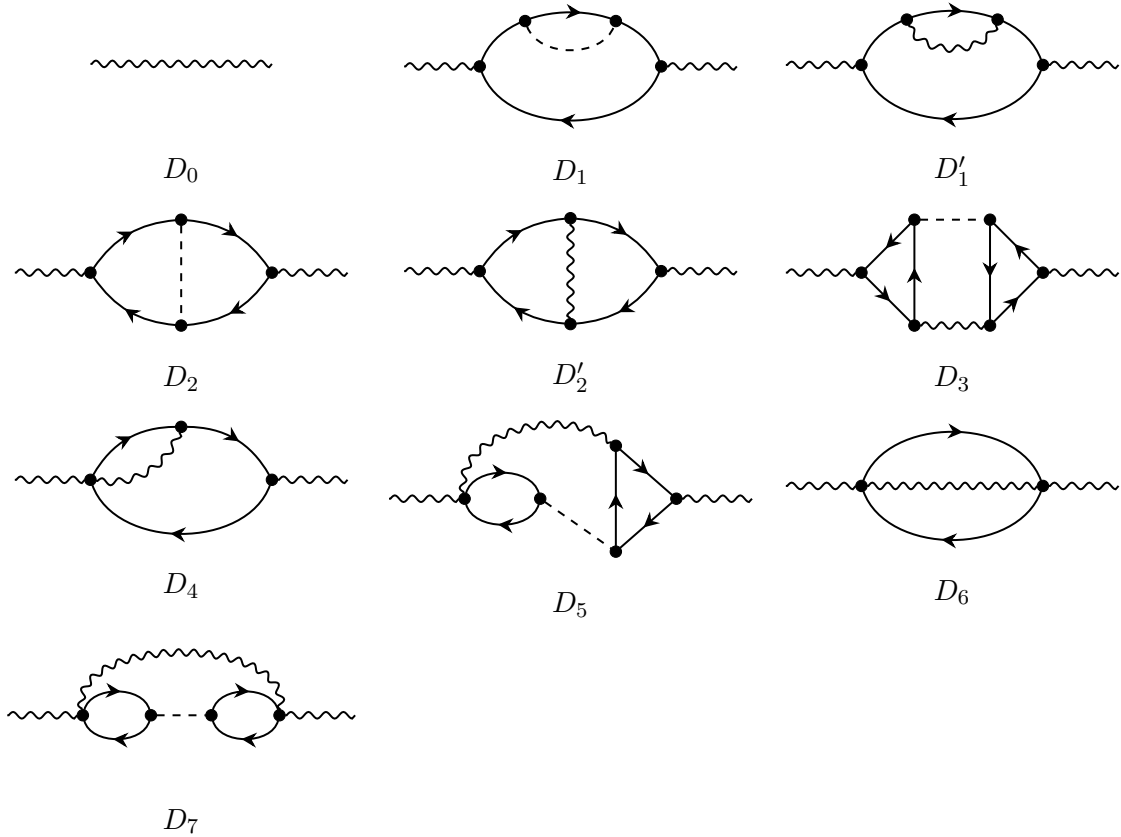
\begin{figure}[H]
    \centering
    \renewcommand{\arraystretch}{4}
\begin{tabular}{ccc}

\begin{tikzpicture}[baseline={(current bounding box.center)}]
  \begin{feynman}
    \vertex (in) at (-1.2,0);
    \vertex (out) at (1.2,0);
    \vertex (phantom) at (0, 0.75) {};
    \diagram* {
      (in) -- [photon] (out)
    };
    \node at (0, -1.4) {$D_0$};
  \end{feynman}
\end{tikzpicture}

&

\begin{tikzpicture}[baseline={(current bounding box.center)}]
  \begin{feynman}
    \vertex (in) at (-1,0);
    \vertex[dot] (a) at (0,0) {};
    \vertex[dot] (b) at (2.4,0) {};
    \vertex[dot] (v1) at (0.6, 0.6) {};
    \vertex[dot] (v2) at (1.8, 0.6) {};
    \vertex (out) at (3.4,0);
    \diagram* {
      (in) -- [photon] (a),
      (a) -- [fermion, bend left=70] (b) -- [fermion, bend left=70] (a),
      (v1) -- [scalar, bend right=70] (v2),
      (b) -- [photon] (out)
    };
    \node at (1.2, -1.4) {$D_1$};
  \end{feynman}
\end{tikzpicture}

&

\begin{tikzpicture}[baseline={(current bounding box.center)}]
  \begin{feynman}
    \vertex (in) at (-1,0);
    \vertex[dot] (a) at (0,0) {};
    \vertex[dot] (b) at (2.4,0) {};
    \vertex[dot] (v1) at (0.6, 0.6) {};
    \vertex[dot] (v2) at (1.8, 0.6) {};
    \vertex (out) at (3.4,0);
    \diagram* {
      (in) -- [photon] (a),
      (a) -- [fermion, bend left=70] (b) -- [fermion, bend left=70] (a),
      (v1) -- [photon, bend right=70] (v2),
      (b) -- [photon] (out)
    };
    \node at (1.2, -1.4) {$D'_1$};
  \end{feynman}
\end{tikzpicture}

\\

\begin{tikzpicture}[baseline={(current bounding box.center)}]
  \begin{feynman}
    \vertex (in) at (-1,0);
    \vertex[dot] (a) at (0,0) {};
    \vertex[dot] (b) at (2.4,0) {};
    \vertex[dot] (v1) at (1.2, 0.7) {};
    \vertex[dot] (v2) at (1.2, -0.7) {};
    \vertex (out) at (3.4,0);
    \diagram* {
      (in) -- [photon] (a),
      (a) -- [fermion, bend left=25] (v1) -- [fermion, bend left=25] (b) -- [fermion, bend left=25] (v2) -- [fermion, bend left=25] (a),
      (v1) -- [scalar] (v2),
      (b) -- [photon] (out)
    };
    \node at (1.2, -1.4) {$D_2$};
  \end{feynman}
\end{tikzpicture}

&

\begin{tikzpicture}[baseline={(current bounding box.center)}]
  \begin{feynman}
    \vertex (in) at (-1,0);
    \vertex[dot] (a) at (0,0) {};
    \vertex[dot] (b) at (2.4,0) {};
    \vertex[dot] (v1) at (1.2, 0.7) {};
    \vertex[dot] (v2) at (1.2, -0.7) {};
    \vertex (out) at (3.4,0);
    \diagram* {
      (in) -- [photon] (a),
      (a) -- [fermion, bend left=25] (v1) -- [fermion, bend left=25] (b) -- [fermion, bend left=25] (v2) -- [fermion, bend left=25] (a),
      (v1) -- [photon] (v2),
      (b) -- [photon] (out)
    };
    \node at (1.2, -1.4) {$D'_2$};
  \end{feynman}
\end{tikzpicture}

&

\begin{tikzpicture}[baseline={(current bounding box.center)}]
  \begin{feynman}
    \vertex (in) at (-1,0);
    \vertex[dot] (a) at (0,0) {};
    \vertex[dot] (tl) at (0.7, 0.7) {};
    \vertex[dot] (bl) at (0.7, -0.7) {};
    \vertex[dot] (tr) at (1.7, 0.7) {};
    \vertex[dot] (br) at (1.7, -0.7) {};
    \vertex[dot] (b) at (2.4,0) {};
    \vertex (out) at (3.4,0);
    \diagram* {
      (in) -- [photon] (a),
      (a) -- [fermion] (bl) -- [fermion] (tl) -- [fermion] (a),
      (b) -- [fermion] (tr) -- [fermion] (br) -- [fermion] (b),
      (tl) -- [scalar] (tr),
      (bl) -- [photon] (br),
      (b) -- [photon] (out)
    };
    \node at (1.2, -1.4) {$D_3$};
  \end{feynman}
\end{tikzpicture}

\\

\begin{tikzpicture}[baseline={(current bounding box.center)}]
  \begin{feynman}
    \vertex (in) at (-1,0);
    \vertex[dot] (a) at (0,0) {};
    \vertex[dot] (b) at (2.4,0) {};
    \vertex[dot] (t1) at (1.2, 0.7) {};
    \vertex (out) at (3.4,0);
    \diagram* {
      (in) -- [photon] (a),
      (a) -- [fermion, bend left=30] (t1) -- [fermion, bend left=30] (b) -- [fermion, bend left=70] (a),
      (a) -- [photon, bend right=40] (t1),
      (b) -- [photon] (out)
    };
    \node at (1.2, -1.4) {$D_4$};
  \end{feynman}
\end{tikzpicture}

&

\begin{tikzpicture}[baseline={(current bounding box.center)}]
  \begin{feynman}
    \vertex (in) at (-1,0);
    \vertex[dot] (a) at (0,0) {};
    \vertex[dot] (m1) at (1.0, 0) {};
    \vertex[dot] (t2) at (2.0, 0.7) {};
    \vertex[dot] (b2) at (2.0, -0.7) {};
    \vertex[dot] (b) at (2.8, 0) {};
    \vertex (out) at (3.8,0);
    \diagram* {
      (in) -- [photon] (a),
      (a) -- [fermion, bend left=75] (m1) -- [fermion, bend left=75] (a),
      (b) -- [fermion] (b2) -- [fermion] (t2) -- [fermion] (b),
      (a) -- [photon, out=100, in=140] (t2),
      (m1) -- [scalar] (b2),
      (b) -- [photon] (out)
    };
    \node at (1.4, -1.4) {$D_5$};
  \end{feynman}
\end{tikzpicture}

&

\begin{tikzpicture}[baseline={(current bounding box.center)}]
  \begin{feynman}
    \vertex (in) at (-1,0);
    \vertex[dot] (a) at (0,0) {};
    \vertex[dot] (b) at (2.4,0) {};
    \vertex (out) at (3.4,0);
    \diagram* {
      (in) -- [photon] (a),
      (a) -- [fermion, bend left=70] (b) -- [fermion, bend left=70] (a),
      (a) -- [photon] (b),
      (b) -- [photon] (out)
    };
    \node at (1.2, -1.4) {$D_6$};
  \end{feynman}
\end{tikzpicture}

\\

\begin{tikzpicture}[baseline={(current bounding box.center)}]
  \begin{feynman}
    \vertex (in) at (-1,0);
    \vertex[dot] (a) at (0,0) {};
    \vertex[dot] (m1) at (0.9, 0) {};
    \vertex[dot] (m2) at (1.7, 0) {};
    \vertex[dot] (b) at (2.6, 0) {};
    \vertex (out) at (3.6,0);
    \diagram* {
      (in) -- [photon] (a),
      (a) -- [fermion, bend left=75] (m1) -- [fermion, bend left=75] (a),
      (m2) -- [fermion, bend left=75] (b) -- [fermion, bend left=75] (m2),
      (m1) -- [scalar] (m2),
      (a) -- [photon, out=100, in=80] (b),
      (b) -- [photon] (out)
    };
    \node at (1.3, -1.4) {$D_7$};
  \end{feynman}
\end{tikzpicture}

\end{tabular}
    \caption{Diagrams associated with $\<A_\m(p)A_\n(-p)\>$ to order $1/N^2$.
    For some topologies, we have not shown the complex scalar loops with different directions, while their contributions need to be included in (\ref{JJtop}).}
    \label{Jtop diagrams sQED}
\end{figure}

The definition of the $U(1)$ topological current $J^\m_{\text{top}}$ in $d=3$ is again given by \eqref{Jtop def}.
The diagrams associated with the two-point function of $J^\m_{\text{top}}$ are shown in Fig. \ref{Jtop diagrams sQED}.
Besides the two-loop diagrams with the same structures as those in Fig. \ref{JJ diagrams sQED}, there are also three-loop diagrams contributing to $\<A_\m(p)A_\n(-p)\>$.
The results for these three-loop diagrams are presented in Appendix \ref{JJ TT JJ top sQED appendix}.
From the Feynman integrals in Fig. \ref{JJ diagrams sQED}, we get
\begin{align}
    \<J_{\rm{top}}^\m(p)J_{\rm{top}}^\n(-p)\>
    =\frac{1}{N}\frac{4}{\p^2}\[1+\frac{1}{N}\(4-\frac{272}{3\p^2}\)+O(1/N^2)\]\(\frac{p^\m p^\n}{p^2}-\d^{\m\n}\)(p^2)^{\frac 1 2}
    \,, \label{JJtop}
\end{align}
which leads to the $1/N$ expansion of $C_J^{\text{top}}$ shown in \eqref{CJtop sQED}.
The $\D$ poles, the gauge dependent terms, and the $\log p^2$ terms coming from the three-loop diagrams cancel among themselves, so the above result is of the expected form for the conserved current two-point function.
The result \eqref{CJtop sQED} differs from that in \cite{Huh:2013vga} due to the discrepancy discussed around \eqref{D1p compare}.

\section{The two-flavor theories in $d=3$}
\label{The two-flavor theories in d=3}

In this section, we focus on the cases relevant for the $SO(5)$ DQCP and the conjectured duality  between the two-flavor ($N'_f=2$) QED$_3$-GNY$_+$ model and two-flavor scalar QED$_3$.
It is convenient to rescale the central charges as
\begin{align}
    c_J=S^2_d\, C_J\,, \qquad c_T=S^2_d\, C_T
    \,.
\end{align}
In the QED$_3$-GNY$_\pm$ model, the results are
\begin{equation}
    \begin{aligned}
    c_J\big|_{d=3}&=2\[1+\frac{1}{N'_f}\(\frac{112}{3\p^2}-4\)+O(1/N_f^{\prime 2})\]\,,\\
    c_T\big|_{d=3}&=3N'_f\[1+\frac{1}{N'_f}\(\frac{2116}{45\p^2}-4\)+O(1/N_f^{\prime 2})\]
    \,.
\end{aligned}
\end{equation}
For the two-flavor case, we have\footnote{The two-flavor case of $C_J^{\text{top}}$ is negative due to the $1/N'_f$ correction in \eqref{CJtop GNY+}, so this result does not make sense for $N'_f=2$.
This is also the case for \eqref{CJtop sQED} with $N=2$.}
\begin{equation}
    \begin{aligned}
    c_J\big|_{d=3,\,N'_f=2}\approx 0.89\, c_{J}^\text{free fermions} \,,\qquad
    c_T\big|_{d=3,\,N'_f=2}\approx 1.38\, c_{T}^\text{free fermions}
    \,,\label{cJ cT GNY}
\end{aligned}
\end{equation}
where $c_{J}^\text{free fermions}=2$ and $c_{T}^\text{free fermions}=6$ are the central charges for 2 two-component Dirac free fermions.
In scalar QED$_3$, the central charges are
\begin{equation}
    \begin{aligned}
    c_J\big|_{d=3}&=2\[1+\frac{1}{N}\(\frac{208}{3\p^2}-4\)+O(1/N^2)\]\,,\\
    c_T\big|_{d=3}&=3N\[1+\frac{1}{N}\(\frac{2476}{45\p^2}-4\)+O(1/N^2)\]
    \,.
\end{aligned}
\end{equation}
The two-flavor case corresponds to
\begin{align}
    c_J\big|_{d=3,\,N=2}\approx 2.51\, c_{J}^\text{free scalars}\,, \qquad
    c_T\big|_{d=3,\,N=2}\approx 1.79\, c_{T}^\text{free scalars}
    \,,\label{cJ cT scalar}
\end{align}
where $c_{J}^\text{free scalars}=2$ and $c_{T}^\text{free scalars}=6$ are the central charges for 2 free complex scalars.
The results \eqref{cJ cT GNY} and \eqref{cJ cT scalar} should be used with caution, as the large $N$ expansion is not guaranteed to work well for small $N$.

Now we compare our perturbative results with the nonperturbative estimates of the $SO(5)$ DQCP. The $SO(5)$ DQCP has been simulated using fuzzy sphere regularization \cite{Zhou:2023qfi} and abundant CFT data of the theory has been obtained. The fuzzy sphere results including the central charges show surprising consistency with a particular bootstrap solution obtained in \cite{li2026bootstrap}. In particular,
the fuzzy sphere data  \cite{Zhou:2023qfi}
suggests the central charges
\begin{align}
  \text{\textbf{Fuzzy sphere:}}~~~  c_J \approx 1.68=0.84\times 2 \,, \qquad
    c_T \approx 6.36=1.06\times 6
    \,.
\end{align}
It should be reminded that there are typos in the $C_T$ and related OPE coefficient in the current version of \cite{Zhou:2023qfi}. Here we {\it assume} the number $0.121$ in Table \uppercase\expandafter{\romannumeral 3} of \cite{Zhou:2023qfi} is the square of the OPE coefficient $f_{\phi\phi\mathscr{T}}^2$, but this needs to be confirmed by the authors of \cite{Zhou:2023qfi}. The conformal bootstrap results indicate that the central charges are \cite{li2026bootstrap}
\begin{align}
   \text{\textbf{Conformal bootstrap:}}~~~  c_J \approx 1.71\approx0.85\times 2 \,, \qquad
    c_T \approx 6.95\approx1.16\times 6
    \,.
\end{align}
The above nonperturbative results are reasonably close to the large $N$ expansion of the QED$_3$-GNY$_\pm$ model \eqref{cJ cT GNY}, but show notable difference compared with the large $N$ result for the scalar QED$_3$ (\ref{cJ cT scalar}). 

\section{Discussion} \label{sec5}
We have computed the $1/N$ corrections to the central charges $C_J$, $C_T$, and the scaling dimensions of the lowest $SU(N)$ adjoint bilinear operators in the QED$_d$-GNY model and scalar QED$_d$ theory.
We have obtained the $1/N$ corrections to the central charge $C_{J}^{\text{top}}$ for the topological $U(1)$ currents in $d=3$ dimensions.
The results are summarized in Section \ref{Summary of results}.
The large $N$  expansions of the central charges $C_J$ and $C_T$ in $d=3$ are compared with the fuzzy sphere and bootstrap results for the $SO(5)$ DQCP, and we have found reasonable agreements between the perturbative results of the QED$_3$-GNY$_+$ model and the nonperturbative CFT data. This indicates the large $N$ expansion of the QED$_3$-GNY$_+$ model can provide instructive information for the $SO(5)$ DQCP.

Besides the duality between $N=2$ scalar QED$_3$ and the $N'_f=2$ QED$_3$-GNY$_+$ model, a web of boson-fermion dualities has been proposed for Abelian gauge theories with Chern-Simons interactions denoted by $U(1)_k$, namely the QED$_3$-CS theories  \cite{Karch:2016sxi,Seiberg:2016gmd,Kachru:2016rui,Wang:2017txt}. 
The monopole spectrum of the QED$_3$-CS theories has been studied using the large $N$ expansion in \cite{Chester:2017vdh,Chester:2021drl}, and it would be interesting to apply the large $N$ expansion to study the spectrum which is neutral under the topological $U(1)$ symmetry and the central charges in these theories. 
The 3D duality web also extends to the non-Abelian gauge theories. In particular, a duality web generalizing the level-rank duality \cite{Naculich:1990pa,Mlawer:1990uv,Nakanishi:1990hj,Naculich:2007nc} has been proposed in \cite{Aharony:2015mjs,Hsin:2016blu,Aharony:2016jvv} for gauge theories with non-Abelian gauge groups associated with Chern-Simons couplings, e.g., $U(N)_k$.
The duality has been verified in the planar limit $N,k\ra \infty$ with fixed $\lambda=N/k$ by computing the current correlators  \cite{Giombi:2011kc,Aharony:2011jz,Maldacena:2011jn,Aharony:2012nh,Maldacena:2012sf}, which correspond to the leading terms in the large $N$ expansion.
These dualities are conjectured to hold even at finite $N$, for which the subleading terms in the large $N$ expansion are indispensable.
It would be interesting to develop the large $N$ diagrammatic techniques for the 3D $U(N)_k$ theories and verify the extended duality web at finite $N$.

It is important to apply the large $N$ diagrammatic technique to the current three-point correlators. The current three-point correlators, e.g., $\langle JJT\rangle$ are strongly restricted by the conformal symmetry. Moreover, the parameters in the three-point correlators are subject to the constraints from the unitarity and causality, the so-called conformal collider bounds \cite{Hofman:2008ar,Chowdhury_2013,Hofman:2016awc}. The general current three-point correlators $\langle JJ O\rangle$ and $\langle TT O\rangle$ provide important ingredients to determine the charge conductivity and viscosity near the quantum critical points at finite temperature \cite{Witczak-Krempa:2013nua,Katz:2014rla,Witczak-Krempa:2015pia,Lucas:2016fju,Lucas:2017dqa}. The critical theories are usually strongly coupled for which the large $N$ expansion could provide useful information, like the large $N$ expansion of the QED$_3$-GNY model studied in this work. 
For example, near quantum critical points the three-point function  $\<JJS\>$  consisting of the spin-$1$ conserved current $J$ and the lowest singlet scalar $S$ is related to  the higher order correction to the conductivity at finite temperature \cite{Katz:2014rla}.
The leading $1/N$ correction to $\<JJS\>$ in the O($N$) CFT has been obtained recently in \cite{Li:2025glx}.
It would be interesting to extend the result to other cases such as conformal QED$_3$ and QED$_3$-GNY model, which describe the DQCPs and the current three-point functions could help to study the deconfined critical dynamics at finite temperature. 

Recently, there have been interesting attempts to study the higher-point conformal correlation functions in momentum space, see e.g., \cite{Bzowski:2013sza,Corian_2013, Isono:2018rrb, Isono:2019ihz, Bzowski:2019kwd,Gillioz:2019lgs,Gillioz:2020mdd,Jain:2020puw,Jain:2020rmw,Bzowski:2020kfw, Coriano:2020ees}. 
This has a close connection to the diagrammatic methods, as Feynman diagrams are usually evaluated in momentum space.
An inspiring example is provided in \cite{Li:2025glx}, which shows that the convergence of OPE in momentum space is precisely captured by the subgraph expansions of the Feynman diagrams in the limits with small/large external momenta. The convergence and associativity of the OPE in higher-point conformal correlation functions are the key ingredients for the conformal bootstrap in momentum space, while the subgraph expansions are  extremely useful to compute the conformal correlation functions perturbatively. Studies along this direction are expected to improve our understanding of the strongly coupled CFTs from both perturbative and non-perturbative aspects.


\section*{Acknowledgments}
This research was supported by the Startup Funding  4007022314 of the Southeast University and the National Natural Science Foundation of China funding No. 12375061.
\appendix

\section{$\g^\rm{QED-GNY}$ in QED$_d$-GNY model}
\label{gamma QED-GNY appendix}

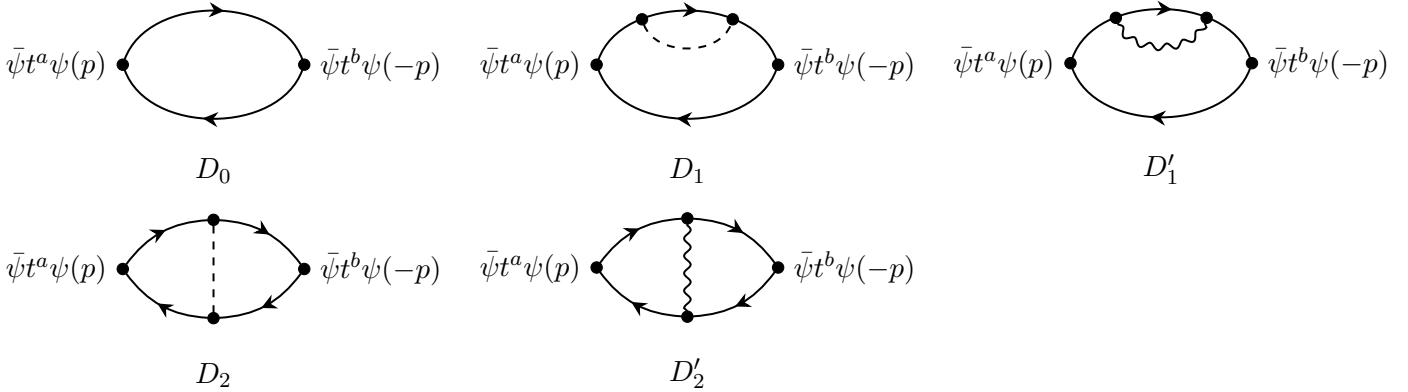
\begin{figure}[H]
    \centering
    \renewcommand{\arraystretch}{4} 
\begin{tabular}{ccc}

\hspace{-12mm}
\begin{tikzpicture}[baseline={(current bounding box.center)}]
  \begin{feynman}
    \vertex[dot, label=left:$\bar{\ps}t^a\ps(p)$] (a) at (0,0) {};
    \vertex[dot, label=right:$\bar{\ps}t^b\ps(-p)$] (b) at (2.4,0) {};
    \diagram* {
      (a) -- [fermion, bend left=70] (b) -- [fermion, bend left=70] (a)
    };
    \node at (1.2, -1.4) {$D_0$};
  \end{feynman}
\end{tikzpicture}

&

\hspace{-3mm}
\begin{tikzpicture}[baseline={(current bounding box.center)}]
  \begin{feynman}
    \vertex[dot, label=left:$\bar{\ps}t^a\ps(p)$] (a) at (0,0) {};
    \vertex[dot, label=right:$\bar{\ps}t^b\ps(-p)$] (b) at (2.4,0) {};
    \vertex[dot] (v1) at (0.6, 0.6) {};
    \vertex[dot] (v2) at (1.8, 0.6) {};
    \diagram* {
      (a)  -- [fermion, bend left=70] (b) -- [fermion, bend left=70] (a),
      (v1) -- [scalar, bend right=70] (v2)
    };
    \node at (1.2, -1.4) {$D_1$};
  \end{feynman}
\end{tikzpicture}

&

\hspace{-3mm}
\begin{tikzpicture}[baseline={(current bounding box.center)}]
  \begin{feynman}
    \vertex[dot, label=left:$\bar{\ps}t^a\ps(p)$] (a) at (0,0) {};
    \vertex[dot, label=right:$\bar{\ps}t^b\ps(-p)$] (b) at (2.4,0) {};
    \vertex[dot] (v1) at (0.6, 0.6) {};
    \vertex[dot] (v2) at (1.8, 0.6) {};
    \diagram* {
      (a)  -- [fermion, bend left=70] (b) -- [fermion, bend left=70] (a),
      (v1) -- [photon, bend right=70] (v2)
    };
    \node at (1.2, -1.4) {$D'_1$};
  \end{feynman}
\end{tikzpicture}

\\

\hspace{-12mm}
\begin{tikzpicture}[baseline={(current bounding box.center)}]
  \begin{feynman}
    \vertex[dot, label=left:$\bar{\ps}t^a\ps(p)$] (a) at (0,0) {};
    \vertex[dot, label=right:$\bar{\ps}t^b\ps(-p)$] (b) at (2.4,0) {};
    \vertex[dot] (v1) at (1.2, 0.65) {};
    \vertex[dot] (v2) at (1.2, -0.65) {};
    \diagram* {
      (a) -- [fermion, bend left=25] (v1) -- [fermion, bend left=25] (b) -- [fermion, bend left=25] (v2) -- [fermion, bend left=25] (a),
      (v1) -- [scalar] (v2)
    };
    \node at (1.2, -1.4) {$D_2$};
  \end{feynman}
\end{tikzpicture}

&

\hspace{-3mm}
\begin{tikzpicture}[baseline={(current bounding box.center)}]
  \begin{feynman}
    \vertex[dot, label=left:$\bar{\ps}t^a\ps(p)$] (a) at (0,0) {};
    \vertex[dot, label=right:$\bar{\ps}t^b\ps(-p)$] (b) at (2.4,0) {};
    \vertex[dot] (v1) at (1.2, 0.65) {};
    \vertex[dot] (v2) at (1.2, -0.65) {};
    \diagram* {
      (a) -- [fermion, bend left=25] (v1) -- [fermion, bend left=25] (b) -- [fermion, bend left=25] (v2) -- [fermion, bend left=25] (a),
      (v1) -- [photon] (v2)
    };
    \node at (1.2, -1.4) {$D'_2$};
  \end{feynman}
\end{tikzpicture}

\end{tabular}
    \caption{Diagrams corresponding to the two-point function of the adjoint fermion bilinear operator $\bar{\ps}_j(t^a)^j_k\ps^k$.}
    \label{adj diagrams}
\end{figure}

We compute the anomalous dimension of the adjoint fermion bilinear operator by considering its two-point function.
The diagrams are shown in Fig. \ref{adj diagrams}.

Here, we list the results for Fig \ref{adj diagrams}:
\begin{align}
D_0 &=\frac{\G(1-\frac d 2)\G(\frac d 2)^2\rm{tr}\bf 1}{(4\p)^{\frac d 2}\G(d-1)}\rm{tr}(t^a t^b)\, p^{2(\frac d 2-1)} \,,\nn
D_1 &= \frac{D_0\,\gamma^{\text{QED-GNY}}}{N} \left[
  \frac{(d-2)^2}{2(d-1)d}
  \left(\frac{1}{\Delta} - \log p^2 + \Psi(d)\right)
  + \frac{(d-2)(5d-4)}{2(d-1)d^2}
\right] \,,\nn
D'_1 &= \frac{D_0\,\gamma^{\text{QED-GNY}}}{N} \left[
  \left(\frac{(d-4)(d-1)}{2d} + \frac{\xi}{2}\right)
  \left(\frac{1}{\Delta} - \log p^2 + \Psi(d)\right)
  + \frac{(d-1)(5d^2 - 20d + 16)}{2(d-2)d^2}
  + \frac{\xi}{2(d-2)}
\right] \,,\nn
D_2 &= \frac{D_0\,\gamma^{\text{QED-GNY}}}{N} \left[
  \frac{d-2}{2(d-1)}
  \left(\frac{1}{\Delta} - \log p^2 + \Psi(d)\right)
  + \frac{1}{2(d-1)}
\right] \,,\nn
D'_2 &= \frac{D_0\,\gamma^{\text{QED-GNY}}}{N} \left[
  \left(-\frac{d-1}{2} - \frac{\xi}{2}\right)
  \left(\frac{1}{\Delta} - \log p^2 + \Psi(d)\right)
  + \frac{3(d-2)}{4}\,\Theta(d)
  - \frac{d-1}{2(d-2)}
  - \frac{\xi}{2(d-2)}
\right]
\,.
\end{align}
The renormalized two-point function is given by
\begin{align}
    Z_{\bar{\ps}t^a\ps}\<(\bar{\ps}t^a\ps)(p)(\bar{\ps}t^b\ps)(-p)\>=C_{\bar{\ps}t^a\ps}\,\rm{tr}(t^a t^b)\, p^{2(\D_{\bar{\ps}t^a\ps}-\frac d 2)}
    \,,
\end{align}
where $\<(\bar{\ps}t^a\ps)(p)(\bar{\ps}t^b\ps)(-p)\>=D_0+D_1+D'_1+D_2+D'_2$.
We obtain the anomalous dimension in \eqref{gamma GNY}, the renormalization factor
\begin{align}
    Z_{\bar{\ps}t^a\ps}=1+\frac{1}{N}\frac{\g^{\text{QED-GNY}}}{\D}+O(1/N^2)
    \,,\label{Zadj}
\end{align}
and the coefficient
\begin{align}
    C_{\bar{\ps}t^a\ps}=\frac{\G(1-\frac d 2)\G(\frac d 2)^2\rm{tr}\bf 1}{(4\p)^{\frac d 2}\G(d-1)}\Bigg[1+\frac{\g^{\text{QED-GNY}}}{N}\(\frac{3(d-2)}{4}\Th(d)-\Ps(d)+\frac{2d^2-9d+8}{(d-2)d}\)+O(1/N^2)\Bigg]
    \,.\label{Cadj}
\end{align}

\section{$\g^\rm{Scalar\; QED}$ in scalar QED$_d$}
\label{gamma Scalar appendix}

The anomalous dimension of the adjoint scalar operator $\f^*_j(t^a)^j_k\f^k$ can be found by studying its two-point function.
The Feynman diagrams have the same topologies as those in Fig. \ref{adj diagrams}.
The corresponding results are
\begin{align}
D_0 &=\frac{\G(2-\frac d 2)\G(\frac d 2 -1)^3}{(4\p)^{\frac d 2}\G(d-2)}\rm{tr}(t^a t^b)\, p^{2(\frac d 2-2)}\,,\nn
D_1 &= \frac{D_0\,\gamma^{\text{Scalar QED}}}{N}\left[
  -\frac{d-4}{2(d-2)\,d}
  \left(\frac{1}{\Delta} - \log p^2 + \Psi(d)\right)
  - \frac{3d^2 - 12d + 8}{2(d-2)^2\,d^2}
\right]\,,\nn
D'_1 &= \frac{D_0\,\gamma^{\text{Scalar QED}}}{N}\left[
  \left(
    -\frac{(d-1)^2}{(d-2)\,d}
    + \frac{(d-1)\,\xi}{4(d-2)}
  \right)
  \left(\frac{1}{\Delta} - \log p^2 + \Psi(d)\right)
  - \frac{2(d-1)^2}{(d-2)\,d^2}
  - \frac{(d-1)\,\xi}{2(d-2)^2}
\right]\,,\nn
D_2 &= \frac{D_0\,\gamma^{\text{Scalar QED}}}{N}\left[
  \frac{1}{4(d-2)}
  \left(\frac{1}{\Delta} - \log p^2 + \Psi(d)\right)
\right]\,,\nn
D'_2 &= \frac{D_0\,\gamma^{\text{Scalar QED}}}{N}\left[
  -\frac{(d-1)\,\xi}{4(d-2)}
  \left(\frac{1}{\Delta} - \log p^2 + \Psi(d)\right)
  + \frac{3(d-1)}{8}\,\Theta(d)
  + \frac{(d-1)\,\xi}{2(d-2)^2}
\right]
\,.
\end{align}
We define the renormalized two-point function as
\begin{align}
    Z_{\f^*t^a\f}\<(\f^*t^a\f)(p)(\f^*t^b\f)(-p)\>=C_{\f^*t^a\f}\,\rm{tr}(t^a t^b)\, p^{2(\D_{\f^*t^a\f}-\frac d 2)}
    \,,
\end{align}
where $\<(\f^*t^a\f)(p)(\f^*t^b\f)(-p)\>=D_0+D_1+D'_1+D_2+D'_2$.
The result for the anomalous dimension is given in \eqref{gamma scalar}.
The renormalization factor is
\begin{align}
    Z_{\f^*t^a\f}=1+\frac{1}{N}\frac{\g^{\text{Scalar QED}}}{\D}+O(1/N^2)
    \,,\label{Zadj sQED}
\end{align}
and the coefficient reads
\begin{align}
    C_{\f^*t^a\f}=\frac{\G(2-\frac{d}{2})\G(\frac{d}{2}-1)^2}{(4\p)^{\frac{d}{2}}\G(d-2)}
    \[1+\frac{\g^\text{Scalar}}{N}\(\frac{3(d-1)}{8}\,\Th(d)-\Ps(d)-\frac{4d^2-13d+8}{2(d-2)^2 d}\)+O(1/N^2)\]
    \,.\label{Cadj sQED}
\end{align}

\section{$Z_T$ in QED$_d$-GNY model}
\label{ZT QED-GNY appendix}

We consider the three-point function $Z_T Z_{\bar{\ps}t^a\ps}\<T_{\m\n}(0)(\bar{\ps}t^a\ps)(p)(\bar{\ps}t^b\ps)(-p)\>$, whose coefficient is fixed as
\begin{align}
    Z_T Z_{\bar{\ps}t^a\ps}\<T_{\m\n}(0)(\bar{\ps}t^a\ps)(p)(\bar{\ps}t^b\ps)(-p)\>=(d-2\D_{\bar{\ps}t^a\ps})\frac{C_{\bar{\ps}t^a\ps}\,\rm{tr}(t^a t^b)}{p^{2(\frac d 2-\D_{\bar{\ps}t^a\ps})}}\(\frac{p_\m p_\n}{p^2}-\frac{\d_{\m\n}}{d}\)
    \,,\label{ZT matching}
\end{align}
due to the Ward identity for the stress-energy tensor.
The renormalization factor $Z_{\bar{\ps}t^a\ps}$, the two-point function coefficient $C_{\bar{\ps}t^a\ps}$, and the scaling dimension $\D_{\bar{\ps}t^a\ps}$ are given by \eqref{Zadj}, \eqref{Cadj}, and \eqref{Delta adj}.
The Feynman diagrams contributing to the three-point function are shown in Fig. \ref{ZT diagrams}, and the results are given below.
Using
\begin{align}
    \<T_{\m\n}(0)(\bar{\ps}t^a\ps)(p)(\bar{\ps}t^b\ps)(-p)\>&=D_0+D_1+D'_1+D_2+D'_2+D_3+D'_3+D_4+D'_4\nn
    &\hspace{0.5cm}+D_5+D'_5+D_6+D'_6+D_7+D_8+D_9+D_{10}
    \,,
\end{align}
the condition \eqref{ZT matching} yields the $Z_T$ factor \eqref{ZT QED-GNY}.
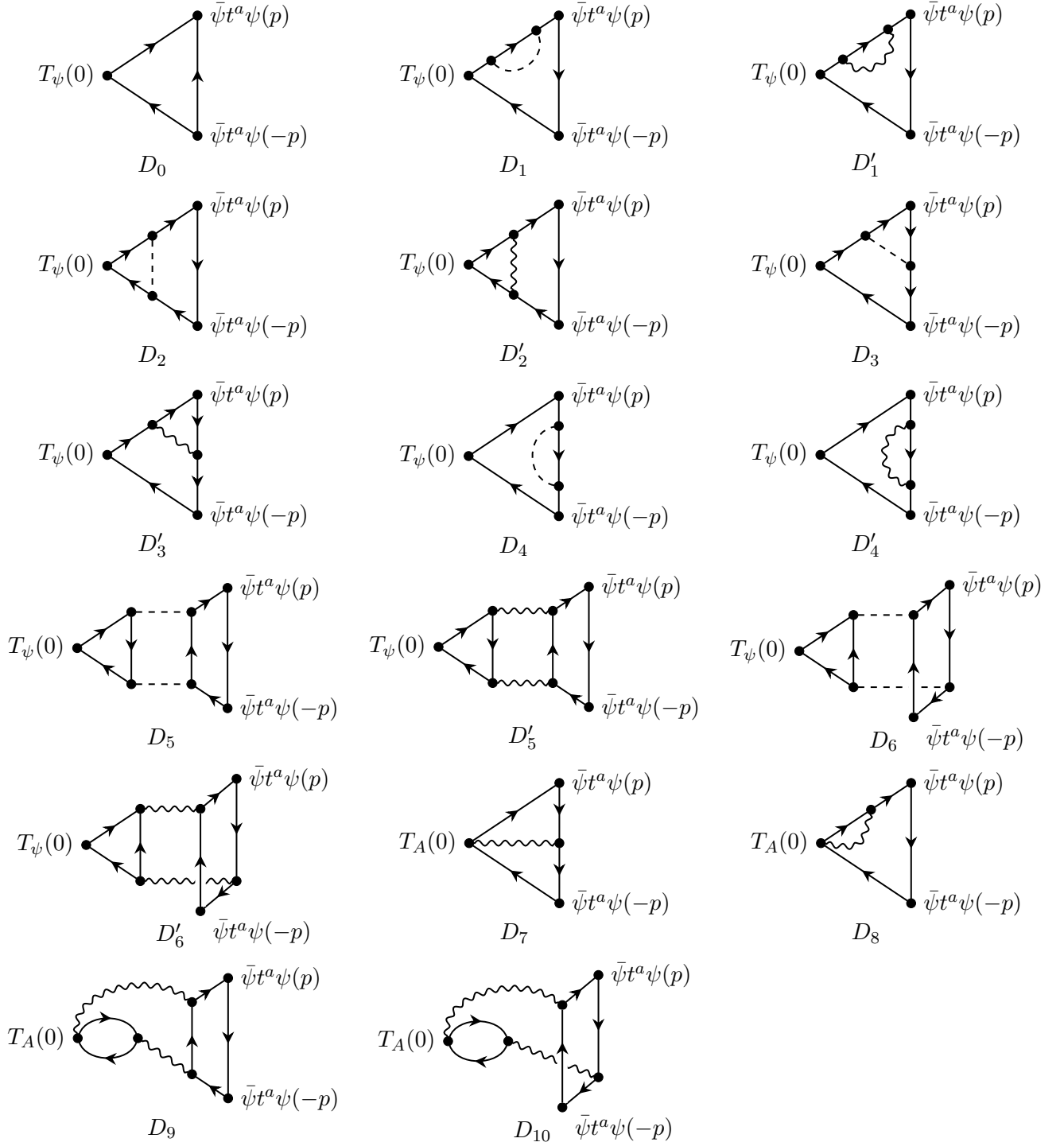
\begin{figure}[H]
    \centering
    \renewcommand{\arraystretch}{4}
\begin{tabular}{ccc}

\hspace{-9mm}
\begin{tikzpicture}[baseline={(current bounding box.center)}]
  \begin{feynman}
    \vertex[dot, label=left:$T_\ps(0)$] (a) at (0,0) {};
    \vertex[dot, label=right:$\bar{\ps}t^a\ps(p)$] (t) at (1.5, 1) {};
    \vertex[dot, label=right:$\bar{\ps}t^a\ps(-p)$] (b) at (1.5, -1) {};
    \diagram* {
      (a) -- [fermion] (t),
      (b) -- [fermion] (a),
      (b) -- [fermion] (t)
    };
    \node at (0.75, -1.5) {$D_0$};
  \end{feynman}
\end{tikzpicture}

&

\hspace{-3mm}
\begin{tikzpicture}[baseline={(current bounding box.center)}]
  \begin{feynman}
    \vertex[dot, label=left:$T_\ps(0)$] (a) at (0,0) {};
    \vertex[dot, label=right:$\bar{\ps}t^a\ps(p)$] (t) at (1.5, 1) {};
    \vertex[dot, label=right:$\bar{\ps}t^a\ps(-p)$] (b) at (1.5, -1) {};
    \vertex[dot] (v1) at (3/8, 1/4) {};
    \vertex[dot] (v2) at (9/8, 3/4) {};
    \diagram* {
      (a) -- [fermion] (t),
      (b) -- [fermion] (a),
      (t) -- [fermion] (b),
      (v1) -- [scalar, bend right=80, looseness=1.3] (v2)
    };
    \node at (0.75, -1.5) {$D_1$};
  \end{feynman}
\end{tikzpicture}

&

\hspace{-3mm}
\begin{tikzpicture}[baseline={(current bounding box.center)}]
  \begin{feynman}
    \vertex[dot, label=left:$T_\ps(0)$] (a) at (0,0) {};
    \vertex[dot, label=right:$\bar{\ps}t^a\ps(p)$] (t) at (1.5, 1) {};
    \vertex[dot, label=right:$\bar{\ps}t^a\ps(-p)$] (b) at (1.5, -1) {};
    \vertex[dot] (v1) at (3/8, 1/4) {};
    \vertex[dot] (v2) at (9/8, 3/4) {};
    \diagram* {
      (a) -- [fermion] (t),
      (b) -- [fermion] (a),
      (t) -- [fermion] (b),
      (v1) -- [photon, bend right=80, looseness=1.3] (v2)
    };
    \node at (0.75, -1.5) {$D'_1$};
  \end{feynman}
\end{tikzpicture}

\\

\hspace{-9mm}
\begin{tikzpicture}[baseline={(current bounding box.center)}]
  \begin{feynman}
    \vertex[dot, label=left:$T_\ps(0)$] (a) at (0,0) {};
    \vertex[dot, label=right:$\bar{\ps}t^a\ps(p)$] (t) at (1.5, 1) {};
    \vertex[dot, label=right:$\bar{\ps}t^a\ps(-p)$] (b) at (1.5, -1) {};
    \vertex[dot] (v1) at (0.75, 0.5) {};
    \vertex[dot] (v2) at (0.75, -0.5) {};
    \diagram* {
      (a) -- [fermion] (v1) -- [fermion] (t),
      (b) -- [fermion] (v2) -- [fermion] (a),
      (t) -- [fermion] (b),
      (v1) -- [scalar] (v2)
    };
    \node at (0.75, -1.5) {$D_2$};
  \end{feynman}
\end{tikzpicture}

&

\hspace{-3mm}
\begin{tikzpicture}[baseline={(current bounding box.center)}]
  \begin{feynman}
    \vertex[dot, label=left:$T_\ps(0)$] (a) at (0,0) {};
    \vertex[dot, label=right:$\bar{\ps}t^a\ps(p)$] (t) at (1.5, 1) {};
    \vertex[dot, label=right:$\bar{\ps}t^a\ps(-p)$] (b) at (1.5, -1) {};
    \vertex[dot] (v1) at (0.75, 0.5) {};
    \vertex[dot] (v2) at (0.75, -0.5) {};
    \diagram* {
      (a) -- [fermion] (v1) -- [fermion] (t),
      (b) -- [fermion] (v2) -- [fermion] (a),
      (t) -- [fermion] (b),
      (v1) -- [photon] (v2)
    };
    \node at (0.75, -1.5) {$D'_2$};
  \end{feynman}
\end{tikzpicture}

&

\hspace{-3mm}
\begin{tikzpicture}[baseline={(current bounding box.center)}]
  \begin{feynman}
    \vertex[dot, label=left:$T_\ps(0)$] (a) at (0,0) {};
    \vertex[dot, label=right:$\bar{\ps}t^a\ps(p)$] (t) at (1.5, 1) {};
    \vertex[dot, label=right:$\bar{\ps}t^a\ps(-p)$] (b) at (1.5, -1) {};
    \vertex[dot] (v1) at (0.75, 0.5) {};
    \vertex[dot] (v2) at (1.5, 0) {};
    \diagram* {
      (a) -- [fermion] (v1) -- [fermion] (t),
      (t) -- [fermion] (v2) -- [fermion] (b),
      (b) -- [fermion] (a),
      (v1) -- [scalar] (v2)
    };
    \node at (0.75, -1.5) {$D_3$};
  \end{feynman}
\end{tikzpicture}

\\

\hspace{-9mm}
\begin{tikzpicture}[baseline={(current bounding box.center)}]
  \begin{feynman}
    \vertex[dot, label=left:$T_\ps(0)$] (a) at (0,0) {};
    \vertex[dot, label=right:$\bar{\ps}t^a\ps(p)$] (t) at (1.5, 1) {};
    \vertex[dot, label=right:$\bar{\ps}t^a\ps(-p)$] (b) at (1.5, -1) {};
    \vertex[dot] (v1) at (0.75, 0.5) {};
    \vertex[dot] (v2) at (1.5, 0) {};
    \diagram* {
      (a) -- [fermion] (v1) -- [fermion] (t),
      (t) -- [fermion] (v2) -- [fermion] (b),
      (b) -- [fermion] (a),
      (v1) -- [photon] (v2)
    };
    \node at (0.75, -1.5) {$D'_3$};
  \end{feynman}
\end{tikzpicture}

&

\hspace{-3mm}
\begin{tikzpicture}[baseline={(current bounding box.center)}]
  \begin{feynman}
    \vertex[dot, label=left:$T_\ps(0)$] (a) at (0,0) {};
    \vertex[dot, label=right:$\bar{\ps}t^a\ps(p)$] (t) at (1.5, 1) {};
    \vertex[dot, label=right:$\bar{\ps}t^a\ps(-p)$] (b) at (1.5, -1) {};
    \vertex[dot] (v1) at (1.5, -0.5) {};
    \vertex[dot] (v2) at (1.5, 0.5) {};
    \diagram* {
      (a) -- [fermion] (t),
      (b) -- [fermion] (a),
      (t) -- [fermion] (b),
      (v1) -- [scalar, bend left=80, looseness=1.3] (v2)
    };
    \node at (0.75, -1.5) {$D_4$};
  \end{feynman}
\end{tikzpicture}

&

\hspace{-3mm}
\begin{tikzpicture}[baseline={(current bounding box.center)}]
  \begin{feynman}
    \vertex[dot, label=left:$T_\ps(0)$] (a) at (0,0) {};
    \vertex[dot, label=right:$\bar{\ps}t^a\ps(p)$] (t) at (1.5, 1) {};
    \vertex[dot, label=right:$\bar{\ps}t^a\ps(-p)$] (b) at (1.5, -1) {};
    \vertex[dot] (v1) at (1.5, -0.5) {};
    \vertex[dot] (v2) at (1.5, 0.5) {};
    \diagram* {
      (a) -- [fermion] (t),
      (b) -- [fermion] (a),
      (t) -- [fermion] (b),
      (v1) -- [photon, bend left=80, looseness=1.3] (v2)
    };
    \node at (0.75, -1.5) {$D'_4$};
  \end{feynman}
\end{tikzpicture}

\\

\hspace{-9mm}
\begin{tikzpicture}[baseline={(current bounding box.center)}]
  \begin{feynman}
    \vertex[dot, label=left:$T_\ps(0)$] (a) at (0,0) {};
    \vertex[dot] (t1) at (0.9, 0.6) {};
    \vertex[dot] (b1) at (0.9, -0.6) {};
    \vertex[dot] (t2) at (1.9, 0.6) {};
    \vertex[dot] (b2) at (1.9, -0.6) {};
    \vertex[dot, label=right:$\bar{\ps}t^a\ps(p)$] (t3) at (2.5, 1) {};
    \vertex[dot, label=right:$\bar{\ps}t^a\ps(-p)$] (b3) at (2.5, -1) {};
    \diagram* {
      (a) -- [fermion] (t1),
      (b1) -- [fermion] (a),
      (t1) -- [fermion] (b1),
      (t1) -- [scalar] (t2),
      (b1) -- [scalar] (b2),
      (b2) -- [fermion] (t2),
      (t2) -- [fermion] (t3),
      (b3) -- [fermion] (b2),
      (t3) -- [fermion] (b3)
    };
    \node at (1.4, -1.5) {$D_5$};
  \end{feynman}
\end{tikzpicture}

&

\hspace{-3mm}
\begin{tikzpicture}[baseline={(current bounding box.center)}]
  \begin{feynman}
    \vertex[dot, label=left:$T_\ps(0)$] (a) at (0,0) {};
    \vertex[dot] (t1) at (0.9, 0.6) {};
    \vertex[dot] (b1) at (0.9, -0.6) {};
    \vertex[dot] (t2) at (1.9, 0.6) {};
    \vertex[dot] (b2) at (1.9, -0.6) {};
    \vertex[dot, label=right:$\bar{\ps}t^a\ps(p)$] (t3) at (2.5, 1) {};
    \vertex[dot, label=right:$\bar{\ps}t^a\ps(-p)$] (b3) at (2.5, -1) {};
    \diagram* {
      (a) -- [fermion] (t1),
      (b1) -- [fermion] (a),
      (t1) -- [fermion] (b1),
      (t1) -- [photon] (t2),
      (b1) -- [photon] (b2),
      (b2) -- [fermion] (t2),
      (t2) -- [fermion] (t3),
      (b3) -- [fermion] (b2),
      (t3) -- [fermion] (b3)
    };
    \node at (1.4, -1.5) {$D'_5$};
  \end{feynman}
\end{tikzpicture}

&

\hspace{-3mm}
\begin{tikzpicture}[baseline={(current bounding box.center)}]
  \begin{feynman}
    \vertex[dot, label=left:$T_\ps(0)$] (a) at (0,0) {};
    \vertex[dot] (t1) at (0.9, 0.6) {};
    \vertex[dot] (b1) at (0.9, -0.6) {};
    \vertex[dot] (t2) at (1.9, 0.6) {};
    \vertex[dot, label=-3:$\bar{\ps}t^a\ps(-p)$] (b2) at (1.9, -1.1) {};
    \vertex[dot, label=right:$\bar{\ps}t^a\ps(p)$] (t3) at (2.5, 1.1) {};
    \vertex[dot] (b3) at (2.5, -0.6) {};
    \diagram* {
      (a) -- [fermion] (t1),
      (b1) -- [fermion] (a),
      (b1) -- [fermion] (t1),
      (t1) -- [scalar] (t2),
      (b1) -- [scalar] (b3),
      (b3) -- [fermion] (b2),
      (t2) -- [fermion] (t3),
      (t3) -- [fermion] (b3),
      (b2) -- [fermion, preaction={draw, white, line width=5pt}] (t2)
    };
    \node at (1.4, -1.5) {$D_6$};
  \end{feynman}
\end{tikzpicture}

\\

\hspace{-9mm}
\begin{tikzpicture}[baseline={(current bounding box.center)}]
  \begin{feynman}
    \vertex[dot, label=left:$T_\ps(0)$] (a) at (0,0) {};
    \vertex[dot] (t1) at (0.9, 0.6) {};
    \vertex[dot] (b1) at (0.9, -0.6) {};
    \vertex[dot] (t2) at (1.9, 0.6) {};
    \vertex[dot, label=-3:$\bar{\ps}t^a\ps(-p)$] (b2) at (1.9, -1.1) {};
    \vertex[dot, label=right:$\bar{\ps}t^a\ps(p)$] (t3) at (2.5, 1.1) {};
    \vertex[dot] (b3) at (2.5, -0.6) {};
    \diagram* {
      (a) -- [fermion] (t1),
      (b1) -- [fermion] (a),
      (b1) -- [fermion] (t1),
      (t1) -- [photon] (t2),
      (b1) -- [photon] (b3),
      (b3) -- [fermion] (b2),
      (t2) -- [fermion] (t3),
      (t3) -- [fermion] (b3),
      (b2) -- [fermion, preaction={draw, white, line width=5pt}] (t2)
    };
    \node at (1.4, -1.5) {$D'_6$};
  \end{feynman}
\end{tikzpicture}

&

\hspace{-3mm}
\begin{tikzpicture}[baseline={(current bounding box.center)}]
  \begin{feynman}
    \vertex[dot, label=left:$T_A(0)$] (a) at (0,0) {};
    \vertex[dot, label=right:$\bar{\ps}t^a\ps(p)$] (t) at (1.5, 1) {};
    \vertex[dot, label=right:$\bar{\ps}t^a\ps(-p)$] (b) at (1.5, -1) {};
    \vertex[dot] (v) at (1.5, 0) {};
    \diagram* {
      (a) -- [fermion] (t),
      (t) -- [fermion] (v) -- [fermion] (b),
      (b) -- [fermion] (a),
      (a) -- [photon] (v)
    };
    \node at (0.75, -1.5) {$D_7$};
  \end{feynman}
\end{tikzpicture}

&

\hspace{-3mm}
\begin{tikzpicture}[baseline={(current bounding box.center)}]
  \begin{feynman}
    \vertex[dot, label=left:$T_A(0)$] (a) at (0,0) {};
    \vertex[dot, label=right:$\bar{\ps}t^a\ps(p)$] (t) at (1.5, 1) {};
    \vertex[dot, label=right:$\bar{\ps}t^a\ps(-p)$] (b) at (1.5, -1) {};
    \vertex[dot] (v1) at (15/18, 10/18) {};
    \diagram* {
      (a) -- [fermion] (v1) -- [fermion] (t),
      (b) -- [fermion] (a),
      (t) -- [fermion] (b),
      (a) -- [photon, bend right=43, looseness=1.7] (v1)
    };
    \node at (0.75, -1.5) {$D_8$};
  \end{feynman}
\end{tikzpicture}

\\

\hspace{-9mm}
\begin{tikzpicture}[baseline={(current bounding box.center)}]
  \begin{feynman}
    \vertex[dot, label=left:$T_A(0)$] (a) at (0,0) {};
    \vertex[dot] (v1) at (1, 0) {};
    \vertex[dot] (t2) at (1.9, 0.6) {};
    \vertex[dot] (b2) at (1.9, -0.6) {};
    \vertex[dot, label=right:$\bar{\ps}t^a\ps(p)$] (t3) at (2.5, 1) {};
    \vertex[dot, label=right:$\bar{\ps}t^a\ps(-p)$] (b3) at (2.5, -1) {};
    \diagram* {
      (a) -- [fermion, bend left=70] (v1),
      (v1) -- [fermion, bend left=70] (a),
      (a) -- [photon, out=100, in=140] (t2),
      (v1) -- [photon] (b2),
      (b2) -- [fermion] (t2),
      (t2) -- [fermion] (t3),
      (b3) -- [fermion] (b2),
      (t3) -- [fermion] (b3)
    };
    \node at (1.4, -1.5) {$D_9$};
  \end{feynman}
\end{tikzpicture}

&

\hspace{-3mm}
\begin{tikzpicture}[baseline={(current bounding box.center)}]
  \begin{feynman}
    \vertex[dot, label=left:$T_A(0)$] (a) at (0,0) {};
    \vertex[dot] (v1) at (1, 0) {};
    \vertex[dot] (t2) at (1.9, 0.6) {};
    \vertex[dot, label=-3:$\bar{\ps}t^a\ps(-p)$] (b2) at (1.9, -1.1) {};
    \vertex[dot, label=right:$\bar{\ps}t^a\ps(p)$] (t3) at (2.5, 1.1) {};
    \vertex[dot] (b3) at (2.5, -0.6) {};
    \diagram* {
      (a) -- [fermion, bend left=70] (v1),
      (v1) -- [fermion, bend left=70] (a),
      (a) -- [photon, out=100, in=140] (t2),
      (v1) -- [photon] (b3),
      (b3) -- [fermion] (b2),
      (t2) -- [fermion] (t3),
      (t3) -- [fermion] (b3),
      (b2) -- [fermion shifted=0.65, preaction={draw, white, line width=5pt}] (t2)
    };
    \node at (1.4, -1.5) {$D_{10}$};
  \end{feynman}
\end{tikzpicture}

&

\end{tabular}
    \caption{Diagrams for computing $Z_T$ to order $1/N$.
    For some topologies, the diagrams with different fermion loop directions are not explicitly drawn, but they need to be included in the calculation.}
    \label{ZT diagrams}
\end{figure}

The results for Fig. \ref{ZT diagrams} are
\begin{align}
D_0 &=-\frac{\G(1-\frac d 2)\G(\frac d 2)^2\rm{tr}\bf 1}{(4\p)^{\frac d 2}\G(d-2)}\rm{tr}(t^a t^b)\(\frac{p_\m p_\n}{p^2}-\frac{\d_{\m\n}}{d}\)p^{2(\frac d 2-1)}\,,\nn
D_1 &= \frac{D_0\,\gamma^{\text{QED-GNY}}}{N} \left[
  \frac{(d-2)^2}{2(d-1)d}
  \left(\frac{1}{\Delta} - \log p^2 + \Psi(d)\right)
  - \frac{(d-4)(d-2)}{2d^2}
\right] \,,\nn
D'_1 &= \frac{D_0\,\gamma^{\text{QED-GNY}}}{N} \left[
  \left(\frac{(d-4)(d-1)}{2d} + \frac{\xi}{2}\right)
  \left(\frac{1}{\Delta} - \log p^2 + \Psi(d)\right)\right.
  \nn
  &\hspace{2.5cm}\left.- \frac{(d-1)(d^3 - 9d^2 + 20d - 16)}{2(d-2)d^2}
  - \frac{(d-1)\xi}{2(d-2)}\right] \,,\nn
D_2 &= \frac{D_0\,\gamma^{\text{QED-GNY}}}{N} \left[
  -\frac{(d-2)^3}{4(d-1)d(d+2)}
  \left(\frac{1}{\Delta} - \log p^2 + \Psi(d)\right)
  + \frac{(d-4)(d-2)^2(2d^3 - 3d + 2)}{4d^2(d^2+d-2)^2}
\right] \,,\nn
D'_2 &= \frac{D_0\,\gamma^{\text{QED-GNY}}}{N} \left[
  \left(-\frac{d^3 - 7d^2 + 10d - 8}{4d(d+2)} - \frac{\xi}{4}\right)
  \left(\frac{1}{\Delta} - \log p^2 + \Psi(d)\right) \right. \nn
  &\qquad\left.
  + \frac{2d^7 - 22d^6 + 69d^5 - 69d^4 - 76d^3 + 220d^2 - 176d + 64}
         {4(d-2)(d-1)d^2(d+2)^2}
  + \frac{(2d^3 - 8d^2 + 13d - 8)\,\xi}{4(d-2)(d-1)d}
\right] \,,\nn
D_3 &= \frac{D_0\,\gamma^{\text{QED-GNY}}}{N} \left[
  \frac{d-2}{2(d-1)}
  \left(\frac{1}{\Delta} - \log p^2 + \Psi(d)\right)
  - \frac{2d^2 - 5d + 4}{2(d-1)^2 d}
\right] \,,\nn
D'_3 &= \frac{D_0\,\gamma^{\text{QED-GNY}}}{N} \left[
  \left(-\frac{d-1}{2} - \frac{\xi}{2}\right)
  \left(\frac{1}{\Delta} - \log p^2 + \Psi(d)\right)
  + \frac{3(d-2)^2}{4(d-1)}\,\Theta(d)\right.
  \nn
  &\hspace{2.5cm}\left.
  + \frac{d^2 + d - 4}{2(d-2)(d-1)d}
  + \frac{(2d^2 - 5d + 4)\,\xi}{2(d-2)(d-1)d}
\right] \,,\nn
D_4 &= \frac{D_0\,\gamma^{\text{QED-GNY}}}{N} \left[
  \frac{(d-2)^2}{4(d-1)d}
  \left(\frac{1}{\Delta} - \log p^2 + \Psi(d)\right)
  + \frac{(d-2)(4d^2 - 9d + 4)}{4(d-1)^2 d^2}
\right] \,,\nn
D'_4 &= \frac{D_0\,\gamma^{\text{QED-GNY}}}{N} \left[
  \left(\frac{(d-4)(d-1)}{4d} + \frac{\xi}{4}\right)
  \left(\frac{1}{\Delta} - \log p^2 + \Psi(d)\right)\right.
  \nn
  &\hspace{2.5cm}\left.
  + \frac{4d^3 - 21d^2 + 36d - 16}{4(d-2)d^2}
  - \frac{\xi}{4(d-2)(d-1)}
\right] \,,\nn
D_5 &= \frac{D_0\,\gamma^{\text{QED-GNY}}}{N} \left[
  -\frac{(d-2)^2}{2(d-1)d(d+2)}
  \left(\frac{1}{\Delta} - 2\log p^2 + 2\Psi(d)\right)\right.
  \nn
  &\hspace{2.5cm}\left.
  + \frac{(d-2)(5d^4 - 12d^3 - 2d^2 + 28d - 16)}{2(d-1)^2 d^2(d+2)^2}
\right] \,,\nn
D'_5 &= \frac{D_0\,\gamma^{\text{QED-GNY}}}{N} \left[
  -\frac{(d-2)^2}{2d(d+2)}
  \left(\frac{1}{\Delta} - 2\log p^2 + 2\Psi(d)\right)\right.
  \nn
  &\hspace{2.5cm}\left.
  + \frac{(d-2)(5d^4 - 11d^3 + 2d^2 + 32d - 16)}{2(d-1)d^2(d+2)^2}
  + \frac{(d-2)\,\xi}{(d-1)d}
\right] \,,\nn
D_6 &= \frac{D_0\,\gamma^{\text{QED-GNY}}}{N}
  \(\frac{(d-2)^2}{2(d-1)^2 d}\) \,,\nn
D'_6 &= \frac{D_0\,\gamma^{\text{QED-GNY}}}{N} \left[
  \frac{3(d-2)}{4(d-1)}\,\Theta(d)
  + \frac{d-2}{2d}
  - \frac{(d-2)\,\xi}{(d-1)d}
\right] \,,\nn
D_7 &= \frac{D_0\,\gamma^{\text{QED-GNY}}}{N} \left[
  \frac{3(d-2)}{4(d-1)}\,\Theta(d)
  + \frac{1-\x}{d}
\right] \,,\nn
D_8 &= \frac{D_0\,\gamma^{\text{QED-GNY}}}{N} \left[
  \frac{d-2}{d}
  \left(\frac{1}{\Delta} - \log p^2 + \Psi(d)\right)
  - \frac{(d-2)^2}{d^2}
  + \frac{\xi}{d}
\right] \,,\nn
D_9 &= \frac{D_0\,\gamma^{\text{QED-GNY}}}{N} \left[
  -\frac{d-2}{2d}
  \left(\frac{1}{\Delta} - 2\log p^2 + 2\Psi(d)\right)
  + \frac{(d-2)(d^2 - 2d + 2)}{(d-1)d^2}
  - \frac{(d-2)\,\xi}{(d-1)d}
\right] \,,\nn
D_{10} &= \frac{D_0\,\gamma^{\text{QED-GNY}}}{N} \left[
  -\frac{3(d-2)}{4(d-1)}\,\Theta(d)
  - \frac{2d-3}{(d-1)d}
  + \frac{(d-2)\,\xi}{(d-1)d}
\right]
\,.
\end{align}

\section{$Z_T$ in scalar QED$_d$}
\label{ZT sQED appendix}

To compute $Z_T$ in the scalar QED$_d$, the procedure is analogous to the QED$_d$-GNY case, with the adjoint fermion bilinear operator replaced by $\f^* t^a \f$.
For the adjoint scalar bilinear operator $\f^* t^a \f$, we have obtained the renormalization factor $Z_{\f^* t^a \f}$, the two-point function coefficient $C_{\f^* t^a \f}$, and the scaling dimension $\D_{\f^* t^a \f}$ in \eqref{Zadj sQED}, \eqref{Cadj sQED}, and \eqref{Delta adj sQED}.
The condition
\begin{align}
    Z_T Z_{\f^* t^a \f}\<T_{\m\n}(0)(\f^* t^a \f)(p)(\f^* t^a \f)(-p)\>=(d-2\D_{\f^* t^a \f})\frac{C_{\f^* t^a \f}\,\rm{tr}(t^a t^b)}{p^{2(\frac d 2-\D_{\f^* t^a \f})}}\(\frac{p_\m p_\n}{p^2}-\frac{\d_{\m\n}}{d}\)
\end{align}
leads to the determination of the $Z_T$ factor \eqref{ZT scalar QED}.
Here we use
\begin{align}
    \<T_{\m\n}(0)(\f^* t^a \f)(p)(\f^* t^a \f)(-p)\>&=D_0+D_1+D'_1+D_2+D'_2+D_3+D'_3+D_4+D'_4\nn
    &\hspace{0.5cm}+D_5+D'_5+D_6+D'_6+D_7+D_8+D_9+D_{10}
    \,,
\end{align}
where the Feynman diagrams have the same topologies as those in Fig. \ref{ZT diagrams}.

The results for the Feynman diagrams are
\begin{align}
D_0 &= -\frac{(d-4)\,\Gamma(2-\frac d 2)\,\Gamma(\frac d 2-1)^2}{(4\p)^{\frac d 2}\Gamma(d-2)}\rm{tr}(t^a t^b)
\left(\frac{p_{\m}p_{\n}}{p^2}- \frac{\delta_{\m\n}}{d}\right)p^{2(\frac d 2-2)}
\,,\nn
D_1 &= \frac{D_0\,\gamma^{\text{Scalar QED}}}{N}\left[
  -\frac{d-4}{4(d-2)d}\!\left(\frac{1}{\Delta}-\log p^2+\Psi(d)\right)
  +\frac{(d-4)^2(d-1)}{4(d-2)^2 d^2}
\right]
\,,\nn
D'_1 &=\frac{D_0\,\gamma^{\text{Scalar QED}}}{N}\left[
  \left(-\frac{(d-1)^2}{(d-2)d}+\frac{(d-1)\xi}{4(d-2)}\right)
  \!\left(\frac{1}{\Delta}-\log p^2+\Psi(d)\right)\right.
  \nonumber\\&\hspace{2.5cm}\left.
  +\frac{(d-1)^2(d^3-5d^2+12d-16)}{(d-4)(d-2)^2 d^2}
  -\frac{(d-1)(d^2-d-8)\,\xi}{4(d-4)(d-2)^2}
\right]
\,,\nn
D_2 &=\frac{D_0\,\gamma^{\text{Scalar QED}}}{N}\left[
  \frac{d-4}{8d(d+2)}\!\left(\frac{1}{\Delta}-\log p^2+\Psi(d)\right)
  -\frac{(d-4)(2d^3-5d^2-6d+8)}{8(d-2)d^2(d+2)^2}
\right]
\,,\nn
D'_2 &=\frac{D_0\,\gamma^{\text{Scalar QED}}}{N}\left[
  \left(\frac{d-1}{(d-2)d(d+2)}-\frac{(d-1)\xi}{8(d-2)}\right)
  \!\left(\frac{1}{\Delta}-\log p^2+\Psi(d)\right)\right.
  \nonumber\\&\quad\left.
  -\frac{(d-1)(d^6-5d^5+4d^4+2d^3+4d^2+16d-64)}{2(d-4)(d-2)^2 d^2(d+2)^2}
  +\frac{(d-1)(2d^3-13d^2+32d-32)\,\xi}{8(d-4)(d-2)^2 d}
\right]
\,,\nn
D_3 &= \frac{D_0\,\gamma^{\text{Scalar QED}}}{N}\left[
  \frac{1}{4(d-2)}\!\left(\frac{1}{\Delta}-\log p^2+\Psi(d)\right)
  -\frac{3d^2-12d+16}{4(d-4)(d-2)^2 d}
\right]
\,,\nn
D'_3 &=\frac{D_0\,\gamma^{\text{Scalar QED}}}{N}\left[
  -\frac{(d-1)\xi}{4(d-2)}\!\left(\frac{1}{\Delta}-\log p^2+\Psi(d)\right)
  +\frac{3(d^2-7d+8)}{8(d-4)}\,\Theta(d)-\frac{d-1}{2(d-4)}
  \right.
  \nonumber\\&\hspace{2.5cm}\left.
  +\frac{(d-1)(7d^2-28d+16)\,\xi}{4(d-4)(d-2)^2 d}
\right]
\,,\nn
D_4 &= \frac{D_0\,\gamma^{\text{Scalar QED}}}{N}\left[
  -\frac{d-4}{8(d-2)d}\!\left(\frac{1}{\Delta}-\log p^2+\Psi(d)\right)
  -\frac{(d-4)(5d-4)}{8(d-2)^2 d^2}
\right]
\,,\nn
D'_4 &=\frac{D_0\,\gamma^{\text{Scalar QED}}}{N}\left[
  \left(-\frac{(d-1)^2}{2(d-2)d}+\frac{(d-1)\xi}{8(d-2)}\right)
  \!\left(\frac{1}{\Delta}-\log p^2+\Psi(d)\right)\right.
  \nonumber\\&\hspace{2.5cm}\left.
  -\frac{(d-1)^2(d^2-12d+16)}{2(d-4)(d-2)^2 d^2}
  -\frac{(d-1)(3d-8)\,\xi}{8(d-4)(d-2)^2}
\right]
\,,\nn
D_5 &=\frac{D_0\,\gamma^{\text{Scalar QED}}}{N}\left[
  \frac{d-4}{4(d-2)d(d+2)}\!\left(\frac{1}{\Delta}-2\log p^2+2\Psi(d)\right)
  -\frac{5d^4-26d^3+24d^2+64d-64}{4(d-2)^2 d^2(d+2)^2}
\right]
\,,\nn
D'_5 &=\frac{D_0\,\gamma^{\text{Scalar QED}}}{N}\left[
  \frac{(d-1)(d^2+d-4)}{4(d-2)d(d+2)}\!\left(\frac{1}{\Delta}-2\log p^2+2\Psi(d)\right)\right.
  \nonumber\\&\hspace{2.5cm}\left.
  -\frac{(d-1)(d^6-15d^4+26d^3-16d^2-64d+128)}{2(d-4)(d-2)^2 d^2(d+2)^2}
  +\frac{(d-1)\xi}{(d-2)d}
\right]
\,,\nn
D_6 &= \frac{D_0\,\gamma^{\text{Scalar QED}}}{N}\(\frac{d-4}{4(d-2)^2 d}\)
\,,\nn
D'_6 &=\frac{D_0\,\gamma^{\text{Scalar QED}}}{N}\left[
  \frac{3(d-2)}{4(d-4)}\,\Theta(d)
  +\frac{d+1}{2(d-4)}
  -\frac{(d-1)\xi}{(d-2)d}
\right]
\,,\nn
D_7 &=\frac{D_0\,\gamma^{\text{Scalar QED}}}{N}\left[
  \frac{3(d-2)}{4(d-4)}\,\Theta(d)
  +\frac{d-1}{(d-4)(d-2)}
  -\frac{(d-1)^2\xi}{(d-2)^2 d}
\right]
\,,\nn
D_8 &=\frac{D_0\,\gamma^{\text{Scalar QED}}}{N}\left[
  \frac{(d-1)(d^2-2)}{(d-2)d(d+2)}\!\left(\frac{1}{\Delta}-\log p^2+\Psi(d)\right)\right.
  \nonumber\\&\hspace{2.5cm}\left.
  -\frac{(d-1)(d^6-4d^5+6d^4+6d^3-36d^2-16d+64)}{(d-4)(d-2)^2 d^2(d+2)^2}
  +\frac{(d-1)^2\xi}{(d-2)^2 d}
\right]
\,,\nn
D_9 &=\frac{D_0\,\gamma^{\text{Scalar QED}}}{N}\left[
  -\frac{(d-1)(d^2-2)}{2(d-2)d(d+2)}\!\left(\frac{1}{\Delta}-2\log p^2+2\Psi(d)\right)\right.
  \nonumber\\&\hspace{2.5cm}\left.
  +\frac{(d-1)(d^6-3d^5+4d^4-6d^3-28d^2+16d+64)}{(d-4)(d-2)^2 d^2(d+2)^2}
  -\frac{(d-1)\xi}{(d-2)d}
\right]
\,,\nn
D_{10} &=\frac{D_0\,\gamma^{\text{Scalar QED}}}{N}\left[
  -\frac{3(d-2)}{4(d-4)}\,\Theta(d)
  -\frac{2(d-1)}{(d-4)d}
  +\frac{(d-1)\xi}{(d-2)d}
\right]
\,.
\end{align}

\section{$\<JJ\>$ and $\<TT\>$ in QED$_d$-GNY model}
\label{JJ TT QED-GNY appendix}

For convenience, we define
\begin{align}
\cal N_{JJ}^{ab}=\frac{\G(1-\frac{d}{2})\G(\frac{d}{2})^2\rm{tr}\bf 1}{(4\p)^{\frac{d}{2}}(d-1)\G(d-2)}\rm{tr}(t^a t^b)
\,.
\end{align}
The results for Fig. \ref{JJ diagrams} are
\begin{align}
D_0 &= \mathcal{N}_{JJ}^{ab}
  \left(\frac{p_\mu p_\nu}{p^2} - \delta_{\mu\nu}\right)
  p^{2(\frac d 2-1)}\,,
\nn
D_1 &= \frac{\mathcal{N}_{JJ}^{ab}\,\gamma^{\text{QED-GNY}}}{N}
  \left[
    \left(
      \frac{(d-2)^2}{2(d-1)d}
      \left(\frac{1}{\Delta} - \log p^2 + \Psi(d)\right)
      + \frac{(d-2)(4d^2 - 9d + 4)}{2(d-1)^2 d^2}
    \right)
    \left(\frac{p_\mu p_\nu}{p^2} - \delta_{\mu\nu}\right) \right.\nn
    &\hspace{2.5cm}\left.- \frac{d-2}{2(d-1)d}\,\delta_{\mu\nu}
  \right]
  p^{2(\frac d 2-1)}\,,
\nn
D'_1 &= \frac{\mathcal{N}_{JJ}^{ab}\,\gamma^{\text{QED-GNY}}}{N}
  \left[
    \left(
      \left(\frac{(d-4)(d-1)}{2d} + \frac{\xi}{2}\right)
      \left(\frac{1}{\Delta} - \log p^2 + \Psi(d)\right)
      + \frac{4d^3 - 21d^2 + 36d - 16}{2(d-2)d^2}
      \right.\right.\nn
    &\hspace{2.5cm}\left.\left.- \frac{\xi}{2(d-2)(d-1)}
    \right)
    \left(\frac{p_\mu p_\nu}{p^2} - \delta_{\mu\nu}\right)
    + \left(
        - \frac{(d-4)(d-1)}{2(d-2)d}
        - \frac{\xi}{2(d-2)}
      \right)\delta_{\mu\nu}
  \right]
  p^{2(\frac d 2-1)}\,,
\nn
D_2 &= \frac{\mathcal{N}_{JJ}^{ab}\,\gamma^{\text{QED-GNY}}}{N}
  \left[
    \left(
      -\frac{(d-2)^2}{2(d-1)d}
      \left(\frac{1}{\Delta} - \log p^2 + \Psi(d)\right)
      + \frac{d-2}{2(d-1)^2 d}
    \right)
    \left(\frac{p_\mu p_\nu}{p^2} - \delta_{\mu\nu}\right)
    \right.\nn
    &\hspace{2.5cm}\left.+ \frac{d-2}{2(d-1)d}\,\delta_{\mu\nu}
  \right]
  p^{2(\frac d 2-1)}\,,
\nn
D’_2 &= \frac{\mathcal{N}_{JJ}^{ab}\,\gamma^{\text{QED-GNY}}}{N}
  \left[
    \left(
      \left(-\frac{(d-4)(d-1)}{2d} - \frac{\xi}{2}\right)
      \left(\frac{1}{\Delta} - \log p^2 + \Psi(d)\right)
      + \frac{3(d-2)}{4}\,\Theta(d)
      + \frac{d - 4}{2(d-2)d}
      \right.\right.\nn
    &\hspace{2.5cm}\left.\left.+ \frac{\xi}{2(d-2)(d-1)}
    \right)
    \left(\frac{p_\mu p_\nu}{p^2} - \delta_{\mu\nu}\right)
    + \left(
        \frac{(d-4)(d-1)}{2(d-2)d}
        + \frac{\xi}{2(d-2)}
      \right)\delta_{\mu\nu}
  \right]
  p^{2(\frac d 2-1)}
  \,.
\end{align}
To write the results for $\<TT\>$ diagrams, we define
\begin{align}
    \cal N_{TT}=\frac{\G(1-\frac d 2)\G(1+\frac d 2)^2\rm{tr}\bf 1}{2(4\p)^{\frac d 2}\G(d+2)}
    \,.
\end{align}
We also define the tensor structures
\begin{align}
\bf T'_{\m\n,\r\l} &=\d_{\m\r}\d_{\n\l}+\d_{\m\l}\d_{\n\r}-\frac 2 d \d_{\m\n}\d_{\r\l}
\,,\nn
\bf T''_{\m\n,\r\l} &=\frac{2 \delta _{\lambda  \mu } \delta _{\nu  \rho }}{d}+\frac{2 \delta _{\lambda  \nu } \delta _{\mu  \rho }}{d}-\frac{4 p_{\lambda } p_{\rho } \delta _{\mu 
   \nu }}{d p^2}-\frac{4 p_{\mu } p_{\nu } \delta _{\lambda  \rho }}{d p^2}+\frac{p_{\nu } p_{\rho } \delta _{\lambda  \mu }}{p^2}+\frac{p_{\mu } p_{\rho }
   \delta _{\lambda  \nu }}{p^2}+\frac{p_{\lambda } p_{\nu } \delta _{\mu  \rho }}{p^2}+\frac{p_{\lambda } p_{\mu } \delta _{\nu  \rho }}{p^2}
   \,.\label{T prime T double prime}
\end{align}
The results for Fig. \ref{TT diagrams} are
\begin{align}
D_0 &= N \left( \mathcal{N}_{TT} \,\bf{T}_{\mu\nu,\rho\lambda}\, (p^2)^{\frac d 2} \right)
\,,\nn
D_1 &= \mathcal{N}_{TT}\, \gamma^{\text{QED-GNY}} \left[ \left( \frac{(d-2)^2}{2(d-1)d} \left( \frac{1}{\Delta} - \log p^2 + \Psi(d) \right) + \frac{(d-2)(5d^3 - 8d^2 - 3d + 4)}{2(d-1)^2 d^2 (d+1)} \right) \bf{T}_{\mu\nu,\rho\lambda}\right.\nn
&\hspace{2.5cm}\left.- \frac{(d-2)(2d^2 - d + 2)}{2(d-1)d^3} \bf{T}'_{\mu\nu,\rho\lambda} + \frac{d-2}{2(d-1)d} \bf{T}''_{\mu\nu,\rho\lambda} \right] (p^2)^{\frac d 2}
\,,\nn
D'_1 &= \mathcal{N}_{TT}\, \gamma^{\text{QED-GNY}} \left[ \left( \left( \frac{(d-4)(d-1)}{2d} + \frac{\xi}{2} \right) \left( \frac{1}{\Delta} - \log p^2 + \Psi(d) \right) + \frac{5d^4 - 24d^3 + 29d^2 + 12d - 16}{2(d-2)d^2(d+1)}
\right.\right.\nn
&\hspace{2.5cm}\left.\left.+ \frac{(d^2 - 4d + 1)\xi}{2(d-2)(d-1)(d+1)} \right) \bf{T}_{\mu\nu,\rho\lambda} + \left( -\frac{(d-4)(d-1)(2d^2 - d + 2)}{2(d-2)d^3}
\right.\right.\nn
&\hspace{2.5cm}\left.\left.- \frac{(2d^2 - d + 2)\xi}{2(d-2)d^2} \right) \bf{T}'_{\mu\nu,\rho\lambda} + \left( \frac{(d-4)(d-1)}{2(d-2)d} + \frac{\xi}{2(d-2)} \right) \bf{T}''_{\mu\nu,\rho\lambda} \right] (p^2)^{\frac d 2}
\,,\nn
D_2 &= \mathcal{N}_{TT}\, \gamma^{\text{QED-GNY}} \left[ \left( -\frac{(d-2)^3}{2(d-1)d(d+2)} \left( \frac{1}{\Delta} - \log p^2 + \Psi(d) \right) - \frac{(d-2)^2(3d^2 - 4d - 1)}{2(d-1)^2 d(d+1)(d+2)} \right) \bf{T}_{\mu\nu,\rho\lambda}
\right.\nn
&\hspace{2.5cm}\left.+ \frac{(d-2)^2(2d^2 - d + 2)}{2(d-1)d^3(d+2)} \bf{T}'_{\mu\nu,\rho\lambda} - \frac{(d-2)^2}{2(d-1)d(d+2)} \bf{T}''_{\mu\nu,\rho\lambda} \right] (p^2)^{\frac d 2}
\,,\nn
D'_2 &= \mathcal{N}_{TT}\, \gamma^{\text{QED-GNY}} \left[ \left( \left( -\frac{d^3 - 7d^2 + 10d - 8}{2d(d+2)} - \frac{\xi}{2} \right) \left( \frac{1}{\Delta} - \log p^2 + \Psi(d) \right) + \frac{3(d-2)}{4} \Theta(d) 
\right.\right.\nn
&\hspace{2.5cm}\left.\left.- \frac{d^6-17 d^5+63 d^4-89 d^3+50 d^2+32 d-32}{2 (d-2) (d-1) d^2 (d+1) (d+2)} - \frac{(3d^3 - 8d^2 - d + 4)\xi}{2(d-2)(d-1)d(d+1)} \right) \bf{T}_{\mu\nu,\rho\lambda} 
\right.\nn
&\hspace{2.5cm}\left.+ \left( \frac{(d-1)(4d^4 - 13d^3 + 4d^2 - 28d + 16)}{2(d-2)d^3(d+2)} + \frac{(2d^2 - d + 2)\xi}{2(d-2)d^2} \right) \bf{T}'_{\mu\nu,\rho\lambda} 
\right.\nn
&\hspace{2.5cm}\left.+ \left( -\frac{d^4 - 4d^3 + 2d^2 - 2d + 4}{(d-2)d^2(d+2)} - \frac{\xi}{2(d-2)} \right) \bf{T}''_{\mu\nu,\rho\lambda} \right] (p^2)^{\frac d 2}
\,,\nn
D_3 &= \mathcal{N}_{TT}\, \gamma^{\text{QED-GNY}} \left[ \left( -\frac{(d-2)^2}{(d-1)d(d+2)} \left( \frac{1}{\Delta} - 2\log p^2 + 2\Psi(d) \right)+\frac{(d-2)^2}{(d-1)d(d+2)}\Ps(d) \right.\right.\nn
&\hspace{2.5cm}\left.\left.- \frac{(d-2)(d^3 + 2d^2 - 10d + 1)}{(d-1)^2 d(d+1)(d+2)} \right) \bf{T}_{\mu\nu,\rho\lambda} 
\right.\nn
&\hspace{2.5cm}\left.+ \frac{2(d-2)(2d^2 - d + 2)}{(d-1)d^3(d+2)} \bf{T}'_{\mu\nu,\rho\lambda} - \frac{2(d-2)}{(d-1)d(d+2)} \bf{T}''_{\mu\nu,\rho\lambda} \right] (p^2)^{\frac d 2}
\,,\nn
D'_3 &= \mathcal{N}_{TT}\, \gamma^{\text{QED-GNY}} \left[ \left( -\frac{(d-2)^2}{d(d+2)} \left( \frac{1}{\Delta} - 2\log p^2 + 2\Psi(d) \right) + \frac{2(d-2)}{d+2} \Psi(d) 
\right.\right.\nn
&\hspace{2.5cm}\left.\left.- \frac{(d-4)(d-2)^2 d}{16(d-3)(d-1)^5(d+1)} \,\mathcal{G}(d) - \frac{2(d-2)(d^3 + 2d^2 - 6d - 1)}{(d-1)d(d+1)(d+2)} + \frac{\xi}{d} \right) \bf{T}_{\mu\nu,\rho\lambda} 
\right.\nn
&\hspace{2.5cm}\left.+ \left( -\frac{(d-4)(d-2)^3(d+2)}{16(d-3)(d-1)^5 d^2} \,\mathcal{G}(d) - \frac{(d-2)(d-1)(d^2 + 4)}{d^3(d+2)} \right) \bf{T}'_{\mu\nu,\rho\lambda} 
\right.\nn
&\hspace{2.5cm}\left.+ \left( \frac{(d-4)(d-2)^3}{16(d-3)(d-1)^5 d} \,\mathcal{G}(d) + \frac{(d-2)(d^2 - d + 2)}{2d^2(d+2)} \right) \bf{T}''_{\mu\nu,\rho\lambda} \right] (p^2)^{\frac d 2}
\,,\nn
D_4 &= \mathcal{N}_{TT}\, \gamma^{\text{QED-GNY}} \left[ \left( \frac{2(d-2)}{d} \left( \frac{1}{\Delta} - \log p^2 + \Psi(d) \right) - \frac{2(d-2)(d^3 - 3d^2 - d + 1)}{(d-1)d^2(d+1)} + \frac{2\xi}{d} \right) \bf{T}_{\mu\nu,\rho\lambda} 
\right.\nn
&\hspace{2.5cm}\left.- \frac{2(d^3 + d^2 - 3d + 2)}{d^3} \bf{T}'_{\mu\nu,\rho\lambda} + \frac{(d-1)(d+2)}{d^2} \bf{T}''_{\mu\nu,\rho\lambda} \right] (p^2)^{\frac d 2}
\,,\nn
D_5 &= \mathcal{N}_{TT}\, \gamma^{\text{QED-GNY}} \left[ \left( -\frac{d-2}{d} \left( \frac{1}{\Delta} - 2\log p^2 + 2\Psi(d) \right) + \frac{(d-4)(d-2)^2 d}{8(d-3)(d-1)^5(d+1)} \,\mathcal{G}(d) 
\right.\right.\nn
&\hspace{2.5cm}\left.\left.+ \frac{2(d-2)(d^3 - 3d^2 - d + 1)}{(d-1)d^2(d+1)} - \frac{2\xi}{d} \right) \bf{T}_{\mu\nu,\rho\lambda} 
\right.\nn
&\hspace{2.5cm}\left.+ \left( \frac{(d-4)(d-2)^3(d+2)}{8(d-3)(d-1)^5 d^2} \,\mathcal{G}(d) + \frac{2(d^3 + d^2 - 3d + 2)}{d^3} \right) \bf{T}'_{\mu\nu,\rho\lambda} 
\right.\nn
&\hspace{2.5cm}\left.+ \left( -\frac{(d-4)(d-2)^3}{8(d-3)(d-1)^5 d} \,\mathcal{G}(d) - \frac{(d-1)(d+2)}{d^2} \right) \bf{T}''_{\mu\nu,\rho\lambda} \right] (p^2)^{\frac d 2}
\,,\nn
D_6 &= \mathcal{N}_{TT}\, \gamma^{\text{QED-GNY}} \left[ -\frac{\xi-1}{d} \bf{T}_{\mu\nu,\rho\lambda} + \frac{(d-2)(d+1)}{d^2} \bf{T}'_{\mu\nu,\rho\lambda} - \frac{(d-2)(d+1)}{2d^2} \bf{T}''_{\mu\nu,\rho\lambda} \right] (p^2)^{\frac d 2}
\,,\nn
D_7 &= \mathcal{N}_{TT}\, \gamma^{\text{QED-GNY}} \left[ \frac{\xi-1}{d} \bf{T}_{\mu\nu,\rho\lambda} - \frac{(d-2)(d+1)}{d^2} \bf{T}'_{\mu\nu,\rho\lambda} + \frac{(d-2)(d+1)}{2d^2} \bf{T}''_{\mu\nu,\rho\lambda} \right] (p^2)^{\frac d 2}
\,,\nn
D_8 &= \mathcal{N}_{TT}\, \gamma^{\text{QED-GNY}} \left[ -\frac{(d-4)(d-2)^2 d}{16(d-3)(d-1)^5(d+1)} \,\mathcal{G}(d)\, \bf{T}_{\mu\nu,\rho\lambda} - \frac{(d-4)(d-2)^3(d+2)}{16(d-3)(d-1)^5 d^2} \,\mathcal{G}(d)\, \bf{T}'_{\mu\nu,\rho\lambda} 
\right.\nn
&\hspace{2.5cm}\left.+ \frac{(d-4)(d-2)^3}{16(d-3)(d-1)^5 d} \,\mathcal{G}(d)\, \bf{T}''_{\mu\nu,\rho\lambda} \right] (p^2)^{\frac d 2}
\,.
\end{align}
Here and below, the function $\cal G(d)$ is defined as
\begin{align}
    \cal G(d)=-\frac{(d-1)^4(d+1)\G(3-\frac d 2)\G(\frac d 2 -2)^2}{8\G(d-4)}
    \,.
\end{align}

\section{$\<JJ\>$, $\<TT\>$, and $\<J^{\rm{top}}J^{\rm{top}}\>$ in scalar QED$_d$}
\label{JJ TT JJ top sQED appendix}

Here we define
\begin{align}
\cal N_{JJ}^{ab}=\frac{\G(2-\frac{d}{2})\G(\frac{d}{2}-1)^2}{(4\p)^{\frac{d}{2}}(d-1)\G(d-2)}\rm{tr}(t^a t^b)
\,.\label{NJJ sQED}
\end{align}
The Fig. \ref{JJ diagrams sQED} results are
\begin{align}
    D_0&=\mathcal{N}_{JJ}^{ab}\, p^{2(\frac d 2-1)}\left(\frac{p_\mu p_\nu}{p^2}-\delta_{\mu\nu}\right)
\,,\nn
D_1&= \frac{\mathcal{N}_{JJ}^{ab}\,\gamma^{\text{Scalar QED}}}{N}
\Bigg[
\left(
-\frac{d-4}{4(d-2)d}\left(\frac{1}{\Delta}-\log p^2+\Psi(d)\right)
-\frac{7d^3-36d^2+48d-16}{4(d-2)^2(d-1)d^2}
\right)\left(\frac{p_\mu p_\nu}{p^2}-\delta_{\mu\nu}\right) \nonumber\\
&\hspace{2.5cm}
+\frac{d-4}{4(d-2)^2 d}\,\delta_{\mu\nu}
\Bigg]p^{2(\frac d 2-1)}
\,,\nn
D'_1&= \frac{\mathcal{N}_{JJ}^{ab}\,\gamma^{\text{Scalar QED}}}{N}
\Bigg[
\Bigg(
\left(
-\frac{(d-1)^2}{(d-2)d}
+\frac{(d-1)\,\xi}{4(d-2)}
\right)
\left(\frac{1}{\Delta}-\log p^2+\Psi(d)\right)
-\frac{(d-1)(3d-2)}{(d-2)d^2} \nn
&\hspace{2.5cm}-\frac{d\,\xi}{4(d-2)^2}
\Bigg)\left(\frac{p_\mu p_\nu}{p^2}-\delta_{\mu\nu}\right)
+\left(
\frac{(d-1)^2}{(d-2)^2 d}
-\frac{(d-1)\,\xi}{4(d-2)^2}
\right)\delta_{\mu\nu}
\Bigg]p^{2(\frac d 2-1)}
\,,\nn
D_2&=\frac{\mathcal{N}_{JJ}^{ab}\,\gamma^{\text{Scalar QED}}}{N}
\Bigg[
\left(
-\frac{d-4}{4(d-2)d}\left(\frac{1}{\Delta}-\log p^2+\Psi(d)\right)
+\frac{3}{4(d-1)d}
\right)\left(\frac{p_\mu p_\nu}{p^2}-\delta_{\mu\nu}\right) \nn
&\hspace{2.5cm}-\frac{d-4}{4(d-2)^2 d}\,\delta_{\mu\nu}
\Bigg]p^{2(\frac d 2-1)}
\,,\nn
D'_2&= \frac{\mathcal{N}_{JJ}^{ab}\,\gamma^{\text{Scalar QED}}}{N}
\Bigg[
\Bigg(
-\frac{(d-1)\,\xi}{4(d-2)}
\left(\frac{1}{\Delta}-\log p^2+\Psi(d)\right)
+\frac{3(d-1)}{8}\Theta(d)
+\frac{(d-2)(d-1)}{2d^2} \nn
&\hspace{2.5cm}-\frac{\xi}{4(d-2)}
\Bigg)\left(\frac{p_\mu p_\nu}{p^2}-\delta_{\mu\nu}\right)
+\left(
\frac{(d-1)^2}{2(d-2)d}
+\frac{(d-1)\,\xi}{4(d-2)^2}
\right)\delta_{\mu\nu}
\Bigg]p^{2(\frac d 2-1)}
\,,\nn
D_3&= \frac{\mathcal{N}_{JJ}^{ab}\,\gamma^{\text{Scalar QED}}}{N}
\Bigg[
\left(
\frac{(d-1)^2}{(d-2)d}
\left(\frac{1}{\Delta}-\log p^2+\Psi(d)\right)
-\frac{(d-1)(d^3-5d^2+9d-4)}{(d-2)^2 d^2}
+\frac{(d-1)\,\xi}{(d-2)^2}
\right) \nn
&\hspace{2.5cm}
\times \left(\frac{p_\mu p_\nu}{p^2}-\delta_{\mu\nu}\right)
-\frac{(d-1)^2}{(d-2)^2}\,\delta_{\mu\nu}
\Bigg]p^{2(\frac d 2-1)}
\,,\nn
D_4&=\frac{\mathcal{N}_{JJ}^{ab}\,\gamma^{\text{Scalar QED}}}{N}\Bigg[
\left(
\frac{d-1}{2(d-2)^2}
-\frac{(d-1)\,\xi}{2(d-2)^2}
\right)\left(\frac{p_\mu p_\nu}{p^2}-\delta_{\mu\nu}\right)
+\frac{(d-1)^2}{2(d-2)^2}\,\delta_{\mu\nu}
\Bigg]p^{2(\frac d 2-1)}
\,.
\end{align}
For the $\<TT\>$ diagrams, we define
\begin{align}
    \cal N_{TT}=\frac{2\G(1-\frac d 2)\G(1+\frac d 2)^2}{(4\p)^{\frac d 2}(d-1)\G(d+2)}
    \,.
\end{align}
We use the extra tensor structures defined in \eqref{T prime T double prime}, and the results for Fig. \ref{TT diagrams sQED} are
\begin{align}
D_0 &= N\(\mathcal{N}_{TT}\, \bf{T}_{\m\n,\r\l}\, (p^2)^{\frac d 2}\)
\,,\nn
D_1 &= \mathcal{N}_{TT}\,\gamma^{\text{Scalar QED}} \Biggl[
\left(
  -\frac{d-4}{4(d-2)\,d}\Bigl(\frac{1}{\Delta} - \log p^2 + \Psi(d)\Bigr)
  - \frac{8d^4 - 37d^3 + 29d^2 + 28d - 16}{4(d-2)^2(d-1)d^2(1+d)}
\right)\bf{T}_{\m\n,\r\l}
\nn
  &\hspace{2.5cm}
+ \frac{(d-4)(2d^2 - d + 2)}{4(d-2)^2\,d^3}\,\bf{T}'_{\m\n,\r\l}
- \frac{d-4}{4(d-2)^2\,d}\,\bf{T}''_{\m\n,\r\l}
\Biggr]\, (p^2)^{\frac d 2}
\,,\nn
D'_1 &= \mathcal{N}_{TT}\,\gamma^{\text{Scalar QED}} \Biggl[
\left(
  \left(-\frac{(d-1)^2}{(d-2)d} + \frac{(d-1)\xi}{4(d-2)}\right)
  \Bigl(\frac{1}{\Delta} - \log p^2 + \Psi(d)\Bigr)
  - \frac{(d-1)(4d^3 - 9d^2 - 3d + 4)}{(d-2)^2 d^2(1+d)}\right.\nn
  &\hspace{2.5cm}\left.- \frac{(5d-1)\xi}{4(d-2)^2(1+d)}
\right)\bf{T}_{\m\n,\r\l}
+ \left(
  \frac{(d-1)^2(2d^2 - d + 2)}{(d-2)^2 d^3}
  - \frac{(d-1)(2d^2 - d + 2)\xi}{4(d-2)^2 d^2}
\right)\bf{T}'_{\m\n,\r\l}
\nn
  &\hspace{2.5cm}
+ \left(
  -\frac{(d-1)^2}{(d-2)^2 d}
  + \frac{(d-1)\xi}{4(d-2)^2}
\right)\bf{T}''_{\m\n,\r\l}
\Biggr]\, (p^2)^{\frac d 2}
\,,\nn
D_2 &= \mathcal{N}_{TT}\,\gamma^{\text{Scalar QED}} \Biggl[
\left(
  \frac{d-4}{4(d+2)\,d}\Bigl(\frac{1}{\Delta} - \log p^2 + \Psi(d)\Bigr)
  + \frac{6d^3 - 25d^2 + 15d + 16}{4(d-2)(d-1)d(1+d)(d+2)}
\right)\bf{T}_{\m\n,\r\l}
\nn
  &\hspace{2.5cm}
- \frac{(d-4)(2d^2 - d + 2)}{4(d-2)d^3(d+2)}\,\bf{T}'_{\m\n,\r\l}
+ \frac{d-4}{4(d-2)d(d+2)}\,\bf{T}''_{\m\n,\r\l}
\Biggr]\, (p^2)^{\frac d 2}
\,,\nn
D'_2 &= \mathcal{N}_{TT}\,\gamma^{\text{Scalar QED}} \Biggl[
\left(
  \left(\frac{2(d-1)}{(d-2)d(d+2)} - \frac{(d-1)\xi}{4(d-2)}\right)
  \Bigl(\frac{1}{\Delta} - \log p^2 + \Psi(d)\Bigr)
  + \frac{3(d-1)}{8}\,\Theta(d)
\right.\nn
  &\hspace{2.5cm}\left.
  + \frac{d^7-3 d^6-3 d^5+15 d^4+6 d^3-48 d^2-4 d+24}{(d-2)^2 d^2 (d+1) (d+2)^2}
  - \frac{(4d^3 - 9d^2 - 3d + 4)\xi}{4(d-2)^2 d(1+d)}
\right)\bf{T}_{\m\n,\r\l}
\nn
  &\hspace{2.5cm}
+ \left(
  \frac{(d-1)^2(d^3 - d^2 - 3d - 6)}{(d-2)^2 d^2(d+2)}
  + \frac{(d-1)(2d^2 - d + 2)\xi}{4(d-2)^2 d^2}
\right)\bf{T}'_{\m\n,\r\l}
\nn
  &\hspace{2.5cm}
+ \left(
  -\frac{(d-1)^2(d^3 - d^2 - 4d - 4)}{2(d-2)^2 d^2(d+2)}
  - \frac{(d-1)\xi}{4(d-2)^2}
\right)\bf{T}''_{\m\n,\r\l}
\Biggr]\, (p^2)^{\frac d 2}
\,,\nn
D_3 &= \mathcal{N}_{TT}\,\gamma^{\text{Scalar QED}} \Biggl[
\left(
  \frac{d-4}{2(d-2)d(d+2)}\Bigl(\frac{1}{\Delta} - 2\log p^2 + 2\Psi(d)\Bigr)
  - \frac{d-4}{2(d-2)d(d+2)}\,\Psi(d)
  \right.\nn
  &\hspace{2.5cm}\left.+ \frac{d^4 + 12d^3 - 77d^2 + 72d + 40}{4(d-2)^2(d-1)d(1+d)(d+2)}
\right)\bf{T}_{\m\n,\r\l}
- \frac{(d-4)(2d^2 - d + 2)}{(d-2)^2 d^3(d+2)}\,\bf{T}'_{\m\n,\r\l}
\nn
  &\hspace{2.5cm}
+ \frac{d-4}{(d-2)^2 d(d+2)}\,\bf{T}''_{\m\n,\r\l}
\Biggr]\, (p^2)^{\frac d 2}
\,,\nn
D'_3 &= \mathcal{N}_{TT}\,\gamma^{\text{Scalar QED}} \Biggl[
\left(
  \frac{(d-1)(d^2 + d - 4)}{2(d-2)d(d+2)}\Bigl(\frac{1}{\Delta} - 2\log p^2 + 2\Psi(d)\Bigr)
  - \frac{(d-4)\,d}{16(d-3)(d-1)^3(1+d)}\,\mathcal{G}(d)
\right.\nn
  &\hspace{2.5cm}\left.
  + \frac{(d-1)^2}{2(d-2)(d+2)}\,\Psi(d)
  - \frac{3d^7 - 15d^6 - 9d^5 + 79d^4 + 46d^3 - 168d^2 - 16d + 32}{4(d-2)^2 d^2(1+d)(d+2)^2}
  \right.\nn
  &\hspace{2.5cm}\left.
  + \frac{(d-1)^2 \xi}{(d-2)^2 d}
\right)\bf{T}_{\m\n,\r\l}
+ \left(
  -\frac{(d-1)^3(d^3 + 2d^2 + 2d - 4)}{(d-2)^2 d^3(d+2)}
  \right.\nn
  &\hspace{2.5cm}\left.- \frac{(d-4)(d-2)(d+2)}{16(d-3)d^2(d-1)^3}\,\mathcal{G}(d)
\right)\bf{T}'_{\m\n,\r\l}
+ \left(
  \frac{(d-1)^2(d^3 + d^2 - 4)}{2(d-2)^2 d^2(d+2)}
  \right.\nn
  &\hspace{2.5cm}\left.+ \frac{(d-4)(d-2)}{16(d-3)d(d-1)^3}\,\mathcal{G}(d)
\right)\bf{T}''_{\m\n,\r\l}
\Biggr]\, (p^2)^{\frac d 2}
\,,\nn
D_4 &= \mathcal{N}_{TT}\,\gamma^{\text{Scalar QED}} \Biggl[
\left(
  \frac{2(d-1)(d^2 - 2)}{(d-2)d(d+2)}\Bigl(\frac{1}{\Delta} - \log p^2 + \Psi(d)\Bigr)
  \right.\nn
  &\hspace{2.5cm}\left.- \frac{2d^7 - 8d^6 - 7d^5 + 46d^4 + 17d^3 - 74d^2 - 12d + 24}{(d-2)^2 d^2(1+d)(d+2)^2}
  + \frac{2(d-1)^2 \xi}{(d-2)^2 d}
\right)\bf{T}_{\m\n,\r\l}
\nn
  &\hspace{2.5cm}
- \frac{(d-1)^2(2d^3 - 5d + 2)}{(d-2)^2 d^3}\,\bf{T}'_{\m\n,\r\l}
+ \frac{(d-1)^2(d^2 - 2)}{(d-2)^2 d^2}\,\bf{T}''_{\m\n,\r\l}
\Biggr]\, (p^2)^{\frac d 2}
\,,\nn
D_5 &= \mathcal{N}_{TT}\,\gamma^{\text{Scalar QED}} \Biggl[
\left(
  -\frac{(d-1)(d^2 - 2)}{(d-2)d(d+2)}\Bigl(\frac{1}{\Delta} - 2\log p^2 + 2\Psi(d)\Bigr)
  + \frac{(d-4)\,d}{8(d-3)(d-1)^3(1+d)}\,\mathcal{G}(d)
\right.\nn
  &\hspace{2.5cm}\left.
  + \frac{2d^7 - 8d^6 - 7d^5 + 46d^4 + 17d^3 - 74d^2 - 12d + 24}{4(d-2)^2 d^2(1+d)(d+2)^2}
  - \frac{2(d-1)^2 \xi}{(d-2)^2 d}
\right)\bf{T}_{\m\n,\r\l}
\nn
  &\hspace{2.5cm}
+ \left(
  \frac{(d-1)^2(2d^3 - 5d + 2)}{(d-2)^2 d^3}
  + \frac{(d-4)(d-2)(d+2)}{8(d-3)(d-1)^3 d^2}\,\mathcal{G}(d)
\right)\bf{T}'_{\m\n,\r\l}
\nn
  &\hspace{2.5cm}
+ \left(
  -\frac{(d-1)^2(d^2 - 2)}{(d-2)^2 d^2}
  - \frac{(d-4)(d-2)}{8(d-3)d(d-1)^3}\,\mathcal{G}(d)
\right)\bf{T}''_{\m\n,\r\l}
\Biggr]\, (p^2)^{\frac d 2}
\,,\nn
D_6 &= \mathcal{N}_{TT}\,\gamma^{\text{Scalar QED}} \left[
\left(
  \frac{(d-1)^2}{(d-2)^2 d}
  - \frac{(d-1)^2 \xi}{(d-2)^2 d}
\right)\bf{T}_{\m\n,\r\l}
+ \frac{(d-1)^2(1+d)}{(d-2)d^2}\,\bf{T}'_{\m\n,\r\l}
\right.\nn
  &\hspace{2.5cm}\left.- \frac{(d-1)^2(1+d)}{2(d-2)d^2}\,\bf{T}''_{\m\n,\r\l}
\right] (p^2)^{\frac d 2}
\,,\nn
D_7 &= \mathcal{N}_{TT}\,\gamma^{\text{Scalar QED}} \left[
\left(
  -\frac{(d-1)^2}{(d-2)^2 d}
  + \frac{(d-1)^2 \xi}{(d-2)^2 d}
\right)\bf{T}_{\m\n,\r\l}
- \frac{(d-1)^2(1+d)}{(d-2)d^2}\,\bf{T}'_{\m\n,\r\l}
\right.\nn
  &\hspace{2.5cm}\left.+ \frac{(d-1)^2(1+d)}{2(d-2)d^2}\,\bf{T}''_{\m\n,\r\l}
\right] (p^2)^{\frac d 2}
\,,\nn
D_8 &= \mathcal{N}_{TT}\,\gamma^{\text{Scalar QED}} \left[
  - \frac{(d-4)\,d}{16(d-3)(d-1)^3(1+d)}\,\mathcal{G}(d)\,\bf{T}_{\m\n,\r\l}
  - \frac{(d-4)(d-2)(d+2)}{16(d-3)(d-1)^3 d^2}\,\mathcal{G}(d)\,\bf{T}'_{\m\n,\r\l}
  \right.\nn
  &\hspace{2.5cm}\left.+ \frac{(d-4)(d-2)}{16(d-3)d(d-1)^3}\,\mathcal{G}(d)\,\bf{T}''_{\m\n,\r\l}
\right] (p^2)^{\frac d 2}
\,.
\end{align}
Omitting the external legs, the results for the three-loop diagrams in Fig. \ref{Jtop diagrams sQED} are
\begin{align}
D_3&= \mathcal{N}_{JJ}\, \gamma^{\text{Scalar QED}} \Bigg[
\left(
  \frac{(d-1)^2}{2(d-2)d}
  \left(\frac{1}{\Delta} - 2\log p^2 + 2\Psi(d)\right)
  + \frac{(d-1)(d^2 - 10d + 8)}{2(d-2)^2 d^2}
  + \frac{(d-1)\,\xi}{2(d-2)^2}
\right)\nn
&\hspace{2.5cm}\times\left(\frac{p_{\mu} p_{\nu}}{p^2} - \delta_{\mu\nu}\right)
- \frac{(d-1)^2}{2(d-2)^2}\,\delta_{\mu\nu}
\Bigg] p^{2(\frac d 2-1)} 
\,,\nn
D_5&= \mathcal{N}_{JJ}\, \gamma^{\text{Scalar QED}} \Bigg[
\Big(
  -\frac{(d-1)^2}{2(d-2)d}
  \left(\frac{1}{\Delta} - 2\log p^2 + 2\Psi(d)\right)
  + \frac{(d-1)(d^3 - 5d^2 + 9d - 4)}{(d-2)^2 d^2} \nn
  &\hspace{2.5cm}- \frac{(d-1)\,\xi}{(d-2)^2}
\Big)
\left(\frac{p_{\mu} p_{\nu}}{p^2} - \delta_{\mu\nu}\right)
+ \frac{(d-1)^2}{(d-2)^2}\,\delta_{\mu\nu}
\Bigg] p^{2(\frac d 2-1)}
\,,\nn
D_7&= \mathcal{N}_{JJ}\, \gamma^{\text{Scalar QED}} \left[
\left(
  -\frac{d-1}{2(d-2)^2}
  + \frac{(d-1)\,\xi}{2(d-2)^2}
\right)
\left(\frac{p_{\mu} p_{\nu}}{p^2} - \delta_{\mu\nu}\right)
- \frac{(d-1)^2}{2(d-2)^2}\,\delta_{\mu\nu}
\right] p^{2(\frac d 2-1)}.
\end{align}
Here $\cal N_{JJ}$ corresponds to \eqref{NJJ sQED} without the trace,
\begin{align}
\cal N_{JJ}=\frac{\G(2-\frac{d}{2})\G(\frac{d}{2}-1)^2}{(4\p)^{\frac{d}{2}}(d-1)\G(d-2)}
\,.
\end{align}

\bibliographystyle{JHEP}
\bibliography{references}

\end{document}